\documentclass[aps,pra,reprint,amsmath,amssymb,superscriptaddress,onecolumn,longbibliography,nofootinbib,notitlepage]{revtex4-2}

\usepackage{mathtools}
\usepackage{siunitx}
\usepackage{hyperref}
\usepackage{caption}
\usepackage{float}
\usepackage{subcaption}
\usepackage[braket, qm]{qcircuit}
\usepackage{titletoc}
\usepackage{tocloft}
\usepackage{multirow}
\usepackage{footnote}
\usepackage{hhline}
\usepackage{tablefootnote}

\newcommand\chitwo{0.47}
\newcommand\chitwomax{1.1}
\newcommand\wpixel{3.772}
\newcommand\resolution{7.5}

\usepackage{xcolor}

\begin{document}
\title{Programmable on-chip nonlinear photonics}
\author{Ryotatsu~Yanagimoto}
\thanks{Contact: ryotatsu.yanagimoto@ntt-research.com, pmcmahon@cornell.edu}
\affiliation{School of Applied and Engineering Physics, Cornell University, Ithaca, NY 14853, USA}
\affiliation{Physics \& Informatics Laboratories, NTT Research, Inc., Sunnyvale, CA 94085, USA}

\author{Benjamin~A.~Ash}
\affiliation{School of Applied and Engineering Physics, Cornell University, Ithaca, NY 14853, USA}

\author{Mandar~M.~Sohoni}
\affiliation{School of Applied and Engineering Physics, Cornell University, Ithaca, NY 14853, USA}

\author{Martin~M.~Stein}
\thanks{Present address: Department of Applied Physics, Yale University, New Haven, CT 06520, USA}
\affiliation{School of Applied and Engineering Physics, Cornell University, Ithaca, NY 14853, USA}

\author{Yiqi~Zhao}
\affiliation{School of Applied and Engineering Physics, Cornell University, Ithaca, NY 14853, USA}

\author{Federico~Presutti}
\thanks{Present address: Research Laboratory of Electronics, Massachusetts Institute of Technology, Cambridge, MA 02139, USA}
\affiliation{School of Applied and Engineering Physics, Cornell University, Ithaca, NY 14853, USA}

\author{Marc~Jankowski}
\affiliation{Physics \& Informatics Laboratories, NTT Research, Inc., Sunnyvale, CA 94085, USA}
\affiliation{E.\,L. Ginzton Laboratory, Stanford University, Stanford, CA 94350, USA}

\author{Logan~G.~Wright}
\thanks{Present address: Department of Applied Physics, Yale University, New Haven, CT 06520, USA}
\affiliation{School of Applied and Engineering Physics, Cornell University, Ithaca, NY 14853, USA}
\affiliation{Physics \& Informatics Laboratories, NTT Research, Inc., Sunnyvale, CA 94085, USA}

\author{Tatsuhiro~Onodera}
\affiliation{School of Applied and Engineering Physics, Cornell University, Ithaca, NY 14853, USA}
\affiliation{Physics \& Informatics Laboratories, NTT Research, Inc., Sunnyvale, CA 94085, USA}

\author{Peter~L.~McMahon}
\thanks{Contact: ryotatsu.yanagimoto@ntt-research.com, pmcmahon@cornell.edu}
\affiliation{School of Applied and Engineering Physics, Cornell University, Ithaca, NY 14853, USA}
\affiliation{Kavli Institute at Cornell for Nanoscale Science, Cornell University, Ithaca, NY 14853, USA}

\begin{abstract}
Nonlinear photonics uses coherent interactions between optical waves to engineer functionality that is not possible with purely linear optics. Traditionally, the function of a nonlinear-optical device is determined during design and fixed during fabrication, which limits the scope and flexibility of its use. In this paper, we present a photonic device with highly programmable nonlinear functionality: an optical slab waveguide with an arbitrarily reconfigurable two-dimensional distribution of $\chi^{(2)}$ nonlinearity. The nonlinearity is realized using electric-field-induced $\chi^{(2)}$ in a $\chi^{(3)}$ material. The programmability is engineered by massively parallel control of the electric-field distribution within the device using a photoconductive layer and optical programming with a spatial light pattern. To showcase the versatility of our device, we demonstrated spectral, spatial, and spatio-spectral engineering of second-harmonic generation by tailoring arbitrary quasi-phase-matching (QPM) grating structures in two dimensions. Second-harmonic light was generated with programmable spectra, enabled by real-time in situ inverse design of QPM gratings on our prototype device. Flexible spatial control was also achieved, including the generation of complex waveforms such as Airy beams and the simultaneous engineering of spectral and spatial features. This allowed us to create distinct spatial light profiles across multiple wavelengths. The programmability also allowed us to demonstrate in situ, real-time compensation of fluctuations in pump laser wavelength. Our work shows that we can transcend the conventional one-device--one-function paradigm, expanding the potential applications of nonlinear optics in situations where fast device reconfigurability is not merely practically convenient but essential---such as in programmable optical quantum gates and quantum light sources, all-optical signal processing, optical computation, and adaptive structured light for sensing.
\end{abstract}

\maketitle

\section{Introduction}
\label{sec:intro}

Second-harmonic generation (SHG) was first demonstrated in 1961~\cite{Franke1961SHG}. Since then, the scope of nonlinear optics (NLO) has expanded far beyond the frequency doubling of single-frequency lasers. Today, NLO encompasses a diverse range of processes, including sum-frequency, difference-frequency, and parametric generation, and has become an essential tool for accessing ever-wider swaths of the electromagnetic spectrum~\cite{Giordmaine1965OPO,Sutherland2003EFISH-book, Boes2023LN, Herter2023THz}. Complex phenomena have also emerged from simultaneous nonlinear interactions between waves of light across broad bands. These phenomena include a rich variety of solitons~\cite{Hasegawa1973soliton, Kibler2010Peregrine, Buryak2002soliton, Kivshar1998DarkSolitons}, modulation instability, and rogue waves~\cite{Solli2007Rogue}. Leveraging these phenomena has enabled various key optical technologies, including mode-locked lasers~\cite{Haus2000modelocking,Grelu2012mode-lock}, optical frequency combs~\cite{Diddams2020comb, Fortier2019Comb, Shen2020SolitonComb}, and supercontinuum light sources~\cite{Dudley2006SCG}. The coherent nature of NLO can also generate and manipulate quantum states of light~\cite{Nehra2022squeezing, Dutt2015Squeezing, Yanagimoto2024mesoscopic, RoquesCarmes2023OPO}, and has played central roles in fundamental quantum physics~\cite{Hong1987HOM, Bouwmeester1997teleportation, Giustina2015Bell}, optical quantum computation~\cite{Brecht2015temporal, Takeda2019Review, Langford2011CPC, Zhong2020GBS}, and metrology~\cite{Vahlbruch2016Squeezing, Jia2024LIGO}.
 
The NLO processes used in nonlinear photonics are usually \emph{not} accessible in raw materials because, in their natural state, raw materials do not satisfy the conditions for phase-matching, which is essential for an NLO process to be efficient~\cite{Boyd2008nonlinear}. However, the phase-matching condition can be engineered to precisely control how and which wavelengths of light interact---even quantum mechanically~\cite{Ansari2018review, Lu2023Frequency-Bin, Weiss2025QPM-review}. This control of phase-matching is an exceptionally powerful engineering framework in modern NLO, and various techniques for phase-matching have been invented~\cite{Giordmaine1962Birefringence, Du2023modal, Agha2007FWM-phase-matching, Coen2002Supercontinuum-Phase-matching, Gagnon2022RelaxedPM}. A particularly efficacious and flexible technique is quasi-phase-matching (QPM). In QPM, a periodic spatial modulation of $\chi^{(2)}$ optical nonlinearity, which is referred to as a QPM grating, compensates for phase mismatch among interacting light waves~\cite{Hum2006QPM-review, Hu2013QPM-review, Byer1997QPM-review}. A simple QPM grating enables highly efficient coherent-wave mixing while more complicated QPM gratings can realize highly nontrivial nonlinear-optical functions. Many exotic QPM grating structures have been explored, enabling a wide range of functions including high-harmonic generation~\cite{Chen2015HHG,Zhu1997THG}, arbitrary pulse shaping~\cite{Imeshev1998PulseShaping}, quantum-pulse gates~\cite{Laura2023QPM}, and holographic generation of structured light~\cite{Ellenbogen2009Airy, Dolev2010Airy, Fang2020slab}.
 
These functions of an NLO device, including engineering phase-matching, often require special structures to be ``sculpted'' in the raw nonlinear-optical material, and this is usually done via nanofabrication processes~\cite{Koenderink2015Nanophotonics}. For instance, a QPM grating can be formed by periodically inverting the nonlinear susceptibility of a material. Fabricating such a structure requires sophisticated techniques---for example, epitaxial growth of semiconductors on orientation-patterned wafers~\cite{Yoo1996OrientationPatterning} or electric domain inversion of ferroelectric materials via lithographically patterned electrodes~\cite{Boes2021LN}.
Although this sculpting-based paradigm has driven decades of progress in nonlinear photonics, it restricts the design and use of photonic devices because each device is typically optimized for one function that is fixed when the device is made. The device's performance is also very sensitive to fabrication imperfections and environmental drifts~\cite{Chen2023AdaptedPoling}, which lowers the yield of correctly functioning devices. If only one working device is needed, this low yield can be mitigated by testing many devices and post-selecting a suitable one. However, if individual components have low yields, it is prohibitively challenging to produce integrated-photonic systems in which many NLO components must work together on a single chip because the system yield declines exponentially as a function of the number of constituent components.

\begin{figure}[b]
    \centering
    \includegraphics[width=1.0\linewidth]{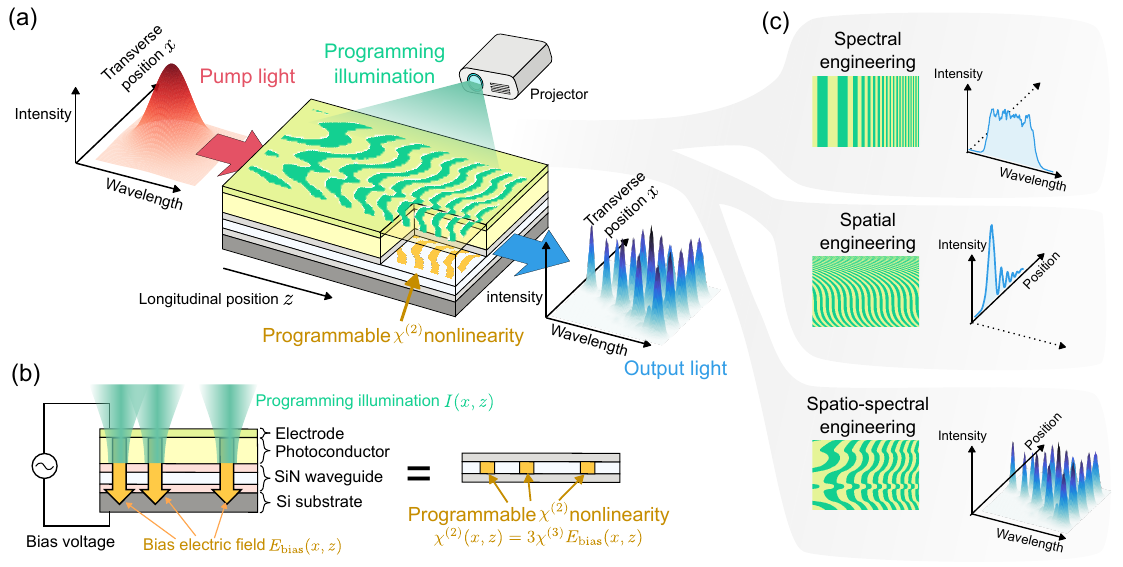}
    \caption{\textbf{Illustration of the working principle and capabilities of a programmable nonlinear waveguide.} (a) Structured light projected onto the surface of a planar waveguide plays the role of a programming illumination $I(x,z)$, inducing the same pattern of $\chi^{(2)}$ nonlinearity, $\chi^{(2)}(x,z)$, which allows versatile control of broadband SHG via quasi-phase-matching. Here, $x$ and $z$ are the transverse and longitudinal positions on the waveguide. (b) The structure and physical working mechanism of a programmable waveguide. The device is composed of a SiN waveguide ($\SI{2.05}{\micro m}$ thick SiN core and $\SI{1}{\micro m}$ thick SiO$_2$ cladding at the top and bottom), photoconductor layer ($\SI{7.5}{\micro m}$ thick silicon-rich silicon nitride), and transparent electrode ($\SI{20}{nm}$ thick indium tin oxide). The photoconductor, when illuminated by green ($\SI{532}{nm}$) laser light, becomes locally conductive, letting the external bias electric field through to the waveguide core. The resulting spatially shaped $E_\text{bias}(x,z)$ induces a spatially programmable $\chi^{(2)}$ nonlinearity according to $\chi^{(2)}(x,z) = 3\chi^{(3)}E_\text{bias}(x,z)$. See Appendix~\ref{appendix:device-fabrication} and \ref{appendix:electric-properties} for details on the device fabrication and electrical characterization of the device, respectively. (c) Varying the longitudinal and transverse structure of quasi-phase-matching gratings enables spectral and spatial control of NLO, respectively. By programming the full two-dimensional structure of $\chi^{(2)}(x,z)$, we can simultaneously engineer the spectral and spatial structure of the generated output light.}
    \label{fig:figure1}
\end{figure}

We present an approach that avoids many of these disadvantages and opens up fundamentally new possibilities: NLO based on a \textit{programmable} nonlinear waveguide. Our proposed device is a planar optical waveguide whose two-dimensional distribution of $\chi^{(2)}$ optical nonlinearity, $\chi^{(2)}(x,z)$, can be arbitrarily programmed, i.e., dynamically set and updated ($x$ and $z$ are the transverse and longitudinal dimensions of the waveguide, respectively; see Fig.~\ref{fig:figure1}(a)). Such programmable $\chi^{(2)}$ nonlinearity allows flexible engineering of QPM gratings to perform various NLO functions with a single device. The programmable $\chi^{(2)}$ nonlinearity is induced by biasing the $\chi^{(3)}$ nonlinearity by an electric field $E_\text{bias}(x,z)$, leading to $\chi^{(2)}(x,z) = 3\chi^{(3)}E_\text{bias}(x,z)$. Electric-field-induced $\chi^{(2)}$ nonlinearity was initially conceived to characterize materials and interfaces~\cite{Maker1965EFISH, Oudar1975EFISH,Lpke1999EFISH} and has recently been used to engineer NLO processes~\cite{Timurdogan2017E-FISH, Heydari2023E-FISH, Nitiss2022Photogalvanic, Lu2020Photogalvanic, Billat2017Photogalvanic, Hickstein2019PhotogalvanicWG, Li2025SPDC-PhotoGalvanic}. In these prior works, the bias field was applied via lithographically patterned electrodes~\cite{Timurdogan2017E-FISH, Heydari2023E-FISH} or all-optically~\cite{Nitiss2022Photogalvanic, Lu2020Photogalvanic, Billat2017Photogalvanic, Hickstein2019PhotogalvanicWG, Li2025SPDC-PhotoGalvanic}. The all-optical approach reconfigured the spatial pattern of nonlinearity depending on how the device was optically pumped, allowing phase-matching to be engineered for a variety of NLO processes. However, completely arbitrary spatial patterns of $\chi^{(2)}$ nonlinearity have not been realized, and this challenge also applies to other approaches to tuning $\chi^{(2)}$ nonlinearity, such as using ferroelectric nematic liquid crystals~\cite{Zhao2022LiquidQPM, Sultanov2024Liquid}. Reconfigurable QPM structures have also been limited to one-dimensional geometries. While two-dimensional control of $\chi^{(2)}$ nonlinearity would enable spatial control of NLO, it is more challenging because there are far more degrees of freedom in two dimensions than in one dimension.

In our planar waveguide device, we use lithography-free photoconductive electrodes and patterned optical illumination (see Sec.~\ref{sec:working-principle}) to program arbitrary spatial patterns of bias fields, $E_\text{bias}(x,z)$. This programmable bias field produces a corresponding programmable $\chi^{(2)}(x,z)$. Patterned optical illumination has previously been used to program the real~\cite{Onodera2024Linear} and imaginary~\cite{Wu2023Gain} parts of the refractive-index distribution of planar waveguides. Here, we demonstrate a programmable $\chi^{(2)}(x,z)$ nonlinearity with a dynamic range (i.e., contrast excluding constant background) of $\SI{\chitwo}{pm/V}$, a spatial resolution of $\SI{\resolution}{\micro m}$, and a functional area ($z \times x$) of approximately $\SI{0.7}{cm}\times \SI{0.4}{cm}$ (corresponding to $\SI{5e5}{}$ programmable pixels), with updates possible every second (see Appendix~\ref{appendix:device-performance} for a summary of the performance of this first prototype device and the potential for future improvements). Using this full two-dimensional programmability, we experimentally realized complex QPM structures and demonstrated highly flexible control over the spectral, spatial, and spatio-spectral dynamics of broadband SHG. Moreover, the real-time reconfigurability of the device enables in situ inverse design and optimization of QPM grating structures, allowing us to engineer very unusual optical spectral and spatial shapes in a way that is robust to experimental imperfections. Notably, all our results were achieved with a single programmable nonlinear waveguide design---a device with this design can arbitrarily switch between all the different demonstrated modes of operation.

\section{Design and operating principle of the device}
\label{sec:working-principle}

Our programmable nonlinear waveguide and how we realized arbitrary two-dimensional distributions of nonlinearity $\chi^{(2)}(x,z)$ is illustrated in Fig.~\ref{fig:figure1}. Our programmable nonlinear waveguide was composed of several layers (Fig.~\ref{fig:figure1}(b), see Appendix~\ref{appendix:device-fabrication} for full details about device fabrication). The waveguide was made on a conductive silicon substrate. On top of the substrate was a silicon nitride (SiN) optical waveguide comprising silicon dioxide ($\text{SiO}_2$) cladding layers and a SiN core layer. On top of the upper cladding layer was a layer of photoconductive material---silicon-rich silicon nitride (SRN). Finally, a transparent electrode was deposited on the photoconductor layer. During operation, a bias electric field was applied across the entire stack by connecting a voltage source to the substrate and the top electrode.

To realize a programmable $\chi^{(2)}(x,z)$, we shone structured light with a spatial intensity pattern $I(x,z)$ onto the top of the device. The photoconductor layer became conductive where light intensity was highest and let the electric field from the bias voltage through to the waveguide SiN core layer.\footnote{This is a very heuristic picture of the mechanism, to convey intuition. A more precise picture can be obtained by modeling the system as a voltage divider; this is given in Appendix~\ref{appendix:electric-properties}.} As a result, the pattern of the programming illumination $I(x,z)$ resulted in a pattern of the bias field $E_\text{bias}(x,z)$ inside the core. This bias field in turn induced an effective $\chi^{(2)}$ optical nonlinearity via $\chi^{(2)}(x,z) = 3\chi^{(3)}E_\text{bias}(x,z)$. The central operating principle of the programmability of our device can be summarized as: patterned illumination $I(x,z)$ on the photoconductor layer induces a spatial pattern of electric field $E_\text{bias}(x,z)$ inside the waveguide core, which then induces a spatial pattern of optical nonlinearity $\chi^{(2)}(x,z)$.

In conventional QPM, $\chi^{(2)}(x,z)$ takes on both positive and negative values---usually of the same magnitude but where the sign is alternated. In our approach, where $\chi^{(2)}(x,z) = 3\chi^{(3)}E_\text{bias}(x,z)$, the induced $\chi^{(2)}$ has the same sign everywhere in space, but we can modulate its magnitude. However, this does not limit our ability to perform QPM because only the spatial variation of $\chi^{(2)}$ contributes to phase matching, and any constant background has a negligible effect on NLO processes performed in the device.

The specific design we used for our device (choice of materials and layer thicknesses) constrained its behavior and performance in various ways (see Appendix~\ref{appendix:electric-properties} for further details, and Sec.~\ref{sec:discussion} for a discussion on how these constraints could be softened or avoided by modifying the device design). The first constraint is that fringing of the electric fields blurs the mapping from $I(x,z)$ to $E_\text{bias}(x,z)$, which limited the smallest feature size we could program to $\SI{\sim \resolution}{\micro\meter}$. The second constraint is that, due to the electrical impedance of the cladding layers, the device has a non-negligible RC time constant that in turn has two effects: the bias voltage has to be AC-modulated to ensure a substantial voltage drop (and hence field $E_\text{bias}(x,z)$) over the waveguide core, and the device has a finite response time to the optical programming pattern, implying a maximum speed at which it is possible to update $\chi^{(2)}(x,z)$. Our device had maximum induced $\chi^{(2)}$ (of $\SI{\chitwo}{pm/V}$) when the bias voltage was modulated at a frequency of ${\approx}\SI{5}{Hz}$, with a drop to 80\% of the maximum when modulating at a frequency of ${\approx}\SI{20}{Hz}$ (see Fig.~\ref{fig:electric} in Appendix~\ref{appendix:electric-properties}). This implies that the maximum update speed of $\chi^{(2)}(x,z)$ for this device design is at least $\SI{20}{Hz}$, but the slow speed of the projector used in our experimental setup limited the update speed of the $\chi^{(2)}$ even more, to ${\sim}\SI{1}{Hz}$.

The nonlinear-optical functionality of our device can be tailored in the spectral, spatial, and spatio-spectral domains by modifying the structure of $\chi^{(2)}(x,z)$ (Fig.~\ref{fig:figure1}(c)). For example, consider narrowband SHG, where light at frequency $\omega_1$ is converted to light at frequency $\omega_2$. In addition to satisfying energy conservation, $\omega_2 = 2\omega_1$, efficient frequency conversion requires the momenta of interacting waves be matched. This requirement is quantified by the native phase mismatch,
\begin{align}
    \Delta k = k_2 - 2k_1,
\end{align}
where $k_j$ is the wave number of light at frequency $\omega_j$ ($j \in \{1,2\}$). Efficient SHG can be achieved when this phase mismatch
is compensated by a QPM grating---that is, by a periodic modulation of $\chi^{(2)}$ with a period designed to offset $\Delta k$, for example, $\chi^{(2)}(x,z) \sim \sin(\Delta k z)$ \cite{Boyd2008nonlinear}.

We performed experiments demonstrating three types of use of our device. First, we engineered the longitudinal (i.e., $z$) structure of $\chi^{(2)}(x,z)$ to control which wavelengths interact efficiently, enabling \textit{spectral-domain engineering} of NLO. Second, we tailored the transverse (i.e., $x$) structure of $\chi^{(2)}(x,z)$, enabling \textit{spatial-domain engineering}. For instance, if we set $\chi^{(2)}(x,z) \sim \sin(\Delta k z + \phi(x))$, with a spatially varying phase term $\phi(x)$, the generated second-harmonic (SH) light acquires a corresponding spatial phase profile, $e^{i\phi(x)}$, thereby shaping the output field in the transverse ($x$) direction. Finally, we used the full two-dimensional programmability of $\chi^{(2)}(x,z)$---in both the longitudinal and transverse directions---enabling simultaneous control of the light in both the spectral and spatial domains, giving rise to \textit{spatio-spectral engineering}. In the following, we present the results of our experiments.

\section{Real-time programmable periodic poling}
\label{sec:periodic-poling}

\begin{figure}[h]
    \centering
    \includegraphics[width=0.6\linewidth]{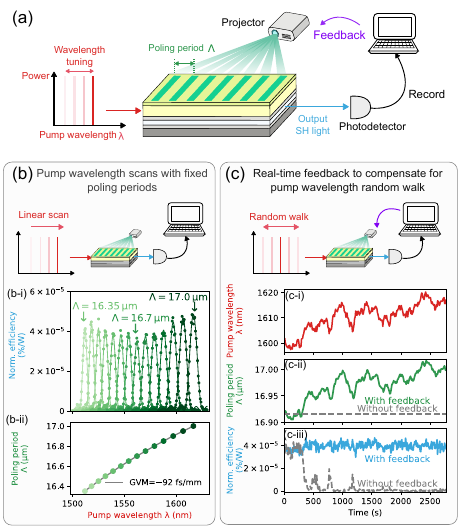}
    \caption{\textbf{Real-time programmable periodic poling with a programmable nonlinear waveguide.} 
    (a) Experimental setup. We pumped a prototype programmable nonlinear waveguide using a CW laser with a tunable wavelength $\lambda$. A grating pattern with period $\Lambda$ was projected onto the waveguide, realizing QPM for an SHG process. The output second-harmonic (SH) power was measured by a photodetector, and the measurements could be used to update $\Lambda$. (b) Nonlinear-optical characterization of the device. (b–i) For various choices of $\Lambda$, we scanned $\lambda$ and measured the SHG conversion efficiency, which we report as an efficiency normalized by input power. (b–ii) The optimal pump wavelength $\lambda$ for each poling period $\Lambda$. The quadratic fit gives the group-velocity mismatch (GVM) between the fundamental and second harmonic at $\SI{1560}{nm}$. The colors of the markers serve as legends for $\Lambda$, corresponding to those in (b–i). (c) Real-time feedback control of $\Lambda$ to compensate for a random walk of the pump wavelength shown in (c–i). To compensate for these fluctuations, we dithered $\Lambda$ and used the measured SHG signal to update $\Lambda$ in a way that optimizes the SHG efficiency. The evolution of $\Lambda$ is shown as a solid green line in (c–ii). In (c–iii), the SHG efficiency with and without such real-time feedback control are shown as blue solid and gray dashed lines, respectively. See Appendix~\ref{appendix:cw-pumped-SHG} for experimental details.}
    \label{fig:periodic-poling}
\end{figure}

First, we characterized the basic nonlinear-optical properties of the device by programming canonical QPM gratings with different poling periods $\Lambda$ and measuring the power of the SH light generated by the device when it was pumped with a continuous-wave (CW) laser that had tunable wavelength between $\lambda=\SI{1500}{nm}$ and $\SI{1630}{nm}$ (Fig.~\ref{fig:periodic-poling}(a)).

For this initial characterization, we fixed the poling period $\Lambda$ and scanned the wavelength of the pump laser to measure the SHG conversion efficiency. Depending on the value of $\Lambda$, different wavelengths of pump light undergo phase-matched SHG (Fig.~\ref{fig:periodic-poling}(b)). The fit between $\Lambda$ and an optimal $\lambda$ yielded a nominal poling period of $\SI{16.65}{\micro m}$ and group-velocity mismatch (GVM) of $\SI{-92}{fs/mm}$ between the fundamental and second harmonic at $\SI{1560}{nm}$. The nonlinearity of the device is proportional to the bias electric field $E_\text{bias}$, which one should not set higher than the breakdown field of the material. With the highest $E_\text{bias}$ we applied, we found the electric field induced $\chi^{(2)}$ nonlinearity of $\chi^{(2)}=\SI{\chitwo}{pm/V}$ (see Appendix~\ref{appendix:estimated-chi2}).

We took advantage of the ability to reprogram the poling period in our initial device characterization, but in these experiments the programming did not have to take place quickly. To showcase the ability of our device to be programmed in real time, we performed an experiment to show that it is possible to compensate for environmental noise and drifts by adjusting the poling period on the timescale of ${\sim}\SI{1}{s}$ (Fig.~\ref{fig:periodic-poling}(c)). To emulate large noise, we artificially modulated the pump wavelength $\lambda$ so that it followed a Gaussian random walk. The compensation task was to dynamically change the poling period $\Lambda$ to maximize the SHG efficiency, without being given information about the random changes in $\lambda$. We used a feedback scheme in which $\Lambda$ was dithered to obtain an error signal, and the error signal was used to update $\Lambda$. The data clearly shows that when the feedback controller was on, $\Lambda$ closely followed the evolution of the pump wavelength $\lambda$, maintaining a high level of SHG efficiency. On the other hand, when the feedback controller was off, the SHG efficiency dropped to near zero relatively quickly.

\section{Spectral engineering}
\label{sec:spectral}
\begin{figure}[h]
    \centering
    \includegraphics[width=1.0\linewidth]{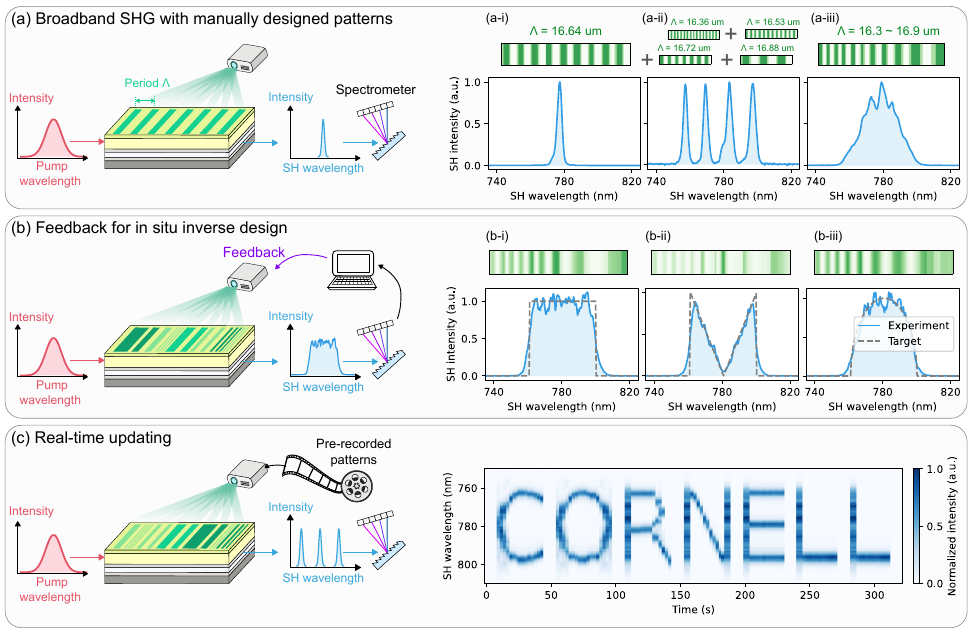}
    \caption{\textbf{Spectral engineering of second-harmonic generation (SHG).} (a) Output second-harmonic (SH) spectrum of broadband SHG pumped by ultrashort pump pulses for various illumination patterns. (a–i) Periodic grating with a period $\Lambda=\SI{16.64}{\micro m}$, (a–ii) superposition of four monotonic grating patterns with different periods, and (a–iii) an adiabatically chirped grating pattern. Due to the rapid spatial oscillations of these quasi-phase-matching (QPM) gratings, displaying the raw illumination patterns is not visually informative. Instead, in the green patterns shown above the results plots, we present the projected QPM grating patterns downsampled to a spatial period of $\SI{17}{\micro m}$ in the longitudinal direction. The same applies to the patterns shown in (b). See Appendix~\ref{appendix:spectral} for the original (un-downsampled) illumination patterns. (b) By constructing a feedback loop based on the measured SH spectrum, we optimized the illumination pattern to obtain various target SH output spectra. Dashed lines represent the target spectrum. (c) The illumination pattern was updated in real-time to output a sequence of SH spectra, using pre-recorded illumination patterns. We show the results for drawing ``CORNELL'' in the SH spectrum, with time as the horizontal axis of the image. See Appendix~\ref{appendix:spectral} for experimental details.}
    \label{fig:spectral}
\end{figure}

In this section, we show how the programmable nonlinear waveguide can be used to manipulate the spectral shape of the generated SH light by programming the $\chi^{(2)}$ nonlinearity with various patterns in the longitudinal ($z$) direction. In each of the experiments we report, we pumped a programmable nonlinear waveguide with an ultrashort pulse laser and measured the output SH spectrum using a spectrometer.

We measured the SHG spectra for various manually designed QPM grating structures (Fig.~\ref{fig:spectral}(a)) to verify that our device could reproduce well-known results in NLO. As a reference case, we programmed a grating pattern with a single period $\Lambda$, which phase-matched SHG for a particular pump wavelength (Fig.~\ref{fig:spectral}(a-i)). This manifested as a single, narrow peak in the SH spectrum. Beyond such a simple grating pattern, we programmed a summation of multiple grating patterns with different periods (Fig.~\ref{fig:spectral}(a-ii)). This super-grating pattern can phase-match various SHG processes at once and can generate multiple wavelengths of SH, which are visible in our experimental data as multiple peaks in the recorded spectrum. Finally, we programmed a chirped grating---a grating in which the period is changed as a function of the longitudinal position---and observed that it resulted in broadband SHG output (Fig.~\ref{fig:spectral}(a-iii)), consistent with prior non-programmable demonstrations of adiabatic SHG~\cite{Margules2021-Adiabatic-SHG}.

Up to this point, the illumination patterns we used to program the waveguide were designed manually, in that we designed them based on standard knowledge of NLO and what $\chi^{(2)}(x,z)$ patterns would yield the desired SHG processes. This approach to designing the poling of nonlinear waveguides is also what is used in conventional NLO, albeit in the conventional setting each pattern typically needs to be fabricated in a new device. To demonstrate the ability to shape the output SHG spectrum in ways that are likely impractical using conventional NLO devices, which don't support real-time reconfiguration, we arbitrarily shaped and dynamically updated the SHG spectrum. We achieved this by constructing a real-time feedback loop between the broadband SH spectral measurement and update of the programming illumination patterns (Fig.~\ref{fig:spectral}(b)). This in situ inverse design can be realized in various ways, which have the same end result but different speeds to converge to an optimal illumination pattern. In this work, we employed an in situ random optimization, where we perturbed the illumination pattern randomly and then accepted the perturbation if it improved the similarity between the SHG output and the target spectrum.

The illumination patterns that are found through inverse design can be stored and later retrieved to program a sequence of nonlinearity distributions in real time. We demonstrated this by drawing ``CORNELL'' in the SH spectrum as a function of time (Fig.~\ref{fig:spectral}(c)); the programming illumination pattern was updated every few seconds. The nonlinearity update speed was limited by the update speed ($\sim \SI{1}{s}$) of the spatial-light modulator (SLM) that we used to pattern the illumination. However, we believe this could be improved to $\SI{20}{Hz}$ by straightforwardly modifying the setup to use a faster SLM, and to ${\sim}\SI{200}{Hz}$ by device improvements (see Appendix~\ref{appendix:device-performance} for a full discussion).

\section{Spatial engineering}
\label{sec:spatial-engineering}
\begin{figure}[h!]
    \centering    \includegraphics[width=0.5\linewidth]{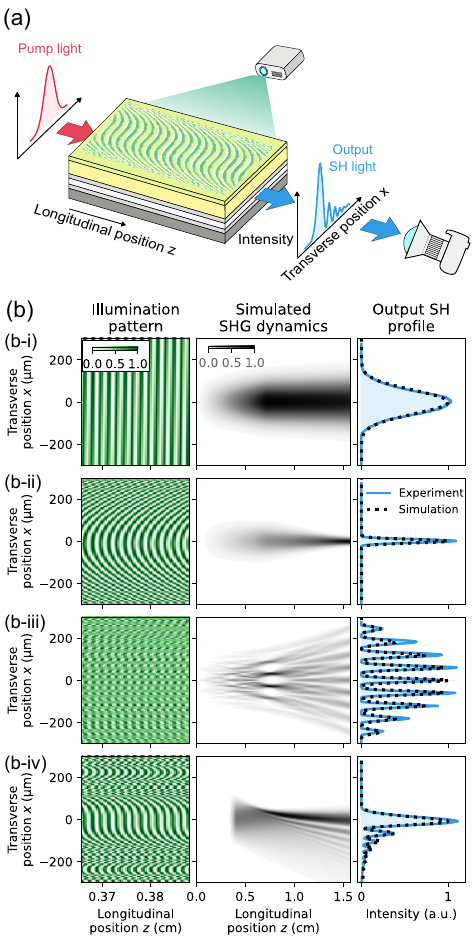}
    \caption{\textbf{Spatial engineering of second-harmonic generation (SHG).}  
     (a) Experimental setup. The waveguide was pumped with a pulsed pump laser with fixed Gaussian spatial profile, and the spatial distribution of the generated SH on a camera was measured. (b) A part of the programming illumination pattern (left column), simulated SHG dynamics within the waveguide (middle column), and a comparison between the experimentally measured and simulated SH spatial profiles. (b–i) Monotonic grating pattern. (b–ii) Quadratically chirped grating pattern. (b–iii) Superposition of nine quadratically chirped grating patterns with transverse offsets. (b–iv) Cubically chirped grating pattern. See Appendix~\ref{appendix:spatial} for experimental details.}
    \label{fig:spatial}
\end{figure}

In this section, we show the ability to engineer the spatial structure of light generated using our programmable nonlinear waveguide by controlling phase-matching conditions in the transverse dimension, as has been demonstrated previously in non-programmable NLO~\cite{Ellenbogen2009Airy, Buono2022StructuredLight, Efremidis2019AiryReview}. Here, we pumped the programmable nonlinear waveguide with a pulse with a fixed Gaussian spatial beam shape (beam waist: $\SI{132}{\micro m}$) and imaged the output SHG profile for various programming illumination patterns, i.e., different distributions of $\chi^{(2)}(x,z)$ (Fig.~\ref{fig:spatial}(a)). The resulting SHG output can be interpreted as a hologram encoded in the spatial structure of $\chi^{(2)}(x,z)$~\cite{Fang2020slab}.

As a reference, we first projected a simple, flat (i.e., constant in the transverse dimension) grating pattern with a period of $\SI{16.75}{\micro m}$, corresponding to phase-matched SHG near $\SI{790}{nm}$. The output SHG also had a Gaussian profile with a large beam waist of $\SI{94}{\micro m}$ (Fig.~\ref{fig:spatial}(b-i)). Then we performed experiments in which the phase of the QPM grating was spatially varied and observed that the generated SH light inherited the phase of the grating, which allowed us to engineer the spatial profiles of the SHG. For instance, by quadratically chirping the phase of the grating oppositely to the spatial diffraction, we were able to focus the generated SH light to the output facet, resulting in a substantially narrower beam waist of $\SI{16}{\micro m}$ (Fig.~\ref{fig:spatial}(b-ii)).

More complex patterns can be produced by superimposing multiple grating structures. For example, nine quadratically curved grating patterns, evenly spaced in the transverse direction, formed an effective microlens array that focused SHG into nine distinct peaks (Fig.~\ref{fig:spatial}(b-iii)). This approach can, in principle, be adopted to generate arbitrary superpositions of Gaussian peaks.

Diffraction-free beams that maintain their spatial profiles during propagation are used in microscopy and imaging~\cite{Efremidis2019AiryReview}. Nonlinear optics can generate Airy beams---one-dimensional non-diffracting beams---by applying a cubic chirp to a QPM grating in the transverse direction. We reproduced the seminal demonstration from Ref.~\cite{Ellenbogen2009Airy} with our programmable platform: our spatially resolved measurement of the waveguide output (Fig.~\ref{fig:spatial}(b-iv)) clearly shows the characteristic asymmetric interference fringes of an Airy beam.

The SH output profiles we recorded are in excellent agreement with theoretical simulations across all measurements; the simulations used only a single fitting parameter, for the overall amplitude. The match between experiment and theory makes it possible to optimize illumination patterns for engineering spatial features entirely in silico.

\section{Spatio-spectral engineering}
\label{sec:spatio-spectral}
\begin{figure}[h]
    \centering
\includegraphics[width=0.52\linewidth]{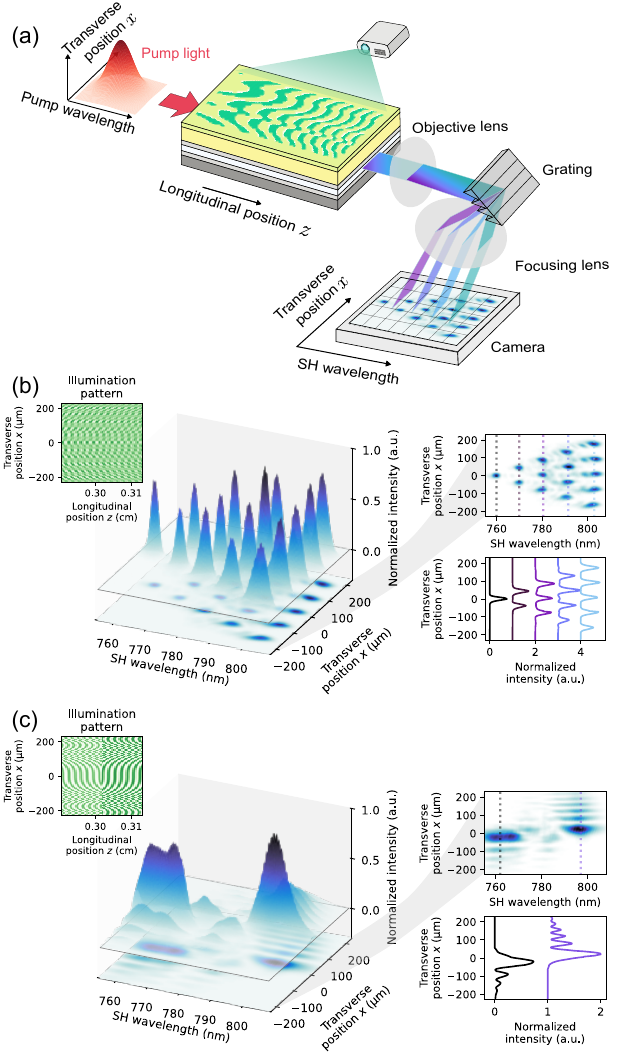}
    \caption{\textbf{Spatio-spectral engineering of second-harmonic generation (SHG).} (a) Experimental setup. We combined a reflective grating with a 4$f$ imaging setup to record spectrally-resolved one-dimensional spatial profiles of the output second-harmonic (SH) light. The waveguide was pumped by pulses with a fixed Gaussian spatial profile. (b) Results for an illumination pattern designed to generate various numbers of spatial peaks at five different wavelengths. (c) Results for an illumination pattern designed to generate oppositely chirped Airy beams at two different wavelengths. In each of (b) and (c), the left inset shows a part of the projected grating pattern. The bottom-right inset shows the spatial distribution of the SH light at various wavelengths, marked with dashed lines in the top-right inset. See Appendix~\ref{appendix:spatio-spectral} for experimental details.}
    \label{fig:spatio-spectral}
\end{figure}

In the previous two sections, we showed independent control of spectral and spatial features of SHG by tailoring the longitudinal and transverse structure of QPM gratings, respectively. In this section, we show that it is possible to leverage the full two-dimensional programmability of the $\chi^{(2)}$ nonlinearity to simultaneously tailor the spatial \emph{and} spectral profiles of the generated light. The experimental setup is illustrated in Fig.~\ref{fig:spatio-spectral}(a). We projected patterns of light onto the programmable waveguide and pumped it with broadband optical pulses. The output SH light was measured by spectrally resolved one-dimensional imaging, where the spectrum was recorded for each transverse spatial position by using a diffraction grating and a camera.

We aimed to obtain a spatio-spectral hologram in which the SHG output has a spatial profile that is a function of the output wavelength. To do this, we superimposed various QPM grating patterns with different longitudinal periods. In our first spatio-spectral experiment, we designed the grating structure to generate $1$, $2$, $3$, $4$, and $5$ spatial peaks at five different wavelengths (Fig.~\ref{fig:spatio-spectral}(b)). As shown in the one-dimensional hyperspectral image captured by the camera, we observed clearly separated Gaussian peaks localized both in space and wavelength. In our second experiment, we took inspiration from the SHG-based hologram proposed in Ref.~\cite{Ellenbogen2009Airy} as a means to generate different Airy beams for different wavelengths. Here, we show that oppositely chirped Airy beams can be generated by combining two grating patterns with different longitudinal periods and opposite cubic spatial chirps. We clearly observed the characteristic asymmetric interference fringes of the Airy beams but in opposite directions for two separate wavelengths (Fig.~\ref{fig:spatio-spectral}(c)).

\section{Discussion and Outlook}
\label{sec:discussion}
\subsection{Summary of the results}

We developed a programmable nonlinear waveguide with an arbitrarily reconfigurable two-dimensional distribution of $\chi^{(2)}$ nonlinearity. By tuning the period of quasi-phase matching (QPM) gratings programmed into the $\chi^{(2)}$ distribution of the waveguide, we demonstrated optimally phase-matched second-harmonic generation (SHG) for any pump wavelength over a range of $>\SI{100}{nm}$. We also showed that even when the pump wavelength fluctuates randomly, optimal phase matching can be robustly maintained through real-time reconfiguration of the grating period via feedback control. Furthermore, by engineering QPM gratings, we demonstrated versatile control over broadband SHG across the spectral, spatial, and spatio-spectral domains. The programmability of our device enabled real-time in situ optimization of QPM grating structures using feedback from experimental measurements. Leveraging this capability for spectral-domain engineering, we showed that the programmable nonlinear waveguide can tailor the output SHG spectrum into desired shapes over a $\SI{50}{nm}$ bandwidth. This was done solely by optimizing the QPM grating structures, without pulse shaping the pump laser. The optimized QPM gratings had highly nontrivial variations in both period and amplitude---to achieve the same results using traditional fixed nonlinear-optical devices would likely require multiple rounds of device fabrication. In contrast, the programmable waveguide allows the gratings to be optimized in situ with a single device. We also demonstrated updating the output SHG spectrum dynamically by modifying the QPM grating structure in real time. For example, we visualized the word ``Cornell'' in the time trace of the SHG spectrum.

We showed that the transverse spatial profile of the output SHG can be engineered by modulating the QPM gratings spatially. For example, we reproduced the seminal demonstration of Ref.~\cite{Ellenbogen2009Airy}, generating an Airy beam with SHG using a cubically chirped QPM grating. Finally, we simultaneously engineered the spatial and spectral shapes of the output second-harmonic (SH) light, demonstrating that the SHG could be spatially focused into a desired number of peaks or formed into distinct Airy beams at different wavelengths. Notably, all of the results reported in this paper were obtained using a single programmable nonlinear waveguide design and the same pulsed laser (except in Fig.~\ref{fig:periodic-poling}(a), where we used a tunable continuous-wave (CW) laser). This highlights the flexibility and multifunctionality of programmable nonlinear waveguides.

\subsection{Limitations of the current device prototype and potential for improvements}
In this study, we demonstrated a broad range of functionalities using programmable nonlinearity in a planar waveguide, emphasizing its potential for diverse nonlinear optics (NLO) applications. However, our experimental prototype has several practical drawbacks that are not fundamental but need to be addressed if the programmable-waveguide approach is to be widely adopted. First and foremost, the estimated maximum programmable $\chi^{(2)}$ nonlinearity of $\SI{\chitwo}{pm/V}$ in our prototype device is low compared with that of conventional nonlinear-optical materials~\cite{Dutt2024review}, limiting the efficiency of NLO processes. Fortunately, there are ways to increase it. For instance, the electric-field contrast within the core layer could be improved by using a thicker photoconductor layer, which reduces the spatially uniform background nonlinearity that does not contribute to nonlinear-optical processes, thereby increasing the dynamic range of the programmable nonlinearity. Electrical characterization of the device suggests that a thicker photoconductor with higher photoconductivity could increase the programmable nonlinearity by up to a factor of $2.3$, which would result in a maximum programmable $\chi^{(2)}$ of $\SI{\chitwomax}{pm/V}$ (see Appendix~\ref{appendix:estimated-chi2}). We chose plasma-enhanced chemical vapor deposition (PECVD)-grown low-index (i.e., nitrogen-rich) SiN as our core material (see Appendix~\ref{appendix:device-fabrication}) because of its commercial availability and large bandgap (making it suitable for SHG pumped at ${\sim}\SI{1.55}{\micro m}$). However, other materials can achieve substantially higher electric-field-induced $\chi^{(2)}$ nonlinearities, such as $\SI{41}{pm/V}$ with Si~\cite{Timurdogan2017E-FISH} and $\SI{22.7}{pm/V}$ with silicon-rich silicon nitride (SRN)~\cite{Lin2019E-FISH}. These materials could increase the magnitude of programmable $\chi^{(2)}$ nonlinearity to near that of conventional state-of-the-art materials like lithium niobate (${\sim}\SI{50}{pm/V}$). However, the bandgaps of Si and SRN are smaller than that of SiN, which limits their operation to longer wavelengths of light. The potential use of Si or SRN as the waveguide core material provides a promising path for developing high-nonlinearity programmable nonlinear waveguides for wavelengths longer than ${\sim}\SI{1}{\micro\meter}$. What are the prospects for making high-nonlinearity devices with transparency windows covering some or all of the visible wavelengths too? A suitable material for the waveguide core should have a large bandgap, high breakdown voltage, and low optical loss, as well as a high $\chi^{(3)}$. Promising candidates for evaluation include diamond, silicon carbide, and aluminum nitride. The relevant $\chi^{(3)}$ tensor element for evaluating the magnitude of possible electric-field-induced $\chi^{(2)}$ in a material is surprisingly understudied in the literature (see Appendix~\ref{appendix:physics-chi2}), leaving open the possibility that these or other candidate materials may enable programmable nonlinear waveguides with nonlinearity comparable to the commonly used native-$\chi^{(2)}$ materials and a transparency window stretching into the ultraviolet.

Another limitation is the low update rate of the programmable nonlinearity, which is currently limited to approximately $\SI{20}{Hz}$ because of the RC time constant of the prototype device (see Appendix~\ref{appendix:electric-properties}). The update rate could be improved by using a material for the photoconductor layer that has higher photoconductivity than SRN. The device is also limited by its need for AC operation; because the $\chi^{(2)}$ nonlinearity is also modulated at the AC frequency, the rest of the system---including the optical inputs---needs to be synchronized to this modulation. This limitation could be eliminated by using a conductive oxide as the cladding, enabling DC operation. 

The use of a conductive cladding would---provided the cladding thicknesses were kept roughly the same---also reduce the percentage of the bias voltage dropped across the cladding layers, allowing more of the supplied bias voltage to drop over the waveguide core where it is useful. This would reduce the bias voltage needed to be applied across the device---the $\SI{\sim 1000}{V}$ used in our experiments (see Appendix~\ref{appendix:electric-properties}) could be reduced to $\SI{\sim 100}{V}$ while achieving the same value of maximum programmable nonlinearity $\chi^{(2)}$.

A final limitation is that the device relies on free-space optics for generating and imaging the illumination pattern for programming, which makes the system bulkier than it likely needs to be. The system could be miniaturized by directly integrating a micro-light-emitting-diode (\textmu LED) display on or near the photoconductor layer. As discussed in Ref.~\cite{Onodera2024Linear}, \textmu LED displays can deliver sufficient optical power for this application, making this a promising approach to compact and robust programmable nonlinear photonic systems.

Appendix~\ref{appendix:device-performance} summarizes the performance of the current devices and the performance that could potentially be achieved through future improvements.

\subsection{Prospective applications}
The ability to realize arbitrary $\chi^{(2)}(x,z)$ distributions makes our device platform very versatile, particularly in enabling devices that can seamlessly switch between performing multiple functions. In quantum technology, engineered QPM gratings are powerful tools for designing quantum light sources, frequency converters, and gates for optical quantum computation and networking~\cite{Laura2023QPM, Lu2023Frequency-Bin, Weiss2025QPM-review, Ansari2018review}. Programmable QPM-based nonlinear operations could, for example, enable a single physical device to perform multiple kinds of quantum gates, or quantum gates on qubits having different wavelengths. Similarly, all-optical signal processing for classical optical communications \cite{willner2013all} could benefit from reconfigurable nonlinear processes, as could classical optical computation \cite{mcmahon2023physics}. Another compelling potential application is in generating structured light for sensing, where diverse spatial and spectral light profiles are often required to probe different features of interest~\cite{Saxena2015structured, Heist2018StructuredSensing, Wang2024StructuredSensing5D}, and sometimes requires adaptivity~\cite{chakrova2016adaptive,li2024adaptive}. The ability to dynamically shape, optimize, and update the spatio-spectral properties of light could give programmable nonlinear waveguides a distinct advantage over conventional nonlinear light sources. Engineered QPM gratings can also facilitate phase-matching for cascaded NLO processes, and this technique has been used to achieve simultaneous harmonic generation~\cite{Chen2015HHG}. A reconfigurable QPM grating could take this further by allowing precise selection and efficient generation of desired harmonics, with the flexibility to modify the output spectrum as needed. The primary obstacle to demonstrating these applications using our current device is its relatively weak optical nonlinearity. However, as we have described in the previous section, there are promising routes to increasing the induced nonlinearity by 10$\times$--100$\times$.

Our approach to achieving programmable nonlinearity is not limited to the SiN slab waveguide that we report in this paper. Since all materials possess a non-zero $\chi^{(3)}$ nonlinearity, one could, in principle, \emph{add} programmable $\chi^{(2)}$ nonlinearity to existing nanophotonic devices by adding a photoconductor layer and bias voltage. Although some devices would not be technically compatible with this approach (e.g., due to the need for a conductive substrate and large enough $\chi^{(3)}$ nonlinearity), it could in other cases allow a novel interplay between the original functionality of the device and the added programmable nonlinearity. For example, an integrated high-quality-factor resonator with programmable nonlinearity could be used to make an efficient light source with a widely tunable output wavelength without a tunable pump---something that is challenging to achieve otherwise (see Appendix~\ref{appendix:applications}). Similarly, simultaneous SHG and supercontinuum generation could be achieved by endowing a dispersion-engineered $\chi^{(3)}$ waveguide with programmable $\chi^{(2)}$ nonlinearity \cite{Porcel2017SCG, Lee2024InverseSCG}, potentially enabling versatile control of a multi-octave-spanning spectrum.

A key benefit of realizing functions on programmable photonic devices---whether in slab-waveguide geometry or other configurations---is that inevitable material and fabrication imperfections when a device is made, as well as environmental fluctuations when it is operated, can be compensated for. As a result, high performance can be achieved with better fabrication yield or less sophisticated fabrication than conventional devices, and with improved robustness to material imperfections and variable operating conditions. For example, nanoscale thickness variations in thin-film lithium niobate waveguides typically place an effective limit on the useful length of a periodically poled waveguide and the maximum achievable conversion efficiency. In Ref.~\cite{Chen2023AdaptedPoling}, the authors reported how they could circumvent this limitation by precisely measuring the thickness distribution and adapting the poling to compensate for the thickness inhomogeneity. Programmable nonlinear photonic devices offer a fundamentally different solution to such challenges. By dynamically optimizing the QPM grating structure for an experimentally measured figure of merit (FOM), programmable devices can be adapted in real time to maximize performance. We demonstrated both robustness to fluctuations in pump wavelength (Sec.~\ref{sec:periodic-poling}), where the FOM was conversion efficiency, and in situ inverse design (Sec.~\ref{sec:spectral}), where the FOM was the similarity between the measured and target SHG spectra. There are many situations in nonlinear optics---supercontinuum generation being a prominent example~\cite{Sylvestre2021SCG-review}---where device behavior is extremely sensitive not only to the device parameters but also to the field profile of the pump light, making it challenging to achieve exact agreement between simulation and experiment. The inverse-design experiments reported in Sec.~\ref{sec:spectral} were performed without prior characterization of the pump; in situ inverse design using programmable devices may ultimately enable the realization of quantitatively correct behavior even for complex nonlinear-optical processes that we don't have accurate simulation models for.

In conclusion, the ability to programmably control nonlinearity has the potential to allow us to circumvent the limitations of the conventional one-device--one-function paradigm. The programmable nonlinear waveguide we have proposed and the demonstrations of reconfigurable SHG we have reported with a prototype device take a step into this new frontier of nonlinear optics.

\section*{Data and code availability}
Experimental data and scripts to replicate the figures in this paper are available at \url{https://doi.org/10.5281/zenodo.15014773}.

\section*{Acknowledgements}
We thank NTT Research for their financial and technical support. We gratefully acknowledge the Air Force Office of Scientific Research for funding under Award Number FA9550-22-1-0378, and the National Science Foundation for funding under Award Number CCF-1918549. This work was performed in part at the Cornell NanoScale Facility, a member of the National Nanotechnology Coordinated Infrastructure (NNCI), which is supported by the National Science Foundation (Grant NNCI-2025233). P.L.M. gratefully acknowledges financial support from a David and Lucile Packard Foundation Fellowship. We thank Claire Bacon and Springer Nature Author Services for English language editing. The authors thank Thibault Chervy, Jeremy Clark, Noah Flemens, Ryan Hamerly, Oscar Jaramillo, Jeffrey Moses, Rajveer Nehra, Edwin Ng, Tadahiro Takahashi, Tianyu Wang, Alan Willner, and Yoshihisa Yamamoto for helpful comments and discussions.

\section*{Author contributions}
R.Y., T.O., L.G.W., and P.L.M. conceived the project. R.Y., B.A.A., M.M.Stein, T.O., and L.G.W. designed the devices. R.Y., T.O., and P.L.M. designed the experiments. R.Y., B.A.A., and Y.Z. fabricated the device with aid and recipe development from M.M.Stein and T.O.. R.Y. and M.M.Sohoni designed and built the imaging setup to program the $\chi^{(2)}$ nonlinearity pattern. R.Y. designed and built the setup for nonlinear-optical experiments with aid from F.P. and M.J.. R.Y. and M.M.Sohoni wrote the code for real-time optimization of the QPM grating. R.Y. performed the experiments, analyzed the results, and produced the figures. R.Y. and P.L.M. wrote the manuscript with input from all authors. P.L.M. supervised the project.

\section*{Competing interests}
M.M.Stein, L.G.W., T.O. and P.L.M. are listed as inventors on a patent application (WO2023220401A1) on 2D-programmable waveguides.

\bibliographystyle{mcmahonlab}
\bibliography{references}

\appendix
\startcontents[Appendices]
\addtocontents{toc}{\protect\setcounter{tocdepth}{0}} 
\section*{Appendices} 
\addtocontents{toc}{\protect\setcounter{tocdepth}{2}}
\printcontents[Appendices]{l}{1}{}

\pdfmapline{-dummy LMRoman12-Regular}

\section{Summary of device performance}
\label{appendix:device-performance}
In Table~\ref{tab:performance}, we summarize the performance of the programmable nonlinear waveguide that was experimentally demonstrated in this work, and we present the projected performance for a future device with enhanced functionalities. More detailed discussions are provided in the corresponding appendix sections.

\begin{table}[h]
    \begin{center}
\begin{tabular}{ |p{7.0cm}||p{4.5cm}|p{4.5cm}| }

 \hline
 Performance & This work &Potential future device\\
 \hhline{|=|=|=|}
 Programmable $\chi^{(2)}$ nonlinearity (material)& $\SI{\chitwo}{pm/V}$ (N-rich SiN~$^\mathrm{a}$)&$\SI{22.7}{pm/V}$ (Si-rich SiN~$^\mathrm{b}$)\\
&&$\SI{41}{pm/V}$ (Si~$^\mathrm{c}$)\\
\hline
Approximate bandgap  wavelength (material)&$\SI{400}{nm}$ (N-rich SiN$~^\mathrm{d}$)&$\SI{600}{nm}$ (Si-rich SiN$~^\mathrm{e}$)\\
&&$\SI{1100}{nm}$ (Si$~^\mathrm{f}$)\\
\hline
Update speed of nonlinearity& $\SI{1}{Hz}~^\mathrm{g}$&$\SI{200}{Hz}~^\mathrm{h}$\\
 \hline 
 Area of programmable nonlinearity ($z\times x$)& $\SI{7.2}{mm}\times\SI{4.5}{mm} ~^\mathrm{i}$&$\SI{1.6}{cm}\times\SI{2.9}{cm}~^\mathrm{j}$ \\
 \hline
Smallest programmable feature size& $\SI{\resolution}{\micro m}~^\mathrm{k}$& $\SI{1}{\micro m}~^\mathrm{l}$\\
 \hline
 Optical loss & $\SI{1}{dB/cm}\sim \SI{5}{dB/cm}~^\mathrm{m}$& $<\SI{1}{dB/cm}~^\mathrm{n}$\\
 \hline
\end{tabular}
\end{center}
\caption{The table below summarizes the performance of the programmable nonlinear waveguide demonstrated in this work, as well as that of a potential future device. ${^\mathrm{a}}$ See Appendix~\ref{appendix:estimated-chi2}. ${^\mathrm{b}}$ Value reported in Ref.~\cite{Lin2019E-FISH}. ${^\mathrm{c}}$ Value reported in Ref.~\cite{Timurdogan2017E-FISH}. ${^\mathrm{d}}$ Inferred from ellipsometry data. ${^\mathrm{e}}$ Corresponding to the value of $\sim \SI{2}{eV}$ reported for the highest-index SRN film in Ref.~\cite{Krckel2017BandgapSRN}. ${^\mathrm{f}}$ Value reported in Ref.~\cite{Meier2007SiBandgap}.  ${^\mathrm{g}}$ Speed limit imposed by the update speed of the SLM. ${^\mathrm{h}}$ A high-speed SLM can achieve update speeds exceeding $\SI{1}{kHz}$. The speed limit is set by the RC time constant of the device (see Appendix~\ref{appendix:lumped-circuit-model}). We assume a $10\times$ increase in the bright-state photoconductivity. ${^\mathrm{i}}$ The current SLM has $1920\times 1200$ pixels, with each pixel mapping to $\SI{\wpixel}{\micro m}$ spot (see Appendix~\ref{appendix:SLM}). ${^\mathrm{j}}$ Assuming a high-resolution SLM, with 8K resolution ($7680\times4320$ pixels)~\cite{Yang2023SLM_Review}. We assume that each pixel of the SLM maps to a $\SI{\wpixel}{\micro m}$ spot on the programmable waveguide. We can further increase the programmable area by combining illumination from multiple SLMs. In this case, the upper bound is imposed by limitations in fabrication, e.g., the size of a wafer. ${^\mathrm{k}}$ Limit imposed by the fringing of electric fields (see Appendix~\ref{appendix:electric-field-fringing}). ${^\mathrm{l}}$ Assuming a programmable nonlinear waveguide with a $\SI{1}{\micro m}$-thick photoconductive core (see Appendix~\ref{appendix:photoconductive-core}). ${^\mathrm{m}}$ The loss is maximal around $\SI{1520}{nm}$, while the typical loss at other wavelengths is $\SI{1}{dB/cm}$ (see Appendix~\ref{appendix:optical-loss}). ${^\mathrm{n}}$ Various materials compatible with programmable nonlinear waveguides have demonstrated this level of loss~\cite{Guidry2020SiC,Ji2021SiN,Jung2021Tantala}.
}
\label{tab:performance}
\end{table}

\section{Device fabrication}
\label{appendix:device-fabrication}

\begin{figure}[h]
    \centering
    \includegraphics[width=1.0\linewidth]{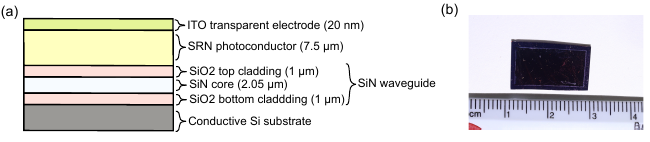}
    \caption{(a) Illustration of the stack structure of a programmable nonlinear waveguide. (b) A programmable nonlinear waveguide that was fabricated and used to produce parts of the results in the main text.}
    \label{fig:layer-stack}
\end{figure}

In this section, we describe the fabrication process for the programmable nonlinear waveguide. As illustrated in Fig.~\ref{fig:layer-stack}(a), the device was composed of a stack of multiple material layers. Silicon Valley Microelectronics (SVM) supplied the substrates, including the bottom cladding and the core layer. The substrate was a conductive, boron-doped Si wafer with a resistivity in the range of $0.01\sim \SI{0.02}{\Omega \cdot cm}$. The bottom cladding consisted of a $\SI{1}{\micro m}$-thick wet thermal oxide layer, onto which approximately $\SI{2}{\micro m}$ of low-stress SiN was deposited via plasma-enhanced chemical vapor deposition (PECVD). Using a Metricon prism coupler, we measured the film thickness as $d_\mathrm{core}=\SI{2.05}{\micro m}$, with a thickness variation of approximately $\SI{50}{nm}$ across a 4-inch wafer. The refractive index of the film was specified as $1.98$ at a wavelength of $\SI{632.8}{nm}$. 

We performed rapid thermal annealing (RTA) on the wafers acquired from SVM at $\SI{650}{C^\circ}$ for $\SI{3}{minutes}$. This RTA process reduced the refractive index of the film and eliminated undesired fluorescence in the near-infrared region when the waveguide was pumped near $\SI{800}{nm}$. Since the results presented in the main text did not depend on this pumping wavelength, the RTA process shifted the phase-matching conditions without causing significant adverse effects.

Next, we deposited a $\SI{1}{\micro m}$-thick layer of $\mathrm{SiO}_{2}$ as the top cladding via our in-house PECVD system (Oxford PlasmaPro 100 PECVD; Oxford Instruments), thereby forming the planar SiN waveguide. To render the waveguide programmable, we further deposited a $\SI{7.5}{\micro m}$ thick layer of silicon-rich silicon nitride (SRN) via PECVD. The SRN was deposited at an RF power of $\SI{200}{W}$ with gas flows of $\text{SiH}_4:\SI{8}{sccm}$, $\text{H}_2:\SI{40}{sccm}$, and $\text{N}_2:\SI{2000}{sccm}$. We note that no $\text{NH}_3$ was used.

At this stage, we cleaved the wafer into rectangular pieces of approximately $\SI{1}{cm} \times \SI{1.5}{cm}$. Although cleaving typically produces facets that are sufficiently clean for light coupling, additional polishing can further improve the beam profile quality. Finally, we deposited a $\SI{20}{nm}$-thick layer of indium tin oxide (ITO) as a transparent electrode via a physical vapor deposition (PVD) system (PVD 75, Kurt J. Lesker). It was important to leave a few millimeters of space between the electrode and the chip edge to prevent electrical breakdown of the air at the boundary. A picture of the resultant programmable waveguides is shown in Fig.~\ref{fig:layer-stack}(b).

\section{Electrical properties of programmable nonlinear waveguides}
\label{appendix:electric-properties}

This section introduces an electric circuit model for a programmable waveguide, demonstrating how photoconductivity can be used to control the electric-field-induced $\chi^{(2)}$. We also discuss the necessary considerations for optimal device operation, the potential for improving the induced nonlinearity, and the resolution limits of programmable $\chi^{(2)}$ nonlinearity.

\subsection{Lumped-element circuit model}
\label{appendix:lumped-circuit-model}

The electrical properties of a programmable waveguide can be approximated by modeling each layer of the device as a lumped-element circuit element with a defined impedance~\cite{Onodera2024Linear}. For the cladding and core layers, we assume that their conductivities are negligible. The impedance of each layer at the frequency $\omega = 2\pi f$ is given by
\begin{align}
    Z_{\alpha} = \frac{1}{i\omega C_\alpha},
\end{align}
where
\begin{align}
    C_\alpha = \frac{\epsilon_0 \epsilon_\alpha A}{d_\alpha}
\end{align}
is the capacitance of the layer. Here, the subscript $\alpha \in \{\text{cladding}, \text{core}\}$ denotes the layer, $\epsilon_0$ is the vacuum permittivity, $\epsilon_\alpha$ is the relative permittivity, $d_\alpha$ is the layer thickness, and $A$ is the area. In this context, the term ``cladding'' refers to the combined stack of the top and bottom cladding layers.

The photoconductor layer, however, can exhibit nonnegligible conductance even in the absence of illumination. Therefore, we account for both its capacitive and conductive contributions to the impedance as
\begin{align}
    Z_{\text{B/D}} = \frac{1}{i\omega C_\text{PC} + 1/R_\text{B/D}},
\end{align}
where
\begin{align}
    C_\text{PC} = \frac{\epsilon_0 \epsilon_\text{PC} A}{d_\text{PC}}
\end{align}
is the capacitance of the photoconductor layer. The subscripts ``B'' and ``D'' denote the bright and dark states, respectively. The resistance $R_\text{B/D}$ is given by
\begin{align}
    R_\text{B/D} = \frac{d_\text{PC}}{\sigma_\text{B/D} A},
\end{align}
where $\sigma_\text{B}$ and $\sigma_\text{D}$ are the conductivities of the photoconductor in the bright and dark states, respectively.

When a total voltage $V_\text{tot}$ is applied, the overall stack acts as a voltage divider. Specifically, the voltage across the core layer in a bright or dark state is
\begin{align}
    V_\text{B/D} = \frac{Z_\text{core}}{Z_\text{core} + Z_{\text{cladding}} + Z_\text{B/D}} V_\text{tot}.
\end{align}

To illustrate how photoconductivity enables control of the electric field within the core, we consider an idealized limit in which the photoconductor layer is infinitely thick ($d_\text{PC} \rightarrow \infty$) and the bright-state photoconductivity is infinitely high (i.e., $d_\text{PC}/\sigma_\text{B} \rightarrow 0$). In this limit, we have $Z_\text{B} \rightarrow 0$ and $Z_\text{D} \rightarrow \infty$, leading to
\begin{align}
    V_\text{B} = \frac{Z_\text{core}}{Z_\text{core} + Z_{\text{cladding}}} V_\text{tot} \quad \text{and} \quad V_\text{D} = 0.
\end{align}
Thus, photoconductivity enables control of the bias electric field over a dynamic range from 
\begin{align}
    V_\text{max} = \frac{Z_\text{core}}{Z_\text{core} + Z_{\text{cladding}}} V_\text{tot}
\end{align}
down to 0. In a realistic device with a finite photoconductor thickness and finite bright-state conductivity, the dynamic range is reduced.

In nonlinear optics, the contrast between the bright and dark states, $V_\text{B} - V_\text{D}$, produces a QPM grating. Since the efficiency of SHG is proportional to the square of the contrast in $\chi^{(2)}$ nonlinearity, the SHG power $P_\text{SH}$ can be expressed as
\begin{align}
\label{eq:frequency-dependence}
    P_\text{SH} \propto \Delta V^2 = |V_\text{B} - V_\text{D}|^2 = \left| \frac{Z_\text{core}}{Z_\text{core} + Z_{\text{cladding}} + Z_\text{B}} - \frac{Z_\text{core}}{Z_\text{core} + Z_{\text{cladding}} + Z_\text{D}} \right|^2 V_\text{tot}^2.
\end{align}

\begin{figure}[h]
    \centering
    \includegraphics[width=1.0\linewidth]{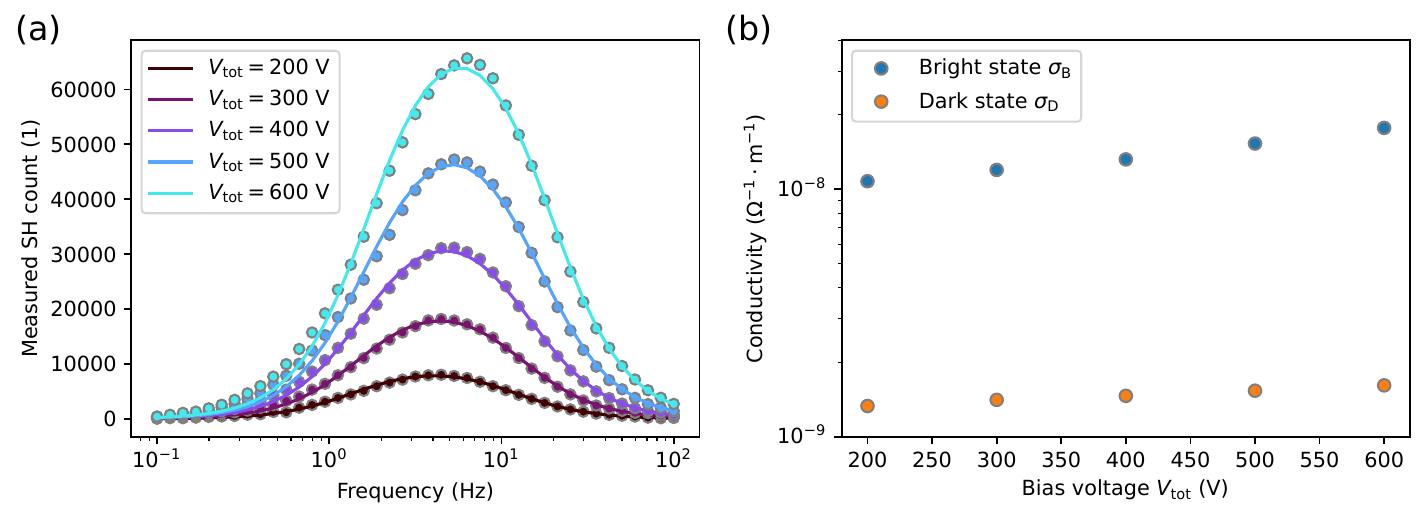}
    \caption{(a) Circles: Experimentally measured SHG power for various bias voltages $V_\text{tot}$ and its frequency. Solid lines: Theoretical fit based on the model \eqref{eq:frequency-dependence}. We assume that $\epsilon_{\text{cladding}}=3.9$ for the silicon dioxide cladding, $\epsilon_{\text{core}}=6.0$ for the silicon nitride core, and $\epsilon_{\text{PC}}=8.0$ for the photoconductor composed of SRN. The thicknesses of the layers are $d_\text{cladding}=\SI{2}{\micro m}$, $d_\text{core}=\SI{2.05}{\micro m}$, and $d_\text{PC}=\SI{7.5}{\micro m}$. See Appendix~\ref{appendix:device-fabrication} for the details of the device fabrication. We note that the dependence on $A$ is canceled in \eqref{eq:frequency-dependence}. (b) Numerically determined bright- and dark-state conductivities of the photoconductor for various total bias voltages $V_\text{tot}$.}
    \label{fig:electric}
\end{figure}

In Fig.~\ref{fig:electric}(a), we present a fit of the model in Eq.~\eqref{eq:frequency-dependence} to our experimental data. In this measurement, we varied both the frequency $\omega$ and the amplitude $V_\text{tot}$ of the bias field, and we recorded the generated SHG power via a spectrometer. The fit showed good agreement between the model and the experiment, indicating that the optimal operating frequency was $f=\omega/2\pi = \SI{5}{Hz}$. For all the experiments reported in this paper, we applied a bias voltage $V_\text{tot}=\SI{1000}{V}$ to our device with frequency $\SI{5}{Hz}$, unless otherwise specified. From the figure, we found the frequency at which the induced $\chi^{(2)}$ nonlinearity became $80\%$ of the optimal value (i.e., where the SHG conversion efficiency dropped to $(80\%)^2=60\%$) was approximately $\SI{20}{\Hz}$. The fit also yielded the conductivities $\sigma_\text{B/D}$ at various $V_\text{tot}$ values, as shown in Fig.~\ref{fig:electric}(b).

On the basis of these values, we estimate the potential improvement in nonlinearity achievable with further optimization of the photoconductor. For this analysis, we use the nominal values $\sigma_\text{B}=\SI{1.8e-8}{\Omega^{-1}m^{-1}}$ and $\sigma_\text{D}=\SI{1.6e-9}{\Omega^{-1}m^{-1}}$ obtained from the fit at $V_\text{tot}=\SI{600}{V}$, along with the optimal operating condition $\omega/2\pi = \SI{5}{Hz}$, to determine the present voltage contrast $\Delta V_\mathrm{present}$ that was achievable with our current waveguide design. 

The theoretical upper bound on the voltage contrast, denoted as $\Delta V_\mathrm{max}$, is reached when the photoconductor exhibits perfect switching characteristics and is infinitely thick. Under these ideal conditions, $\Delta V_\mathrm{max} = V_\mathrm{max}$. The ratio
\begin{align}
\mathcal{R}_\mathrm{max} = \frac{\Delta V_\text{max}}{\Delta V_\mathrm{present}}
\end{align}
quantifies the potential for improvement. Numerically, we find $\mathcal{R}_\mathrm{max}\approx 2.3$, which indicates that the induced $\chi^{(2)}$ could be larger by this factor than that measured in our experiment.

\subsection{Resolution limit imposed by electric field fringing}
\label{appendix:electric-field-fringing}

Notably, the theoretical upper limit assumes an infinitely thick photoconductor, a condition that would lead to significant fringing of the electric field and reduce the resolution. In practice, the use of a reasonably thin photoconductor is preferable to avoid such issues. Under the assumption of perfect switching for a photoconductor of finite thickness (i.e., $\sigma_\text{B}\rightarrow\infty$ and $\sigma_\text{D}=0$), our estimates indicate that the induced $\chi^{(2)}$ nonlinearity could be increased by a factor of approximately $1.2$.

\begin{figure}[b]
    \centering
    \includegraphics[width=1.0\linewidth]{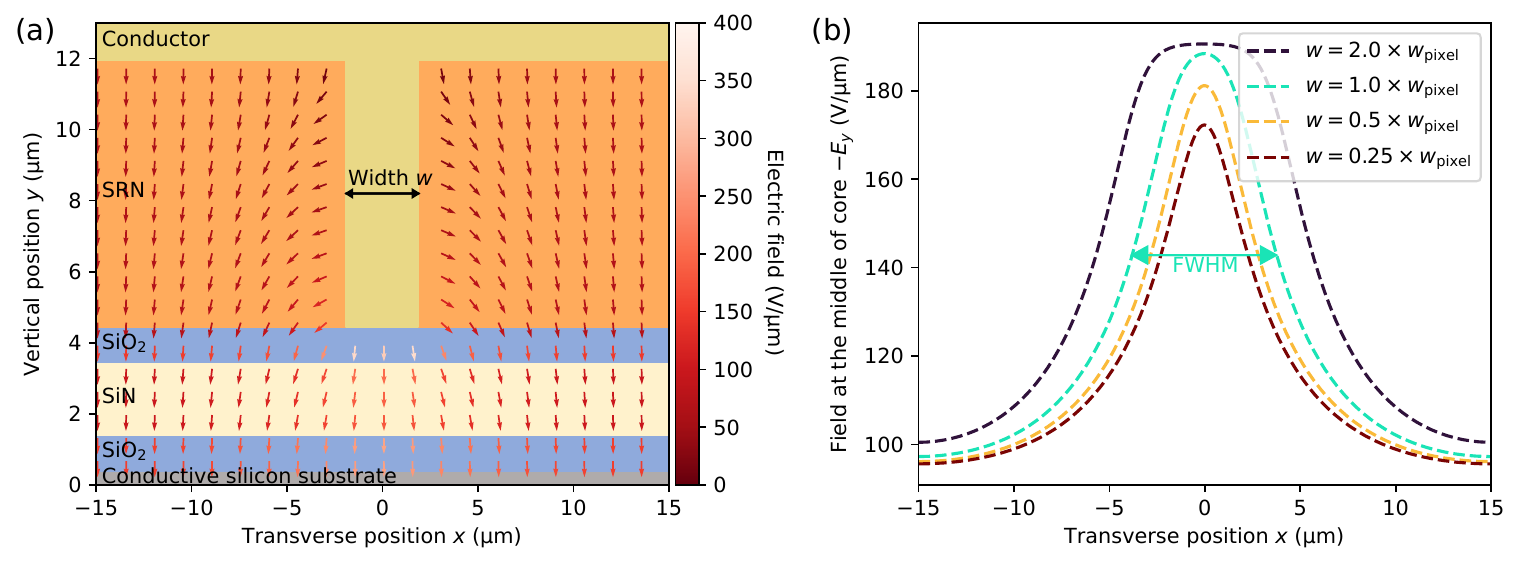}
    \caption{(a) Numerically simulated distribution of the electric field inside a programmable nonlinear waveguide. The simulation is performed by solving \eqref{eq:poisson} via the finite-difference method with the same parameters as in Fig.~\ref{fig:electric}. We assume that $V_\text{tot}=\SI{1000}{V}$ and that there is a periodic boundary condition in the transverse dimension. (b) Vertical component of the electric field in the middle of the core layer for various widths of the conductive region $w$. }
    \label{fig:poisson}
\end{figure}

Thus far, we have assumed uniform programming illumination on a programmable nonlinear waveguide to model its electric properties. In reality, however, the programming illumination can exhibit small spatial variations, which may partially invalidate this assumption. In the following, we study how these spatial variations affect the electric field distribution inside the core by analyzing the resolution limit imposed by electric field ``fringing''.

For this analysis, we consider a simplified model of a programmable nonlinear waveguide, as shown in Fig.~\ref{fig:poisson}(a). In this model, when programming illumination with a width of $w$ is projected onto the photoconductor layer, a vertical pillar of fully conductive material of the same width is created. When a bias voltage $V=V_\text{tot}$ is applied to the top electrode, the electric potential of this conductive region is fixed at $V_\text{tot}$, while the bottom substrate is grounded (i.e., $V=0$). The electric potential distribution between these boundaries is given by the self-consistent solution of the Poisson equation
\begin{align}
\label{eq:poisson}
    \nabla\cdot(\epsilon\nabla V)=0,
\end{align}
where $\epsilon$ denotes the electric permittivity distribution of the medium. This equation can be solved for the specified boundary conditions via the finite-difference method~\cite{Nagel2011} and the biconjugate gradient stabilized algorithm. The resulting electric field distribution, $\mathbf{E}=-\nabla V$, is shown in Fig.~\ref{fig:poisson}(a), revealing that the electric fields fringe inside the medium, which blurs the features.

To quantitatively assess the resolution limit imposed by these fringing effects, Fig.~\ref{fig:poisson}(b) displays the distributions of the vertical electric field inside the core for various feature sizes $w$. In our experimental setup, the smallest feature size we can generate with the programming illumination is $w_\text{pixel}=\SI{\wpixel}{\micro m}$, corresponding to a pixel of the SLM (see Appendix~\ref{appendix:common-part}). According to our simulations, a feature with a width of $w_\text{pixel}$ produces a bias field distribution with a full width at half maximum (FWHM) of $w_\text{FWHM}=\SI{\resolution}{\micro m}$, which defines the smallest feature size possible in our experiment. Notably, owing to the fringing effects, an illumination spot smaller than $w_\text{pixel}$ does not necessarily produce a finer electric field distribution within the core.

\subsection{Photoconductive core for smaller feature sizes }
\label{appendix:photoconductive-core}
In Appendix~\ref{appendix:electric-field-fringing}, it is shown that the minimum feature size for programmable nonlinearity $w_\text{FWHM}=\SI{\resolution}{\micro m}$ is not imposed by the resolution of the programming illumination. Indeed, $w_\text{FWHM}$ is significantly larger than the fundamental diffraction limit at the wavelength of the programming light $\SI{532}{nm}$. Rather, the fringing effects of the electric field inside the photoconductive layer and the top cladding play major roles in blurring the field contrast inside the core layer. 

\begin{figure}[h]
    \centering
    \includegraphics[width=0.85\linewidth]{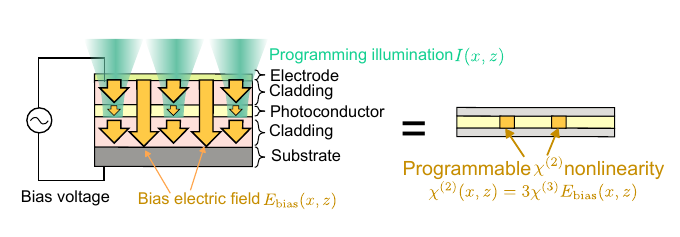}
    \caption{Conceptual illustration of a programmable nonlinear waveguide using a photoconductive core. When the programming illumination is applied to the photoconductive core, the material becomes conductive, locally reducing the bias electric field. Consequently, we can program the distribution of electric-field-induced $\chi^{(2)}$ nonlinearity by reducing its value through illumination with light.}
    \label{fig:photoconductive-core}
\end{figure}

In Fig.~\ref{fig:photoconductive-core}, we show a future device design that could address this challenge to achieve much smaller feature sizes. In this design, we employ a photoconductive material as the core material. An optical waveguide is formed by the photoconductive core layer and cladding layers, and a transparent electrode is deposited on the top cladding. There is no separate photoconductive layer in this design. 

When the programming illumination is applied to the core layer, it locally increases the conductivity. This has the effect of reducing the electric field within this region, decreasing the induced $\chi^{(2)}$ nonlinearity. In contrast, with no programming illumination, the core layer remains highly insulating and experiences a high bias field, thus exhibiting high $\chi^{(2)}$ nonlinearity. In summary, we can dynamically program the distribution of $\chi^{(2)}$ nonlinearity by locally reducing the nonlinearity with illumination, which is the opposite of the design demonstrated in this work. Because the variation in conductivity directly occurs inside the core layer, the aforementioned issue of electric field fringing will be limited to roughly the thickness of the core layer.

\section{Optical properties of a programmable nonlinear waveguide}
\label{appendix:optical-properties}
This section describes the optical properties of a programmable nonlinear waveguide, including the effective index of the guided mode, the phase-matching condition for SHG, and the optical loss.

\subsection{Waveguide mode}
\begin{figure}[h]
    \centering
    \includegraphics[width=0.5\linewidth]{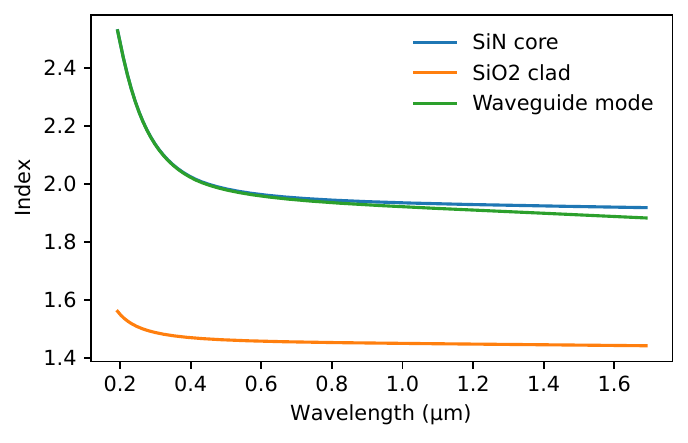}
    \caption{Optical indices of the SiN core (blue lines), $\mathrm{SiO}_\mathrm{2}$ cladding (orange lines), and fundamental TM mode of the optical waveguide (green lines), shown for various wavelengths of light. The index of the SiN core was measured via an ellipsometer, and we use the formula provided in Ref.~\cite{Malitson1965SiO2Index} for the $\mathrm{SiO}_\mathrm{2}$ cladding.}
    \label{fig:waveguide-index}
\end{figure}

We characterize the guided mode of the programmable waveguide by approximating the cladding as infinitely thick and determining the guided modes of the core layer. Deviations from this approximation result in radiation loss and coupling to the photoconductor layer, as discussed in Appendix~\ref{appendix:optical-loss}. The refractive indices of the cladding and core layers are shown in Fig.~\ref{fig:waveguide-index} as functions of wavelength.

In this work, we consider the fundamental transverse magnetic (TM) modes of the waveguide for nonlinear optics. For a TM mode, only the $H_x$, $E_y$, and $E_z$ field components are nonzero. The spatial profile of a TM mode is obtained as a solution to the eigenvalue equation
\begin{align}
\label{eq:TM-eigen}
    \beta^2 H_x(y) = \left( n^2(y)\frac{\partial}{\partial y}\frac{1}{n^2(y)}\frac{\partial}{\partial y} + k_0^2 n^2(y) \right) H_x(y),
\end{align}
where $\beta$ is the effective propagation constant of the mode, $n(y)$ is the refractive index distribution in the vertical direction, and $k_0$ is the wavenumber in a vacuum. The other nonzero field components are related to $H_x$ by
\begin{align}
    E_y &= -\frac{\beta}{\epsilon_0 n^2(y)\omega} H_x, \\
    E_z &= -\frac{i}{\epsilon_0 n^2(y)\omega}\frac{\partial H_x}{\partial y}
\end{align}
with the angular frequency of light $\omega$.

Our slab waveguide is characterized by $n(y)=n_\text{core}$ for $|y|\leq d_\text{core}/2$ and $n(y)=n_\text{cladding}$ elsewhere. In this case, an analytic solution for the fundamental TM mode is given by
\begin{align}
    H_x(y)=\left\{\begin{array}{cc}
         C\cos(k_\text{core}y)&\mathrm{for~}|y|\leq d_\text{core}/2  \\
         C\cos(k_\text{core}d_\text{core}/2)\exp(-\kappa_\text{core}(y-d_\text{core}/2))& \mathrm{for~}y> d_\text{core}/2\\
         C\cos(k_\text{core}d_\text{core}/2)\exp(-\kappa_\text{core}(-y-d_\text{core}/2))& \mathrm{for~}y<- d_\text{core}/2
    \end{array}\right.
\end{align}
where $C$ is a normalization constant. The value of $k_\text{core}$ is determined as the smallest solution of the equation
\begin{align}
\label{eq:index-equation}
    k_\text{core} \tan\left(\frac{k_\text{core} d_\text{core}}{2}\right) = \frac{n_\text{core}^2}{n_\text{cladding}^2} \sqrt{k_0^2\left(n_\text{core}^2 - n_\text{cladding}^2\right)- k_\text{core}^2}.
\end{align}
Equation~\eqref{eq:index-equation} is related to the propagation constants according to
\begin{align}
    -k_\text{core}^2 + n_\text{core}^2 k_0^2 = \kappa^2 + n_\text{cladding}^2 k_0^2 = \beta^2.
\end{align}
Importantly, the effective index of the fundamental TM mode is defined as
\begin{align}
    n_\text{eff} = \frac{\beta}{k_0}.
\end{align}
Figure~\ref{fig:waveguide-index} shows $n_\text{eff}$ for our waveguide structure.

\subsection{Phase-matching conditions}
For the second-harmonic generation (SHG) of pump light with wavelength $\lambda_0$, the phase mismatch is defined as
\begin{align}
    \Delta k = k_2 - 2 k_1,
\end{align}
where $k_1$ and $k_2$ are the wavenumbers of the fundamental and second-harmonic waves, respectively. Since we are considering SHG between the fundamental TM modes of the waveguide, we have
\begin{align}
    k_1 &= \frac{2\pi n_\text{eff}(\lambda_0)}{\lambda_0}, \\
    k_2 &= \frac{4\pi n_\text{eff}(\lambda_0/2)}{\lambda_0}.
\end{align}
Quasi-phase matching (QPM) is achieved when the $\chi^{(2)}$ nonlinearity is modulated in a grating-like pattern with a spatial period
\begin{align}
    \Lambda = \frac{2\pi}{\Delta k}.
\end{align}

Interestingly, the rate at which the optimal QPM grating period changes with wavelength is proportional to the group velocity mismatch (GVM) between the fundamental and second-harmonic waves:
\begin{align}
    \frac{\partial \Lambda}{\partial \lambda_0} = -\frac{8\pi^2 c}{\Delta k^2 \lambda_0^2} \, \mathrm{GVM},
\end{align}
where
\begin{align}
    \mathrm{GVM} = \frac{1}{v_{\text{g,1}}} - \frac{1}{v_{\text{g,2}}}
\end{align}
and where $v_{\text{g,1}}$ and $v_{\text{g,2}}$ are the group velocities of the fundamental and second-harmonic (SH) waves, respectively~\cite{Imeshev1998PulseShaping}.

\begin{figure}
    \centering
    \includegraphics[width=0.55\linewidth]{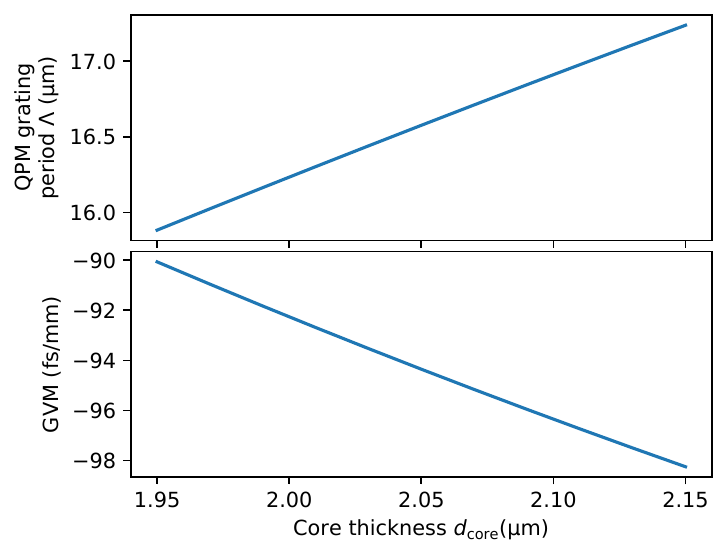}
    \caption{The numerically calculated QPM grating period and GVM for the SHG of pump wavelength $\lambda_0=\SI{1.56}{\micro m}$ as functions of the core thickness $d_\text{core}$. We use the material index data shown in Fig.~\ref{fig:waveguide-index}. }
    \label{fig:phase-matching}
\end{figure}

In Fig.~\ref{fig:phase-matching}, we present the numerically estimated QPM grating period $\Lambda$ and the GVM at $\lambda_0 =\SI{1.56}{\micro m}$ as functions of the core thickness $d_\text{core}$. At the nominal thickness $d_\text{core} = \SI{2.05}{\micro m}$, our simulations yield $\Lambda = \SI{16.58}{\micro m}$ and $\mathrm{GVM} = \SI{-94}{fs/mm}$, which are in good agreement with the experimentally measured values of $\Lambda =\SI{16.69}{\micro m}$ and $\mathrm{GVM} = \SI{-92}{fs/mm}$ (see Sec.~\ref{sec:spectral}). As shown in Fig.~\ref{fig:phase-matching}, the nominal film thickness variations of approximately $\SI{50}{nm}$ can account for these residual discrepancies.

\subsection{Optical loss}
\label{appendix:optical-loss}
Several factors contribute to the optical loss in a programmable nonlinear waveguide. Below, we evaluate the primary contributors in more detail.

\subsubsection{Material absorption}
\label{appendix:material-absorption}

The core material employed for the programmable nonlinear waveguide was PECVD SiN, which can exhibit considerable optical loss at the wavelengths of interest. To characterize the optical loss inside the core, we fabricated several planar waveguides of varying lengths. Laser light was coupled into the fundamental modes of these waveguides, and by comparing the output powers from waveguides of different lengths, we estimated the propagation loss.

\begin{figure}[bth]
    \centering
    \includegraphics[width=0.85\linewidth]{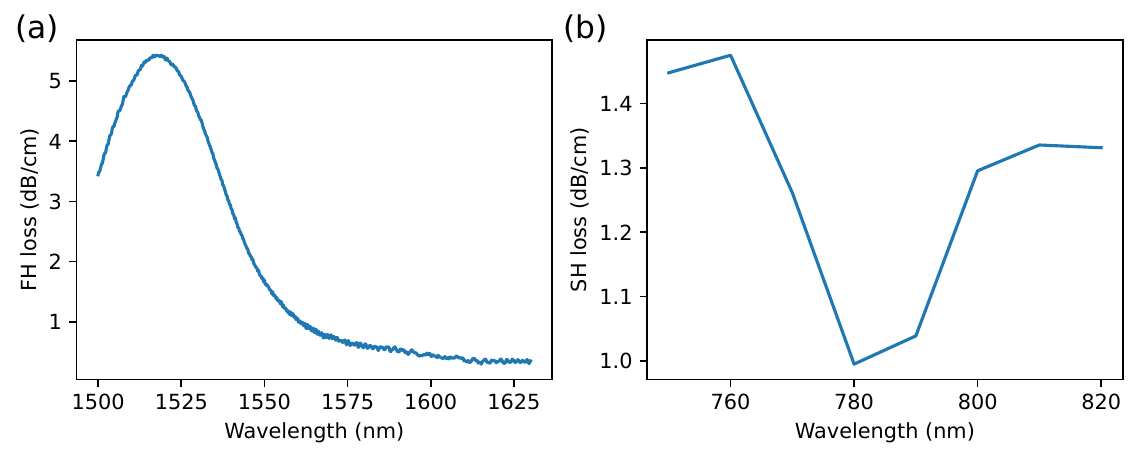}
    \caption{Optical loss in SiN was measured by comparing transmission through two planar waveguides with lengths of $\SI{1.5}{cm}$ and $\SI{3.0}{cm}$. The waveguides consist of a $\SI{1}{\micro m}$ $\mathrm{SiO}_2$ bottom cladding, a $\SI{2.05}{\micro m}$ SiN core, and an air top cladding. (a) The loss of FH light, with wavelengths ranging from $1500$ to $\SI{1630}{nm}$, was measured via a tunable CW laser (TSL-570; Santec) as the light source. (b) The loss of SH light, with wavelengths between $750$ and $\SI{820}{nm}$, was measured via a Ti:sapphire laser.}
    \label{fig:loss}
\end{figure}

Figure~\ref{fig:loss}(a) displays the loss at the fundamental harmonic (FH) wavelength. The peak observed at approximately $\SI{1520}{nm}$ was attributed to the characteristic absorption of PECVD SiN due to N--H bonds~\cite{Wang2018SiN-anneal, Ay2004SiN-loss}. At wavelengths distant from this absorption peak, the loss decreased to approximately $\SI{1}{dB/cm}$. Figure~\ref{fig:loss}(b) shows the loss at the second-harmonic (SH) wavelength, where we observed both lower loss and reduced wavelength dependence.

\subsubsection{Radiation loss to the substrate}
\label{appendix:radiation-loss}

Since the refractive index of the Si substrate was greater than that of the waveguide core, the guided mode was not fully confined, and light could gradually leak into the substrate~\cite{Hu2009LeakyMode}. Here, we numerically analyze the impact of this radiation loss on a programmable nonlinear waveguide.

\begin{figure}[h]
    \centering
    \includegraphics[width=0.6\linewidth]{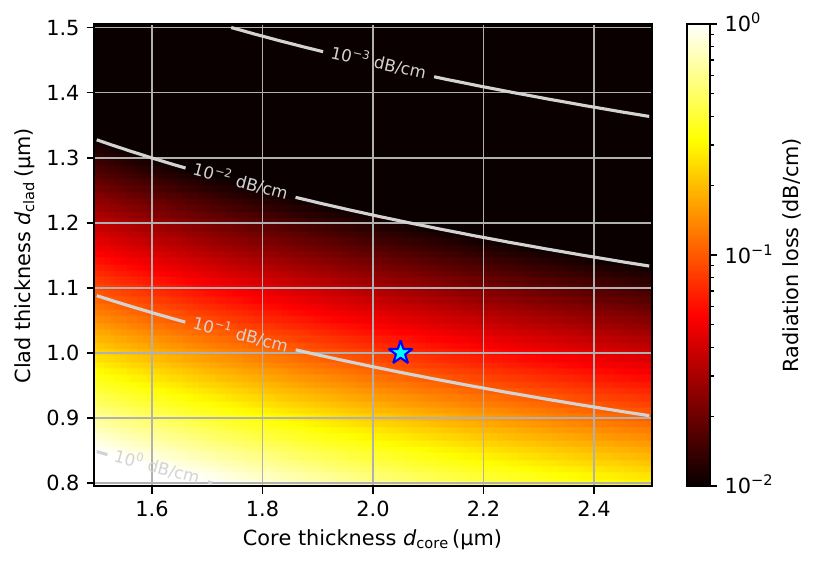}
    \caption{Numerically simulated radiation loss to the Si substrate at a wavelength of $\SI{1.55}{\micro m}$. We use the formalism provided in Ref.~\cite{Hu2009LeakyMode}. The blue star represents the design of our programmable nonlinear waveguide.}
    \label{fig:radiation-loss}
\end{figure}

To model the radiation loss, we represent our system as a symmetric planar waveguide composed of an SiN core with thickness $d_\text{core}=\SI{2.05}{\micro m}$, top and bottom $\mathrm{SiO}_2$ cladding layers with thickness $d_\text{cladding}=\SI{1}{\micro m}$, and Si substrates on both sides of the cladding layers, which are assumed to be infinitely thick. We then employ the formalism in Ref.~\cite{Hu2009LeakyMode} to calculate the radiation loss rate $\kappa$ for this stack and use $\kappa/2$ as a phenomenological estimate for our waveguide, which has an Si substrate only on one side. The simulation results for the loss at $\SI{1.55}{\micro m}$ are shown in Fig.~\ref{fig:radiation-loss}, indicating that the contribution from radiation loss is negligible compared with the material absorption observed in our experiment.

\subsubsection{Coupling to the photoconductor mode}
\label{appendix:photoconductor-mode}

Another loss channel is formed by the coupling between the core mode and the photoconductor mode. Unlike the coupling to the substrate discussed in Appendix~\ref{appendix:radiation-loss}, where the large thickness of the substrate results in a continuum of modes, the finite thickness of the photoconductor layer yields well-resolved discrete modes. Consequently, we observed sharp peaks in the waveguide loss at specific wavelengths where the effective index of the fundamental core mode matched that of a photoconductor mode.

The field distribution of a TM mode in a waveguide satisfies the following equation:
\begin{align}
\label{eq:TM-eigen}
    \beta^2H_x(y)=\left(n^2(y)\partial_y \frac{1}{n^2(y)}\partial_y+k_0^2n^2(y)\right)H_x(y),
\end{align}
where $\beta$ is the propagation constant of the mode, $n(y)$ denotes the refractive index distribution in the vertical direction $y$, and $H_x(y)$ is the magnetic field component in the transverse direction $x$. Thus, for a given planar optical waveguide stack along the $y$-direction, we can numerically diagonalize Eq.~\eqref{eq:TM-eigen} to obtain all the modes of the system.

\begin{figure}[tbh]
    \centering
    \includegraphics[width=0.65\linewidth]{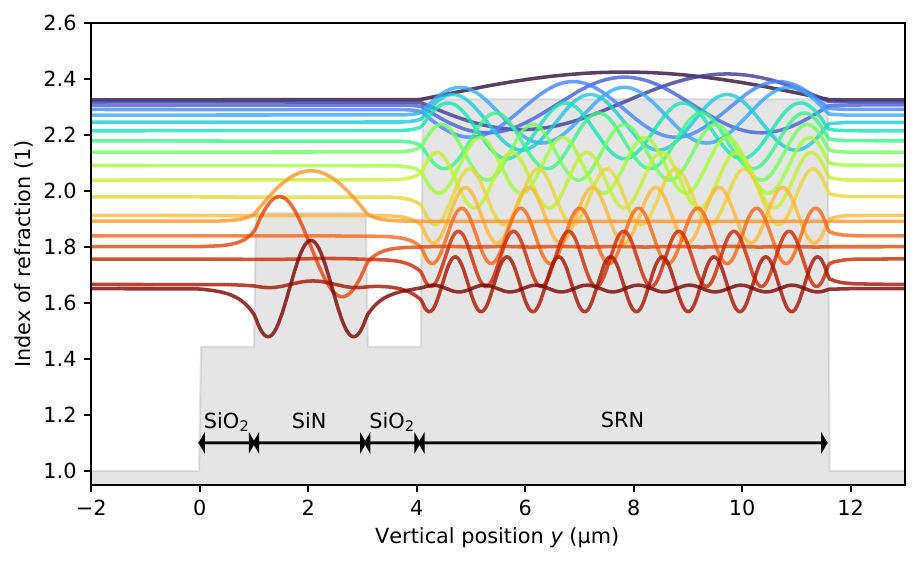}
    \caption{Spatial profiles $H_x(y)$ of the TM modes at $\lambda_0=\SI{1.55}{\micro m}$ calculated via \eqref{eq:TM-eigen}, where we show $18$ modes with the highest effective indices. We use the same index data as in Fig.~\ref{fig:waveguide-index} with the waveguide geometry specified in Appendix~\ref{appendix:device-fabrication}. The gray shaded region represents the distribution of the index of refraction $n(y)$. }
    \label{fig:photoconductor-mode}
\end{figure}

The waveguide modes are shown in Fig.~\ref{fig:photoconductor-mode} for $\lambda_0 = \SI{1.55}{\micro m}$. In this figure, core modes (localized within the SiN layer) and photoconductor modes (localized within the SRN layer) are clearly observed. To focus on the physics of the photoconductor modes, we replace the Si substrate with an air layer in this model.

\begin{figure}[bth]
    \centering
    \includegraphics[width=0.9\linewidth]{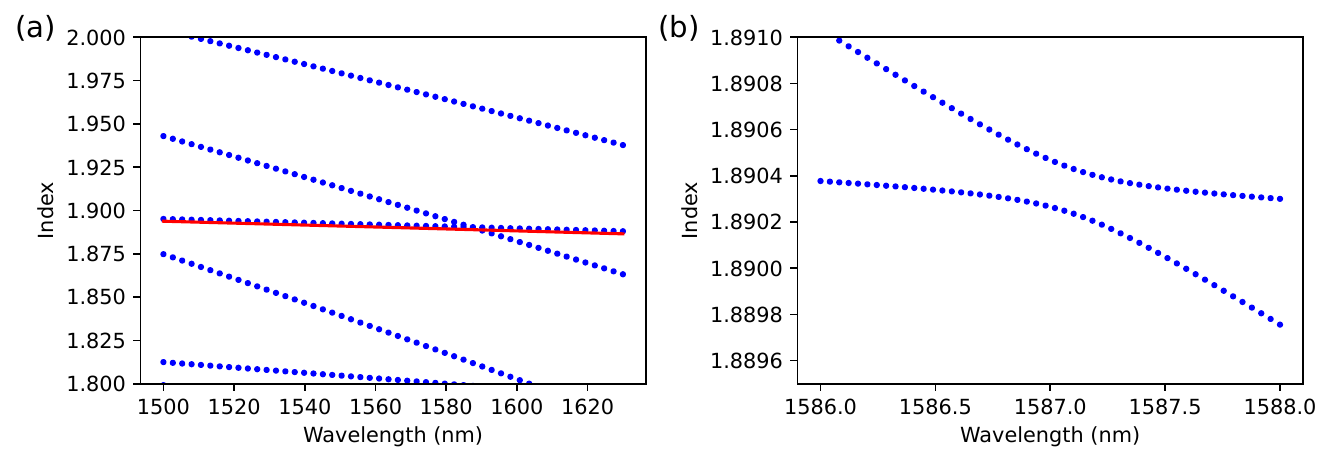}
    \caption{The blue circles represent numerically calculated effective indices of the waveguide modes for various wavelengths $\lambda_0$, where we use the same simulation methods as in Fig.~\ref{fig:photoconductor-mode}. The red solid lines are unperturbed indices of the fundamental TM modes in the absence of the photoconductor layer. (a) and (b) show the same data but with different regions of interest.}
    \label{fig:mode-indices}
\end{figure}

Generally, the effective indices of the core modes and the photoconductor mode vary in different ways with the wavelength $\lambda_0$. Consequently, as $\lambda_0$ changes, their indices may cross over. This behavior is illustrated in Fig.~\ref{fig:mode-indices}(a), which shows the effective indices of the waveguide modes as a function of $\lambda_0$. A closer look at the crossover region in Fig.~\ref{fig:mode-indices}(b) reveals an avoided energy crossing between the modes, a clear manifestation of mode hybridization.

When such hybridization occurs, light in the fundamental core mode can leak into the photoconductor mode via coherent coupling. The strength of this coupling is characterized by the magnitude of the avoided crossing in Fig.~\ref{fig:mode-indices}(b), which is approximately $\Delta n = 10^{-4}$. This implies that a significant portion of the light will be lost to the photoconductor mode over a propagation distance of $\lambda_0/\Delta n \approx \SI{1}{cm}$. Note that this photoconductor-induced loss affects only a narrow band of wavelengths near the crossover point and does not lead to global losses. Furthermore, owing to the coherent nature of the coupling, the light can also be recoupled back into the core mode after sufficient propagation.

\begin{figure}[tbh]
    \centering
    \includegraphics[width=0.5\linewidth]{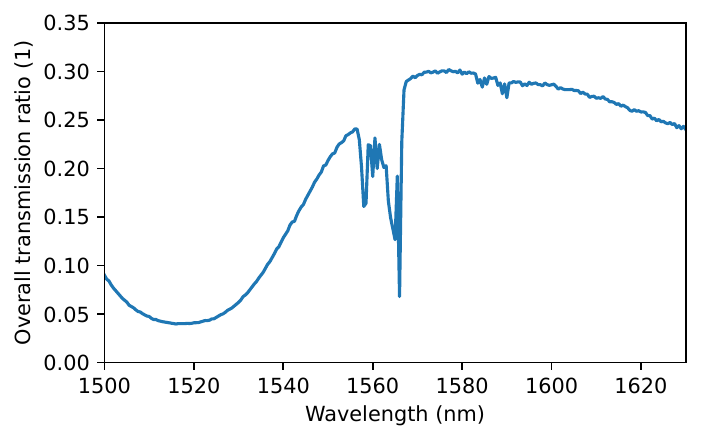}
    \caption{Optical transmission ratio of a programmable nonlinear waveguide measured via a tunable CW laser (TSL-570; Santec). The loss includes incoupling loss to the waveguide, outcoupling loss from the waveguide, and the finite collection efficiency of the light. }
    \label{fig:transmission}
\end{figure}

We present the experimental results for transmission through a programmable nonlinear waveguide in Fig.~\ref{fig:transmission} for various wavelengths of light. Aside from the loss at approximately $\SI{1520}{nm}$ due to material absorption (see Appendix~\ref{appendix:material-absorption}), we observed localized absorption lines near $\SI{1560}{nm}$, which we attributed to loss caused by the photoconductor mode. The effects of the photoconductor mode were also evident in the experimental results shown in Fig.~\ref{fig:spectral}, where the SHG conversion efficiency was reduced around this pump wavelength. Note that the location of the absorption peak is highly sensitive to factors such as the refractive index and thickness of the films, which likely explains the quantitative mismatch between the numerically predicted location of mode crossing and the experimental result.

Finally, we note that the photoconductor-induced loss can be mitigated by employing a thicker top cladding. As the overlap between the core modes and the photoconductor modes decreases exponentially with increasing cladding thickness, even a slight increase in the top cladding can significantly reduce these effects.

\section{Model for SHG in a programmable nonlinear waveguide}
\label{appendix:SHG-model}

In this section, we derive a model for SHG in a programmable nonlinear waveguide by formulating equations of motion for the FH and SH beams. We adapt the formalism from Ref.~\cite{Jankowski2024AOP} to the case of a slab waveguide.

The electric and magnetic field profiles of the fundamental TM modes can be written as functions of the vertical coordinate, i.e., $\mathbf{E}(y)$ and $\mathbf{H}(y)$ (see Appendix~\ref{appendix:optical-properties}). Using Poynting's theorem, we impose the following normalization condition for the mode:
\begin{align}
\label{eq:normalization}
    \frac{1}{2}\int \mathrm{d}y\,\mathrm{Re}\Bigl(\mathbf{E}(y)\times \mathbf{H}^*(y)\Bigr)\cdot \mathbf{z} = W,
\end{align}
where $\mathbf{z}$ is a unit vector in the longitudinal direction and $W$ is a normalization constant with units of one-dimensional power density $[\mathrm{power}\cdot \mathrm{length}^{-1}]$. We then express the field profiles as
\begin{align}
\label{eq:E-field}
    &\mathbf{E}(y) = \sqrt{\frac{2Z_0W}{n_\text{eff}L_\text{mode}}}\,\mathbf{e}(y),&\mathbf{H}(y) = \sqrt{\frac{2n_\text{eff}W}{Z_0L_\text{mode}}}\,\mathbf{h}(y),
\end{align}
where $Z_0=\SI{377}{\Omega}$ is the vacuum impedance and where $\mathbf{e}(y)$ and $\mathbf{h}(y)$ are dimensionless field profiles. The width of the mode in the vertical direction is characterized as
\begin{align}
\label{eq:L-mode}
    L_\text{mode} = \int \mathrm{d}y\,\mathrm{Re}\Bigl(\mathbf{e}(y)\times \mathbf{h}(y)\Bigr)\cdot \mathbf{z}.
\end{align}
Note that Eqs.~\eqref{eq:E-field} and \eqref{eq:L-mode} are defined to be consistent with the normalization condition \eqref{eq:normalization}, which leaves the scaling of $\mathbf{e}(y)$ and $\mathbf{h}(y)$ as a free parameter. Following a convention in NLO, we set the scaling of $\mathbf{e}(y)$ and $\mathbf{h}(y)$ so that the peak value of $\mathrm{Re}\bigl(\mathbf{e}(y)\times \mathbf{h}(y)\bigr)\cdot \mathbf{z}$ is unity.

In the remainder of this section, we assume that only the fundamental TM modes of the FH and SH light, with frequencies $\omega$ and $2\omega$, respectively, are excited. The electromagnetic fields can be parameterized as
\begin{align}
    \mathbf{E}(x,y,z,t) &= \frac{1}{\sqrt{W}} \int \mathrm{d}x\,\Bigl[ \mathbf{E}^{(\omega)}(y)\, a(x,z)\, e^{-i\omega t+ik_1 z}+ \mathbf{E}^{(2\omega)}(y)\, b(x,z)\, e^{-2i\omega t+ik_2 z} \Bigr],
\end{align}
where we explicitly label the waveguide modes by their frequencies. Here, $a(x,z)$ and $b(x,z)$ are the spatial amplitudes of the FH and SH fields, respectively, with units of $[\mathrm{power}^{1/2}\cdot \mathrm{length}^{-1/2}]$.

The evolution of the field amplitudes follows
\begin{align}
\partial_z a(x,z) &= \frac{i}{2k_1}\partial_x^2 a(x,z) - \frac{i\omega}{4\sqrt{W}}\, e^{-ik_1z}\int \mathrm{d}y\,\mathbf{E}^{(\omega)*}(y)\cdot \mathbf{P}_\text{NL}^{(\omega)}(x,y,z),\\[1ex]
\partial_z b(x,z) &= \frac{i}{2k_2}\partial_x^2 b(x,z) - \frac{i\omega}{2\sqrt{W}}\, e^{-ik_2z}\int \mathrm{d}y\,\mathbf{E}^{(2\omega)*}(y)\cdot \mathbf{P}_\text{NL}^{(2\omega)}(x,y,z).
\end{align}
The nonlinear polarizations at the respective frequencies are given by
\begin{align}
    P_{\text{NL},i}^{(\omega)}(x,y,z) &= \frac{2\epsilon_0\, b(x,z)\, a^*(x,z)}{W}\sum_{jk} d_{ijk}\, E_{j}^{(2\omega)}(y)\, E_{k}^{(\omega)*}(y)\, e^{ik_2z-ik_1z},\\[1ex]
    P_{\text{NL},i}^{(2\omega)}(x,y,z) &= \frac{\epsilon_0\, a^2(x,z)}{W}\sum_{jk} d_{ijk}\, E_{j}^{(\omega)}(y)\, E_{k}^{(\omega)}(y)\, e^{2ik_1z},
\end{align}
where the indices $i,j,k$ run over the coordinate axes $x,y,z$. To further evaluate the model, we make the simplifying assumption that only the vertical component of the induced $\chi^{(2)}$ nonlinearity predominantly contributes to SHG between the TM modes. That is, we assume that $d_{ijk}=0$ except when $i=j=k=y$~\cite{Timurdogan2017E-FISH}. The programmed spatial distribution of the nonlinearity is generally denoted as
\begin{align}
    d_{yyy}(x,y,z) = \begin{cases}
         \frac{\chi^{(2)}_{yyy}}{2}\, r(x,z) +\frac{\chi^{(2)}_\text{const}}{2}& \text{for } |y|\leq d_\text{core}/2, \\
         0 & \text{otherwise},
    \end{cases}
\end{align}
with the maximum programmable nonlinearity $\chi^{(2)}_{yyy}$. The function $0 \leq r(x,z) \leq 1$ represents the dynamically programmable distribution of $\chi^{(2)}$ nonlinearity. We assume that any contribution from the cladding to the nonlinearity is negligible. The constant background nonlinearity $\chi^{(2)}_\text{const}$ is caused by the non-zero bias electric field that is present even when the programming illumination is off. While the absolute maximum $\chi^{(2)}$ nonlinearity on the device is $\chi^{(2)}_\text{tot}=\chi^{(2)}_{yyy}+\chi^{(2)}_\text{const}$, only the programmable part with spatial variation can contribute to meaningful nonlinear-optical processes. This is because the contributions from $\chi^{(2)}_\text{const}$ average out over propagation due to phase mismatch. Therefore, we ignore the contribution from $\chi^{(2)}_\text{const}$ below unless otherwise specified. By using a thicker photoconductor layer, we can reduce the background nonlinearity $\chi^{(2)}_\text{const}$ and realize programmable nonlinearity approaching $\chi^{(2)}_\text{tot}$.

Overall, we obtain
\begin{align}
\label{eq:FH-eom}
\partial_z a &= \frac{i}{2k_1}\partial_x^2 a - i\kappa\, e^{i\Delta k\,z}\, r(x,z)\, a^*b,\\[1ex]
\label{eq:SH-eom}
\partial_z b &= \frac{i}{2k_2}\partial_x^2 b - i\kappa\, e^{-i\Delta k\,z}\, r(x,z)\, a^2,
\end{align}
with a nonlinear coupling given by
\begin{align}
\label{eq:kappa-equation}
\kappa = \frac{\epsilon_0\omega \chi^{(2)}_{yyy}}{4 \sqrt{W^3}}\left(\int_\text{core}\mathrm{d}y\, E_y^{(2\omega)*}(E_y^{(\omega)})^2\right)
=\frac{\omega \chi^{(2)}_{yyy}}{2 c\sqrt{L_\text{eff}}}\sqrt{\frac{2Z_0}{n_\text{eff}^{(\omega)2}n_\text{eff}^{(2\omega)}}}.
\end{align}
The effective mode width is defined as
\begin{align}
L_\text{eff} = \frac{\left(L_\text{mode}^{(\omega)}\right)^2 L_\text{mode}^{(2\omega)}}{\left[\int_\text{core}\mathrm{d}y\, e_y^{(2\omega)}\bigl(e_y^{(\omega)}\bigr)^2\right]^2}.
\end{align}
The equations of motion \eqref{eq:FH-eom} and \eqref{eq:SH-eom} are the main results of this section and can be used to simulate the SHG dynamics for a given distribution of $\chi^{(2)}$ nonlinearity, i.e., $r(x,z)$.

\section{Common parts of the experiment}
\label{appendix:common-part}
In this section, we describe the parts of the experimental setup that were commonly used for all the experiments in this work. 
\subsection{Projector setup for programming illumination}
\label{appendix:SLM}

\begin{figure}[h]
    \centering
    \includegraphics[width=0.9\linewidth]{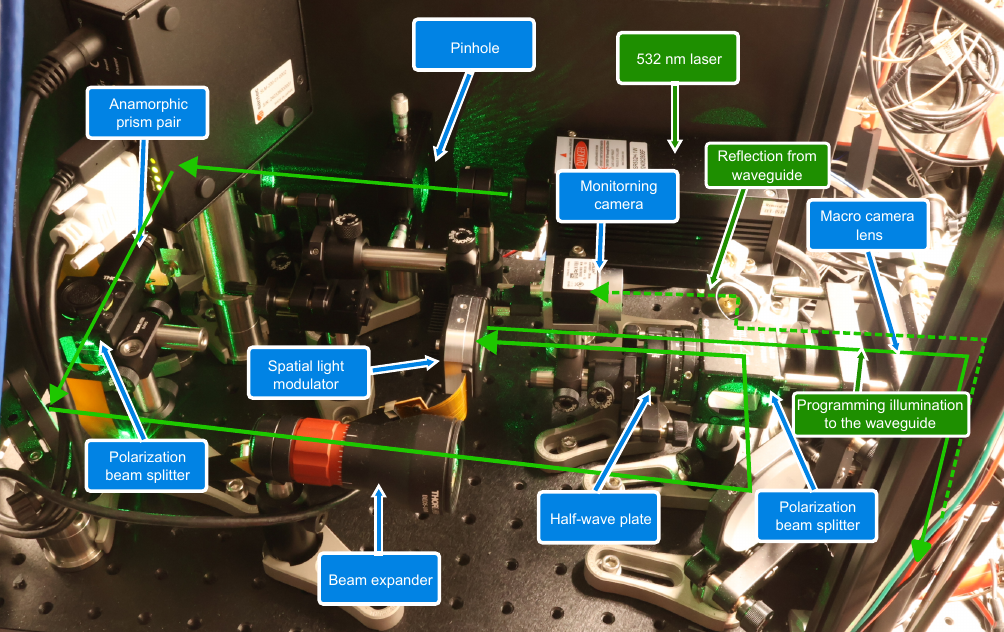}
    \caption{Photograph of the projector setup used to generate the programming illumination. The solid green lines indicate the forward laser path toward the surface of the programmable nonlinear waveguide, and the dashed green lines indicate the reflected light path from the waveguide surface. The light blue boxes and arrows indicate the essential optical components.}
    \label{fig:projector}
\end{figure}

The details of the projector setup used to produce the programming illumination are described here. A photograph of the setup is shown in Fig.~\ref{fig:projector}. The primary light source was a green diode laser with a wavelength of $\SI{532}{nm}$. To clean the spatial mode, the laser output was first focused through a pinhole and then collimated with a lens. The spatial dimensions of the beam were tailored via an anamorphic prism pair and a beam expander. The anamorphic prism pair expanded the beam in the horizontal direction, and the beam expander adjusted the overall beam size. 

The spatial intensity was modulated via a spatial light modulator (SLM-200-01; Santec) in combination with a polarization beam splitter and a half-wave plate. The SLM had a resolution of $1920\times 1200$ pixels, each with a pitch of $\SI{8}{\micro\meter}$, and supported 10-bit grayscale resolution. The SLM pattern was projected onto the surface of the programmable nonlinear waveguide via a macro camera lens (Milvus 100\,mm f/2M Lens; Carl Zeiss). The demagnification ratio of the setup was determined by imaging a test grating with a known period onto a monitoring camera, which was placed at the same distance from the test grating surface as the SLM. Through this calibration process, we measured the demagnification ratio as $2.1209$, meaning that each $\SI{8}{\micro\meter}$ pixel on the SLM corresponds to a feature size of $\SI{\wpixel}{\micro\meter}$ on the waveguide surface. 

The update speed of the illumination pattern was limited by the SLM response time; typically, the system required approximately $\SI{1}{\second}$ to reach a steady state after the SLM pattern was updated. This limitation can be addressed by employing a faster SLM.

\subsection{Electrical and optical coupling to the waveguide}

\begin{figure}[h]
    \centering
    \includegraphics[width=0.85\linewidth]{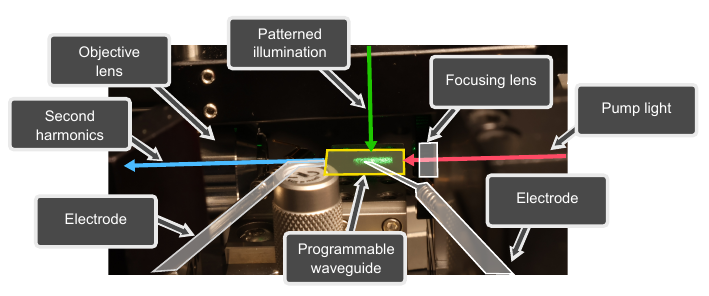}
    \caption{Main part of the experimental setup, where we programmed the nonlinearity, pumped the waveguide, and collected the generated second harmonics. }
    \label{fig:main-experiment}
\end{figure}

As shown in Fig.~\ref{fig:main-experiment}, a programmable nonlinear waveguide was mounted on a micrometer translation stage. The programming illumination from the projector setup (see Appendix~\ref{appendix:SLM}) was applied to the top surface of the waveguide. From the side, a pump laser focused by an aspheric focusing lens (C660TME-C; Thorlabs) was coupled to the waveguide. Bias electric fields were applied via a pair of electrodes, one attached to the top transparent electrode and the other connected to the waveguide substrate. The electrically induced $\chi^{(2)}$ nonlinearity produced the SHG, which was collected by an objective lens. Various objective lenses were used depending on the experimental requirements, as described later.

\section{Programmable periodic poling for CW-pumped SHG}
\label{appendix:cw-pumped-SHG}
This section describes the experimental details of the results presented in Sec.~\ref{sec:periodic-poling}. 

\subsection{Calibration of the experimental setup}
\label{appendix:calibration-cw}
 
\begin{figure}[h]
    \centering
    \includegraphics[width=0.8\linewidth]{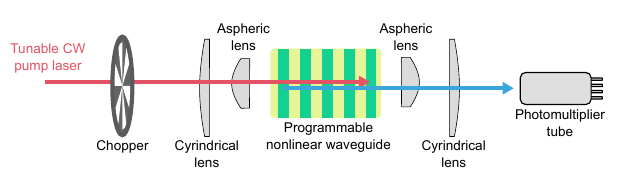}
    \caption{Illustration of the CW-pumped programmable SHG experiment. On the input side, the cylindrical lens had a focal length of $f=\SI{25}{mm}$, and the aspheric lens was a C660TME-C (Thorlabs). For the output side, we used a cylindrical lens and an aspheric lens C330TME-B (Thorlabs). The generated SH signal was detected via a PMT (PMT H10722-20; Hamamatsu Photonics) with a control voltage of $V_\text{control}=\SI{0.5}{V}$.}
    \label{fig:CW-SHG}
\end{figure}

In Fig.~\ref{fig:CW-SHG}, we show an illustration of the experimental setup. The pump light from a CW laser (TSL-570; Santec), with a wavelength tunable between $\lambda=\SI{1500}{nm}\sim\SI{1630}{nm}$, was coupled into a programmable nonlinear waveguide via a pair of focusing lenses. The combination of the first cylindrical lens and a focusing aspheric lens approximately collimated the beam in the horizontal (i.e., $x$-) direction while tightly focusing it in the vertical (i.e., $y$-) direction. The generated SH light was then collimated via an aspheric lens (used as an objective) and a cylindrical lens and was detected via a photomultiplier tube (PMT) through short-pass filters that rejected the pump light.

We applied a total bias voltage of $V_\text{tot}=1000\,\si{V}$ at a frequency of $\SI{5}{Hz}$ unless otherwise specified, which was found to be approximately optimal (see Appendix~\ref{appendix:electric-properties}). During the measurement, an optical chopper wheel modulated the pump light, and we used the difference between the on- and off-state signals as the measurement outcome. This lock-in-like procedure allowed us to reject DC noise in the signal. The raw data were recorded as the signal voltage from the PMT, which we converted to obtain the normalized SHG conversion efficiency. Several calibration and normalization procedures were required for this conversion.

First, we calibrated the sensitivity of the PMT, including the losses incurred by the filters. A measurement using a power reference at $\SI{780}{nm}$ yielded a sensitivity of $\SI{1.34e9}{V/W}$. Using this value, we converted the signal voltage from the PMT to the detected SH power $P_\text{SH}^{\text{detected}}$. To determine the SH power generated inside the chip, we accounted for the collection efficiency, $R^\text{collect}=\SI{75}{\%}$, from the chip to free space, which was independently calibrated. Additionally, the propagation loss of the SH was estimated as 
\begin{align}
R_\text{SH}^\text{prop}=\exp(-\alpha_\text{SH}(L_\text{tot}-L_\text{SHG})),
\end{align}
where $L_\text{tot}=\SI{1.8}{cm}$ is the total length of the waveguide and $L_\text{SHG}=\SI{0.6}{cm}$ is the nominal location of the SHG on the chip measured from the input facet. We use the material loss characterized in Appendix~\ref{appendix:optical-loss} to obtain the attenuation coefficient $\alpha_\text{SH}$. Finally, the $\chi^{(2)}$ nonlinearity was modulated sinusoidally by the AC bias electric field, so the peak SHG power was twice its average. Overall, the calibrated SH power was given by
\begin{align}
P_\text{SH}= \frac{2\,P_\text{SH}^{\text{detected}}}{R^\text{collect}\,R_\text{SH}^\text{prop}}.
\end{align}

\begin{figure}[h]
    \centering
    \includegraphics[width=0.5\linewidth]{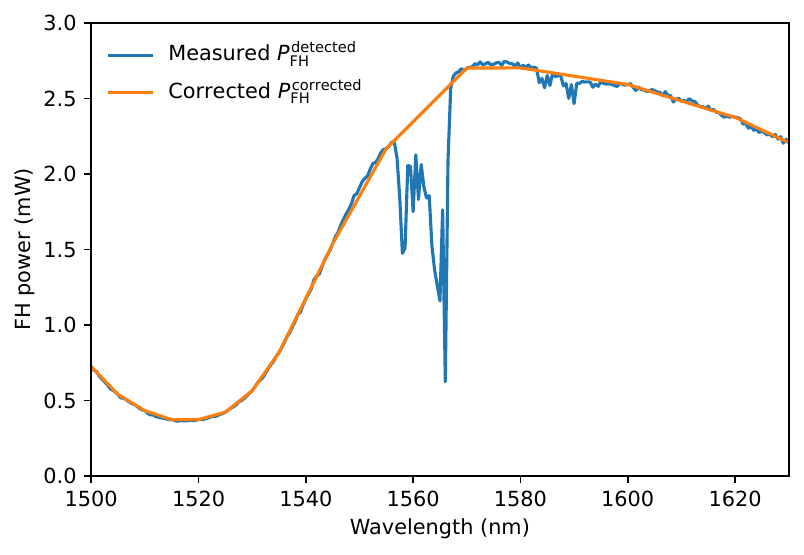}
    \caption{Experimentally measured pump power accounting for the collection efficiency $P_\text{FH}^\text{detected}$(blue lines) and the pump power $P_\text{FH}^\text{corrected}$ corrected via interpolation over the region with photoconductor-induced loss around $\SI{1560}{nm}$ (orange lines).}
    \label{fig:P_FH}
\end{figure}

Second, we estimated the pump power $P_\text{FH}$ that effectively contributed to SHG. To do this, we first measured the FH power $P_\text{FH}^\text{detected}$, accounting for the collection efficiency $R^\text{collect}$, as shown in Fig.~\ref{fig:P_FH}. Notably, the dips around $\SI{1560}{nm}$ were caused by leakage into the photoconductor mode (see Appendix~\ref{appendix:photoconductor-mode}). Owing to the coherent nature of this leakage, we cannot simply apply an exponential decay model to estimate $P_\text{FH}$. Instead, we used smooth interpolation over the dip to obtain $P_\text{FH}^\text{corrected}$, as shown in the figure. We then accounted for the propagation loss 
\begin{align}
R_\text{FH}^\text{prop}=\exp(-\alpha_\text{FH}(L_\text{tot}-L_\text{SHG}))
\end{align}
by using the material absorption described in Appendix~\ref{appendix:material-absorption} for $\alpha_\text{FH}$. Overall, we obtained
\begin{align}
P_\text{FH}= \frac{P_\text{FH}^\text{corrected}}{R_\text{FH}^\text{prop}}.
\end{align}

Combining all the measurements, we obtained our best estimate for the normalized SHG conversion efficiency
\begin{align}
    \eta_\text{norm} = \frac{P_\text{SH}}{P_\text{FH}^2}.
\end{align}
In our programmable nonlinear waveguide, the value of $\eta_\text{norm}$ was lower than that of a ridge waveguide with the same nonlinearity because of the relatively loose transverse confinement inherent to the slab waveguide geometry.

\subsection{Basic nonlinear-optical characterization of the device}

To produce the data shown in Fig.~\ref{fig:periodic-poling}(b), we projected grating patterns with periods ranging from $\Lambda=\SI{16.3}{\micro m}$ to $\SI{16.9}{\micro m}$, and we linearly scanned the pump wavelength, $\lambda$, while measuring the signal with a photodetector. The measured signal was converted to a normalized efficiency via the procedures described in Appendix~\ref{appendix:calibration-cw}. We fit sinc functions to the measured peaks, which allowed us to obtain pairs of poling periods $\Lambda$ and pump wavelengths $\lambda$ at which the SHG was phase-matched. Fitting the relationship between $\Lambda$ and $\lambda$ as a quadratic function around $\lambda=\SI{1560}{nm}$ yielded an optimal poling period of $\Lambda=\SI{16.685}{\micro m}$ and a group velocity mismatch (GVM) between the fundamental and second harmonics of $\SI{-92}{fs/mm}$. Notably, this direct measurement of the GVM on a single device was enabled by the programmability of the poling period, $\Lambda$.

\subsection{Real-time feedback to compensate for random walks in the pump wavelength}

\begin{figure}[h]
    \centering
    \includegraphics[width=0.8\linewidth]{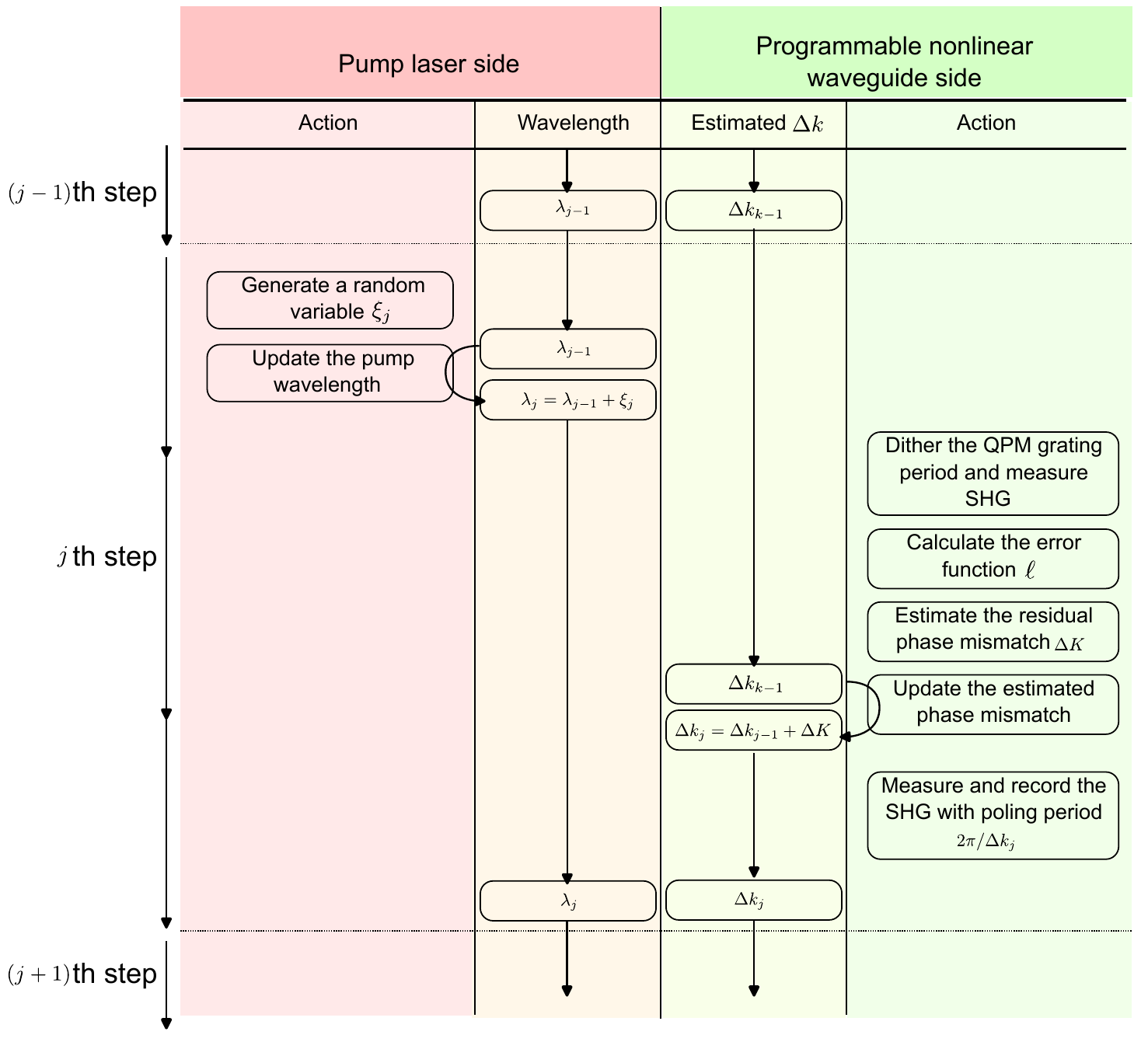}
    \caption{Illustration of the experimental sequence used to implement real-time feedback to compensate for random walks in the pump wavelength.}
    \label{fig:feedback}
\end{figure}

In Fig.~\ref{fig:periodic-poling}(c), we present a proof-of-concept demonstration of the utility of programmable nonlinearity by using programmable poling to compensate for random fluctuations in the pump laser wavelength. As shown in Fig.~\ref{fig:feedback}, the experiment proceeded in discrete iteration steps. At the beginning of the $j$th step, we inherited the pump wavelength $\lambda_{j-1}$ and the best estimate for the phase mismatch $\Delta k_{j-1}$ from the previous $(j-1)$th step.

First, on the side of the pump laser, we applied artificial shifts to the pump wavelength to emulate a random walk. This was achieved by generating a random number $\xi_j$ and updating the pump wavelength to $\lambda_{j} = \lambda_{j-1} + \xi_j$. The random variable $\xi_j$ followed a Gaussian distribution with a standard deviation of $\SI{1}{nm}$, so the sequence $\{\lambda_j\}$ followed a Gaussian random walk process. We denote the phase mismatch for the SHG at $\lambda_j$ as $\tilde{\Delta k_j}$.

\begin{figure}[h]
    \centering
    \includegraphics[width=0.55\linewidth]{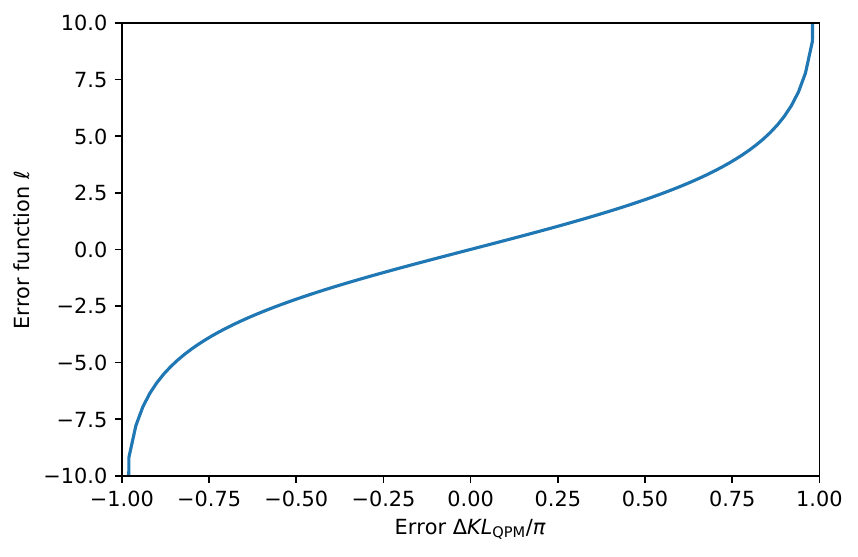}
    \caption{Numerically calculated error function \eqref{eq:error-function}.}
    \label{fig:error_func}
\end{figure}

Then, on the side of the programmable nonlinear waveguide, we leveraged programmability to estimate a poling period that maximized the SHG efficiency for the drifted pump wavelength. Since the programmable nonlinear waveguide did not have knowledge of $\lambda_j$, the value of $\tilde{\Delta k_j}$ was initially unknown. To estimate this value experimentally, we dithered the poling period on the programmable waveguide by measuring the SHG signals at two poling periods, $\Lambda_\pm = 2\pi/(\Delta k_{j-1} \pm \epsilon)$, where $\epsilon$ is a small positive constant. Theoretically, the measured SHG powers were expected to follow
\begin{align}
P_\pm = c\,\mathrm{sinc}^2
\left[\frac{1}{2}(\Delta K \mp \epsilon)L_\text{QPM}\right],
\end{align}
where $c$ is a positive constant, $L_\text{QPM}$ is the length of the phase-matched region, and $\Delta K = \tilde{\Delta k_j} - \Delta k_{j-1}$ represents the error in our estimated phase mismatch. Intuitively, if $P_+ > P_-$ ($P_+ < P_-$), then we were underestimating (overestimating) the value of $\tilde{\Delta k_j}$, and this indicated how the estimate should be updated. We formally defined an error function
\begin{align}
    \ell = \log\frac{P_+}{P_-}
\end{align}
to quantify this imbalance. To maximize the sensitivity of $\ell$, we chose $\epsilon = \pi/L_\text{QPM}$, which yielded
\begin{align}
\label{eq:error-function}
    \ell(\Delta K) = \log\Bigl(\mathrm{sin}^2\bigl(\Delta K L_\text{QPM}/2-\pi/2\bigr)\Bigr) - \log\Bigl(\mathrm{sin}^2\bigl(\Delta K L_\text{QPM}/2+\pi/2\bigr)\Bigr).
\end{align}
Note that the error function \eqref{eq:error-function} depends only on $\Delta K$. Thus, by experimentally measuring $\ell$, we could invert \eqref{eq:error-function} to obtain an estimate for $\Delta K$, and we updated our phase mismatch estimate to $\Delta k_j = \Delta k_{j-1} + \Delta K$. In Fig.~\ref{fig:error_func}, we show a plot of $\ell$. Finally, we measured the SHG efficiency via the updated poling period, $2\pi/\Delta k_j$, and this normalized SHG efficiency is shown in Fig.~\ref{fig:periodic-poling}(c).

In the experiment that produced the data for Fig.~\ref{fig:periodic-poling}(c), we used $L_\text{QPM}=\SI{6500}{\micro m}$. We had 300 iteration steps, and each iteration step took $\SI{9.3}{s}$.

\section{Spectral engineering}
\label{appendix:spectral}
In this section, we present the experimental details of the results shown in Sec.~\ref{sec:spectral}. Specifically, we explain how the data presented in Fig.~\ref{fig:spectral} were experimentally produced.

\subsection{Broadband SHG with manually designed QPM gratings}
\label{appendix:broadband-SHG}

\begin{figure}[h]
    \centering
    \includegraphics[width=0.75\linewidth]{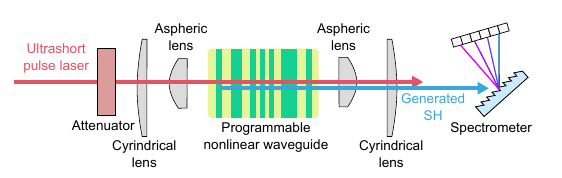}
    \caption{Illustration of the experimental setup used to perform a broadband SHG experiment. The lens systems were the same as those in Fig.~\ref{fig:CW-SHG}. The programmable nonlinear waveguide was pumped by a pulse laser (ELMO-HP; Menlo systems) with a pulse duration of $\approx \SI{45}{fs}$ and an average power of $\SI{180}{mW}$. The pump light was attenuated by a neutral density filter before coupling to the waveguide. The spectrum of the generated SH light was measured with a spectrometer (OCEAN-FX-VIS-NIR-ES; Ocean Optics).}
    \label{fig:broadband-SHG}
\end{figure}

First, we describe how we produced the data in Fig.~\ref{fig:spectral}(a), where manually designed QPM grating structures were used to obtain programmable broadband SHG. An illustration of the experimental setup is shown in Fig.~\ref{fig:broadband-SHG}. As shown in the figure, we pumped a programmable nonlinear waveguide using a pulse laser. On the output side, a spectrometer measured the spectrum of the generated SH light as we varied the structure of the patterned illumination on the waveguide surface. We used the bias voltage of $V_\text{tot}=\SI{500}{V}$ for this experiment. The resulting spectra in Fig.~\ref{fig:spectral}(a) were normalized with respect to the peak value in each plot.

For these broadband SHG experiments, the illumination patterns on the waveguide were uniform in the transverse direction, whereas nontrivial QPM grating patterns were engineered in the longitudinal direction. In other words, the two-dimensional illumination pattern $I(x,z)$ had the form $I(x,z)=I(z)$. The simplest pattern we considered was a monotonic grating pattern 
\begin{align}
    I_\text{monotonic}(z)=\mathrm{H}\left(\sin (2\pi z/\Lambda)\right),
\end{align}
where $\Lambda$ is the spatial period and $H(x)$ is the Heaviside step function. Here, $I$ was discretized in space and became the direct grayscale input to the SLM in the projector setup (see Appendix~\ref{appendix:SLM}). The resulting QPM grating phase matched the SHG for a particular wavelength, yielding a solitary peak in the output SH spectrum, as shown in Fig.~\ref{fig:spectral}(a-i) for $\Lambda=\SI{16.64}{\micro m}$.

To perform multiple SHG processes simultaneously, we superimposed gratings with various periods. The complete grating structure was given by
\begin{align}
    I_\text{multi-peaks}(z)=\mathcal{N}\sum_{j=1} c_j\,\mathrm{H}\left(\sin (2\pi z/\Lambda_j)\right)+\mathcal{C},
\end{align}
where $c_j$ denotes the relative weights. In Fig.~\ref{fig:spectral}(a-ii), we show the results for $\Lambda_1=\SI{16.36}{\micro m}$, $\Lambda_2=\SI{16.53}{\micro m}$, $\Lambda_3=\SI{16.72}{\micro m}$, and $\Lambda_4=\SI{16.88}{\micro m}$, with corresponding weights $c_1=0.4$, $c_2=0.17$, $c_3=0.12$, and $c_4=0.15$. The weights were chosen so that the heights of the peaks in the spectrum were similar. The normalization constant $\mathcal{N}$ and the constant $\mathcal{C}$ were set so that $\min_z I(z)=0$ and $\max_z I(z)=1$.

\begin{figure}[h]
    \centering
    \includegraphics[width=\linewidth]{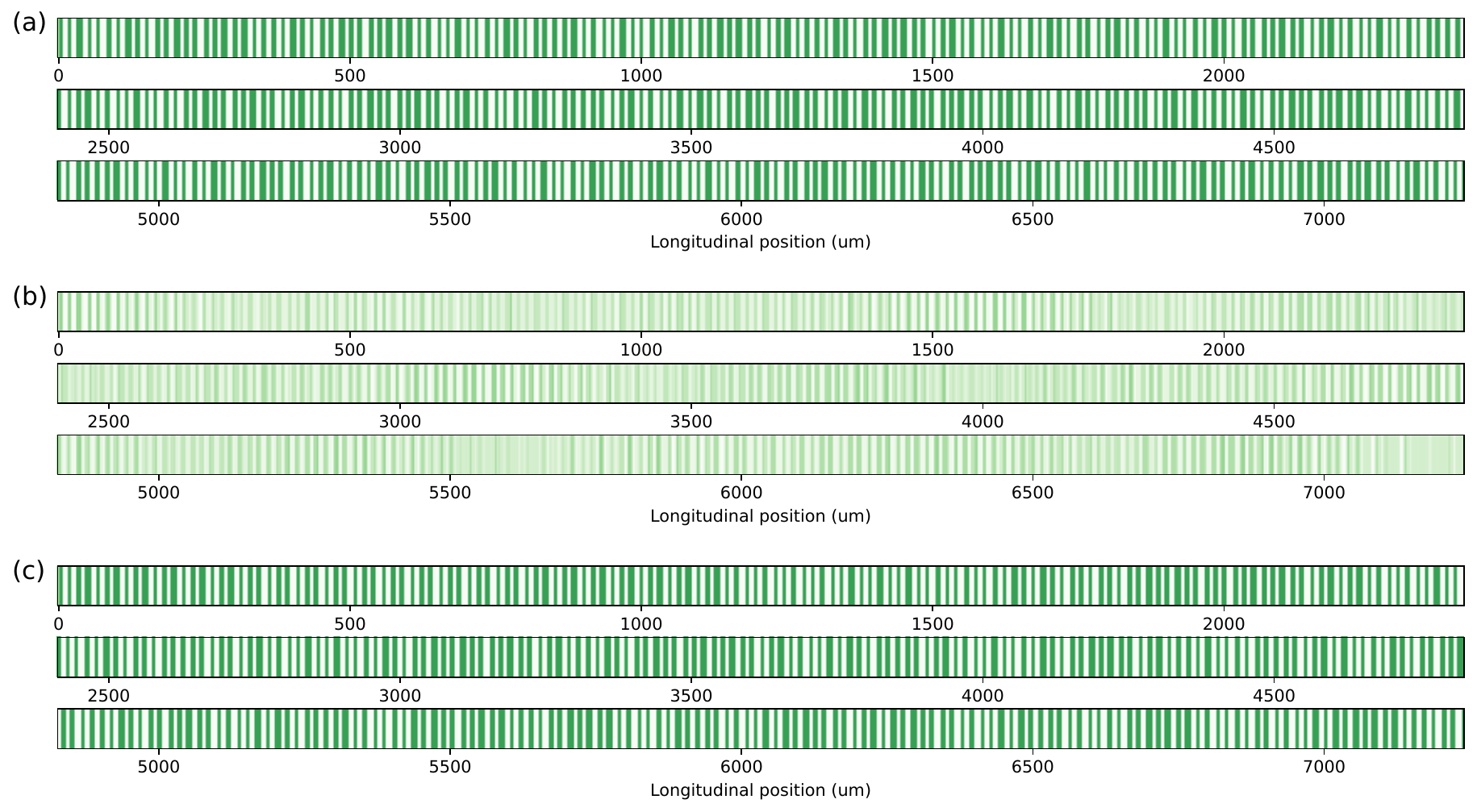}
    \caption{The final illumination patterns $I(z)$ obtained through the optimization process. (a), (b), and (c) indicate Fig.~\ref{fig:spectral}(a-i), \ref{fig:spectral}(a-ii), and \ref{fig:spectral}(a-iii), respectively. }
    \label{fig:full-poling-pattern-analytic}
\end{figure}

We can also achieve simultaneous phase matching for broadband SHG by adiabatically chirping the QPM grating. To generate such an adiabatic AFC, as shown in Fig.~\ref{fig:spectral}(a-iii), we varied the grating period adiabatically from $\Lambda_\text{ini}=\SI{16.3}{\micro m}$ to $\Lambda_\text{fin}=\SI{16.9}{\micro m}$. Quantitatively, the QPM grating structure was given by
\begin{align}
    I_\text{adiabatic}(z)=\mathrm{H}\Bigl(\sin\bigl(\theta_\text{adiabatic}(z)\, z\bigr)\Bigr).
\end{align}
The accumulated grating phase was defined as
\begin{align}
    \theta_\text{adiabatic}(z)=\int_0^{L_\text{img}}\mathrm{d}z'\, k_\text{adiabatic}(z'),
\end{align}
and the local wavenumber of the grating was expressed by
\begin{align}
    k_\text{adiabatic}(z)=\frac{2\pi}{\Lambda_\text{ini}}\frac{L_\text{img}-z}{L_\text{img}} + \frac{2\pi}{\Lambda_\text{fin}}\frac{z}{L_\text{img}}.
\end{align}
Here, $L_\text{img}$ denotes the total distance in the longitudinal dimension over which the programming illumination was projected. As shown in Fig.~\ref{fig:spectral}(b-iii), the resulting SHG spectrum was extremely broad, spanning over $\SI{50}{nm}$ in bandwidth. The full illumination patterns used in these experiments are shown in Fig.~\ref{fig:full-poling-pattern-analytic}.

\subsection{In situ optimization of the QPM grating}
\label{appendix:in-situ-optimization-experiment}

A key capability of our programmable nonlinear waveguide is its ability to modify the QPM grating structure in real time, which allowed us to optimize the grating on the basis of real-time experimental feedback. For example, in Fig.~\ref{fig:spectral}(b), we show how optimizing the QPM grating could shape the SH spectrum into the desired form. Below, we describe the experimental procedure used to achieve these results.

To enable flexible optimization of the QPM grating structure, we parameterized the illumination pattern $I(z)$ using several free parameters. First, we partitioned the entire imaging window of length $L_\text{img}$ into $N_\text{opt}=20$ sections of equal length, $\Delta z_{\text{opt}}=L_\text{img}/N_\text{opt}$. The overall illumination pattern was defined as
\begin{align}
    I(z)=\frac{c_\text{opt}(z)}{2}\left\{\sin\bigl(\theta_\text{opt}(z)z\bigr)+1\right\},
\end{align}
where $c_\text{opt}(z)=c_j$ for $(j-1)\Delta z_{\text{opt}}\leq z<j\Delta z_{\text{opt}}$, meaning that $c_\text{opt}(z)$ takes the value $c_j$ in the $j$th section. The overall phase function $\theta_\text{opt}(z)$ was 
defined as
\begin{align}
    \theta_\text{opt}(z)=\int_0^z\mathrm{d}z'\,k_\text{opt}(z'),
\end{align}
where $k_\text{opt}(z)$ is the local wavenumber of the QPM grating. We parameterized $k_\text{opt}(z)$ so that it increased linearly within each section, where the rate of increase depended on the section. In other words,
\begin{align}
    k_\text{opt}(z)=\int_0^z\mathrm{d}z'\,\mu_\text{opt}(z) + \frac{2\pi}{\Lambda_\text{ini}},
\end{align}
where $\mu_\text{opt}(z)=\mu_j\geq 0$ for $(j-1)\Delta z_{\text{opt}}\leq z<j\Delta z_{\text{opt}}$. We set the initial poling period to $\Lambda_\text{ini}=\SI{16.42}{\micro m}$. Overall, we obtained a monotonically chirped adiabatic QPM grating that was parameterized by $2N_\text{opt}$ parameters, i.e., $\{c_j\}$ and $\{\mu_j\}$ with $j\in\{1,2,\dots,N_\text{opt}\}$.

\begin{figure}[bth]
    \centering
    \includegraphics[width=0.8\linewidth]{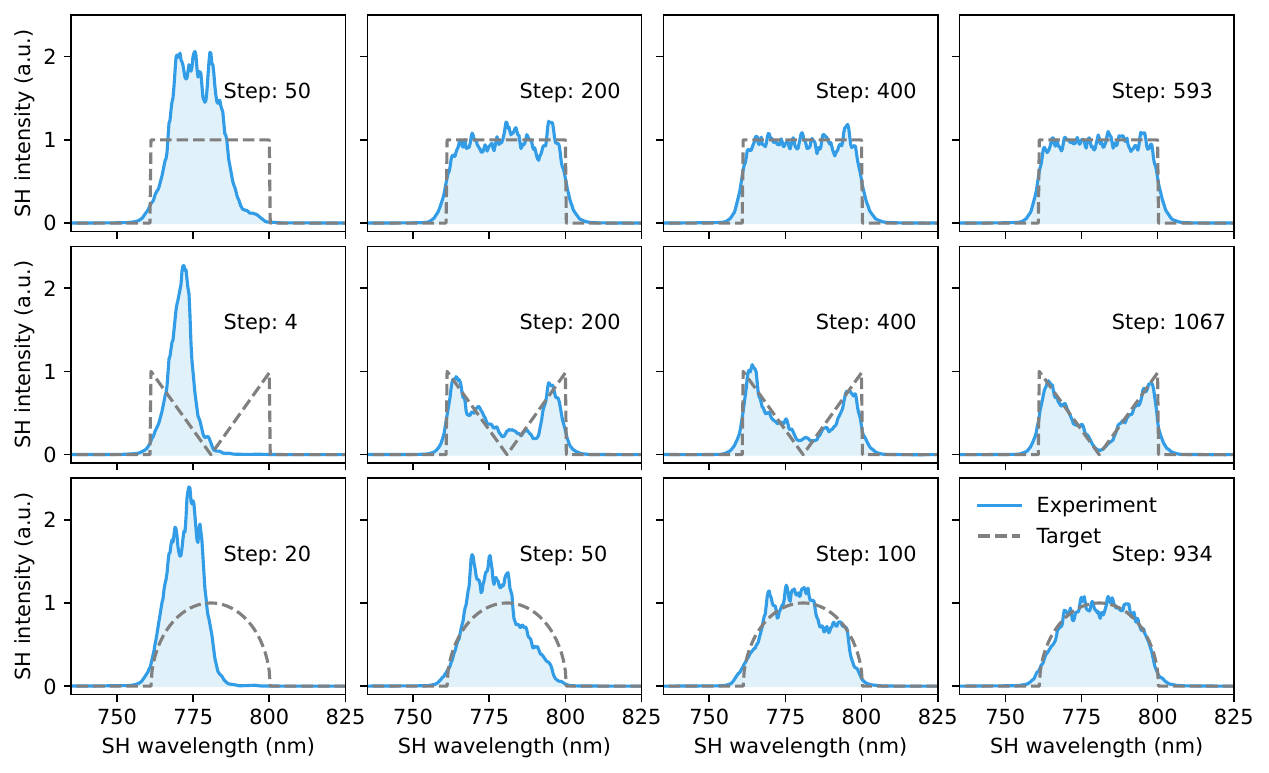}
    \caption{Broadband SH spectra (blue shaded regions) that were experimentally measured during in situ optimization. The gray dashed lines represent the target spectra, which were normalized with respect to their peak values, and the experimental SH spectra were normalized so that the area under each curve matched that of the respective target spectrum. The first, second, and last rows correspond to the optimization steps for the data shown in Fig.~\ref{fig:spectral}(b-i), Fig.~\ref{fig:spectral}(b-ii), and Fig.~\ref{fig:spectral}(b-iii), respectively. The optimization step counts are indicated as integer numbers on the plots. }
    \label{fig:pulse-shaping-evolution}
\end{figure}

\begin{figure}[h]
    \centering
    \includegraphics[width=\linewidth]{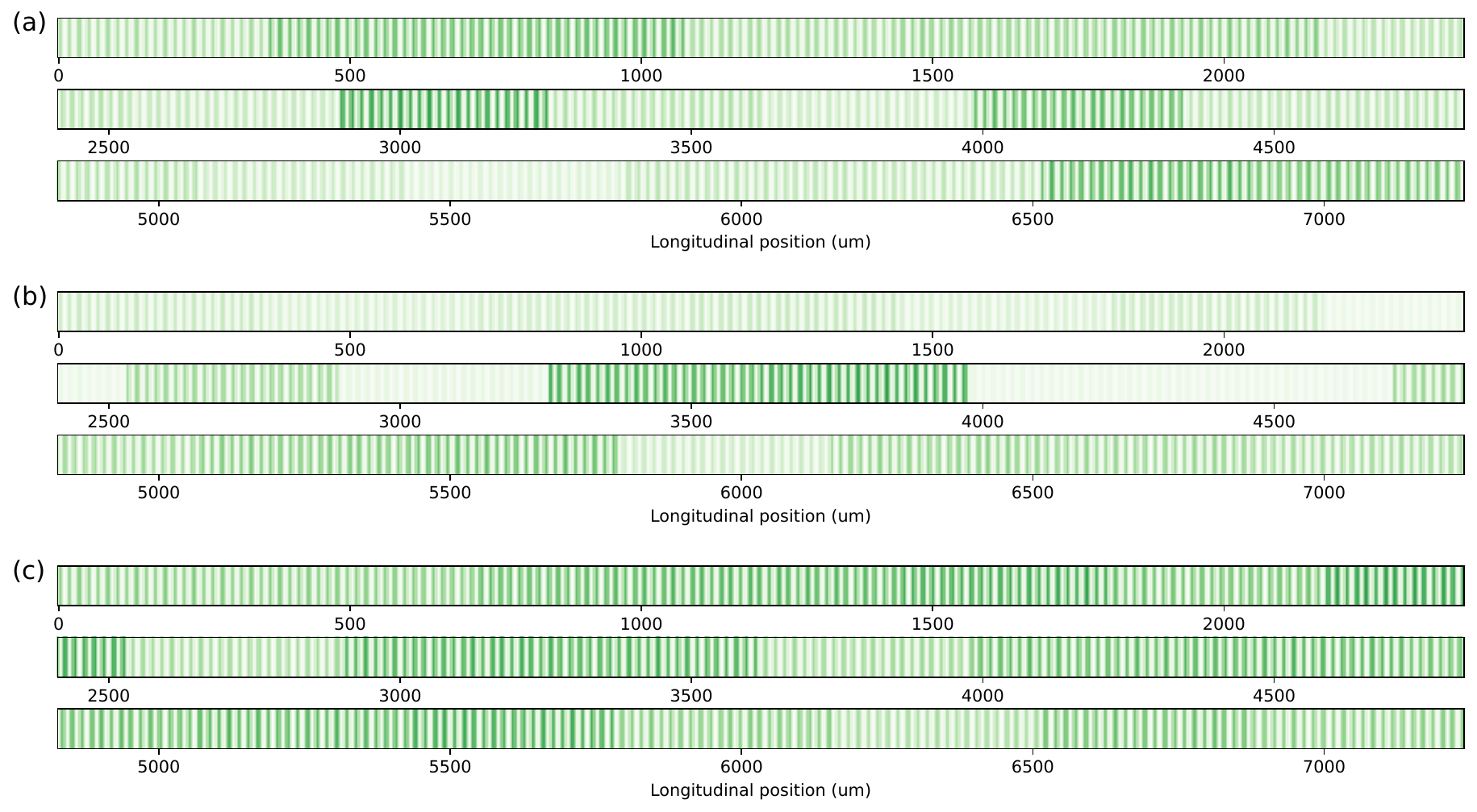}
    \caption{The final illumination patterns $I(z)$ obtained through the optimization process. (a), (b), and (c) indicate Fig.~\ref{fig:spectral}(b-i), \ref{fig:spectral}(b-ii), and \ref{fig:spectral}(b-iii), respectively. }
    \label{fig:full-poling-pattern}
\end{figure}

At each optimization step, we first measured the SH spectrum, $S_\text{measured}(\lambda)$, as a function of the wavelength $\lambda$. We then computed the normalized distance between $S_\text{measured}(\lambda)$ and the target spectrum $S_\text{target}(\lambda)$:
\begin{align}
    \mathcal{D}=\frac{\left(\int \mathrm{d}\lambda\,\bigl|S_\text{measured} - S_\text{target}\bigr|^2\right)^{1/2}}{\left(\int \mathrm{d}\lambda\,S_\text{measured}\right)\left(\int \mathrm{d}\lambda\,S_\text{target}\right)}.
\end{align}
Next, we proposed an update to the QPM grating by applying a small perturbation to the parameters $\{c_j\}$ and $\{\mu_j\}$. If the perturbation decreased $\mathcal{D}$, we accepted the update; otherwise, we rejected it and proceeded to the next step. After many iterations, $S_\text{measured}(\lambda)$ was expected to converge toward the target spectrum. We used the bias voltage of $V_\text{tot}=\SI{800}{V}$ for this experiment. In Fig.~\ref{fig:pulse-shaping-evolution}, we show the evolution of the measured SH spectrum over different numbers of optimization steps. Although some target spectra were more challenging to achieve and required additional iterations, the output SHG spectrum eventually converged close to the target shape. In Fig.~\ref{fig:full-poling-pattern}, we present the full QPM patterns $I(z)$ obtained from the optimization. For clarity in the main text, we downsample $I(z)$ every $\SI{17}{\micro m}$ to provide a concise visualization of the QPM grating structure.

\subsection{Real-time update of the QPM grating}
\label{appendix:long-term}

\begin{figure}[h]
    \centering
    \includegraphics[width=\linewidth]{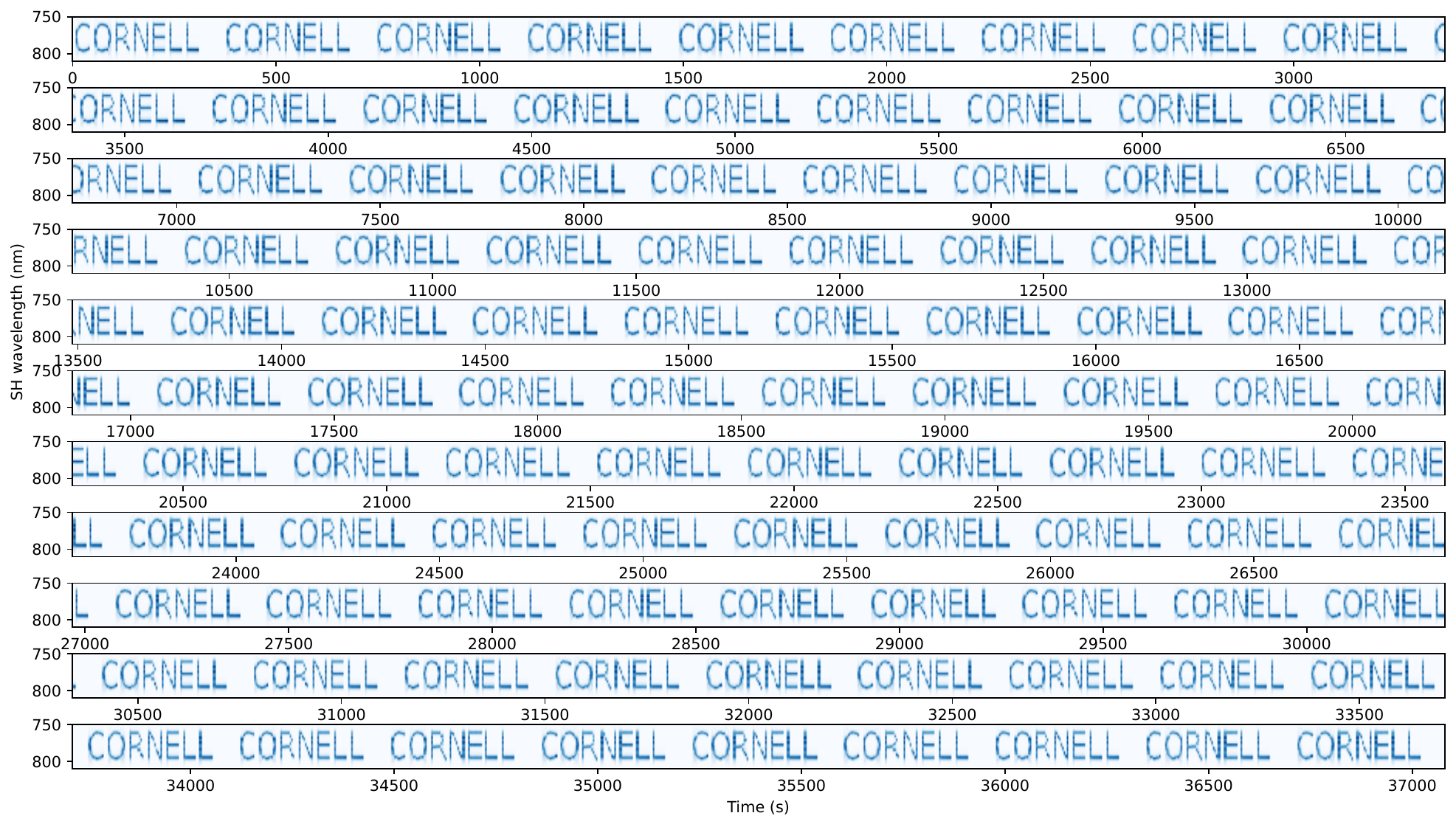}
    \caption{Time trace of the SH spectrum for over $10$ hours of operation, involving $15400$ updates to the QPM grating patterns. The illumination patterns were optimized beforehand to produce the ``Cornell'' pattern in the SH spectrum, which we repeated multiple times.}
    \label{fig:cornell-logo-long}
\end{figure}

In Fig.~\ref{fig:spectral}(d), we updated the programming illumination pattern $I(x,z)$ in real time, effectively projecting a ``movie'' onto the surface of a programmable nonlinear waveguide to achieve dynamic control of the broadband SHG spectrum. Although we show only approximately $\sim\SI{300}{s}$ of the trace in the main text, the operation was highly stable and could continue for much longer periods. In Fig.~\ref{fig:cornell-logo-long}, we present a time trace of the SHG spectrum over $\SI{10}{hours}$ of operation, during which numerous ``Cornell'' patterns were generated. This process involved reconfiguring the $\chi^{(2)}$ nonlinearity 15,400 times, and we did not observe any practical upper bound on the cycle count. These results provide compelling visual evidence of the stability and repeatability of the programmable nonlinear waveguide.

\begin{figure}[h]
    \centering
    \includegraphics[width=0.6\linewidth]{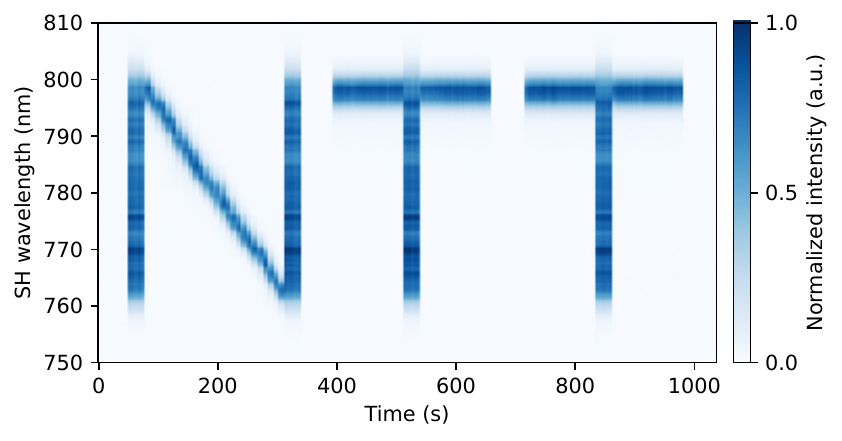}
    \caption{Time trace of the SH spectrum showing an ``NTT'' pattern. We used the same experimental procedure as in Fig.~\ref{fig:cornell-logo-long}.}
    \label{fig:ntt-logo}
\end{figure}

The pattern we could obtain in the trace of the SH spectrum was not limited to the one shown in the main text. In Fig.~\ref{fig:ntt-logo}, we give another demonstration, drawing an ``NTT'' pattern. We used the bias voltage of $V_\text{tot}=\SI{800}{V}$ for the experiments presented in this section.

\section{Spatial engineering}
\label{appendix:spatial}
In this section, we present the experimental details of the results shown in Sec.~\ref{sec:spatial-engineering}. Specifically, we explain how the experimental data and simulation results presented in Fig.~\ref{fig:spatial} were experimentally produced. Throughout the section, we use the notation introduced in Appendix~\ref{appendix:SHG-model}. 

\subsection{Calibration of the experimental setup}
\label{appendix:calibration}

\begin{figure}[h]
    \centering
    \includegraphics[width=0.8\linewidth]{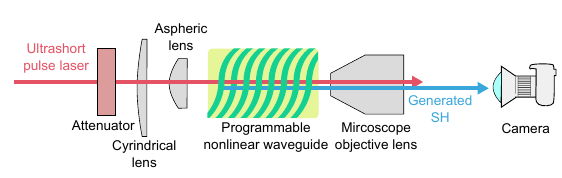}
    \caption{Experimental setup for spatial-domain control of SHG. The optics on the input side were identical to those in Fig.~\ref{fig:broadband-SHG}. On the output side, we used a microscope objective lens (PLN 10X Objective; Olympus) to image the transverse spatial profiles of the generated SH light with a camera (acA1440-220um; Basler).}
    \label{fig:spatial-SHG}
\end{figure}

In Fig.~\ref{fig:spatial-SHG}, we illustrate the experimental setup. Pulse pump light was coupled to the programmable nonlinear waveguide, and the spatial profile of the generated SH light was imaged with a camera. We used the bias voltage of $V_\text{tot}=\SI{600}{V}$ for the experiments presented in this section. First, we calibrated the magnification ratio of this imaging system that measured the SH beam profile. This calibration was performed by translating the programmable waveguide by a known distance via a micrometer stage, capturing images of the output facet, and measuring the displacement of the visible features in the camera image. Overall, we obtained a magnification ratio of $8.67$, meaning that the output SH light was magnified by a factor of $8.67$ when it reached the camera.

To achieve accurate control of the spatial profiles of SHG, the spatial profile of the pump light on the waveguide should be known. This is nontrivial in a conventional waveguide because one cannot simply ``cut open'' the waveguide to measure the field profile. Fortunately, the programmability of our platform offered a unique solution to this challenge. 

\begin{figure}[h]
    \centering
    \includegraphics[width=0.65\linewidth]{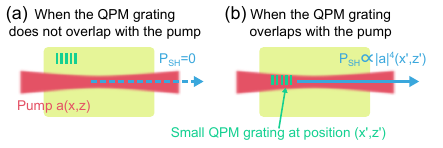}
    \caption{A method of measuring the spatial distribution of the pump intensity $|a|^2(x,z)$ on the waveguide using the programmability of the $\chi^{(2)}$ nonlinearity. (a) As we project the QPM grating pattern only within a small ROI, we observe no SHG when the ROI does not overlap with the pump light. (b) When the ROI overlaps with the pump light, we observe the SHG, whose power is proportional to the square of the local pump intensity around the ROI. }
    \label{fig:mapping}
\end{figure}

Our approach for measuring the pump light intensity distribution, $|a|^2(x,z)$, is illustrated in Fig.~\ref{fig:mapping}. In this measurement, we projected a QPM grating pattern onto the programmable waveguide, but only within a small rectangular region of interest (ROI) centered at the position $(x',z')$. If the ROI does not overlap with the pump light's intensity distribution, no SH light is detected because the SHG process is not phase matched without a QPM grating (see Fig.~\ref{fig:mapping}(a)). Conversely, when the ROI overlaps with the pump light, as shown in Fig.~\ref{fig:mapping}(b), the SHG becomes phase matched, and SH light is generated. Since the SHG power is proportional to the square of the pump intensity in the ROI, i.e., $P_\text{SH}\propto |a|^4(x',z')$, scanning the ROI across the programmable region allowed us to experimentally map out the spatial profile of $|a|^2$.

\begin{figure}[h]
    \centering
    \includegraphics[width=0.9\linewidth]{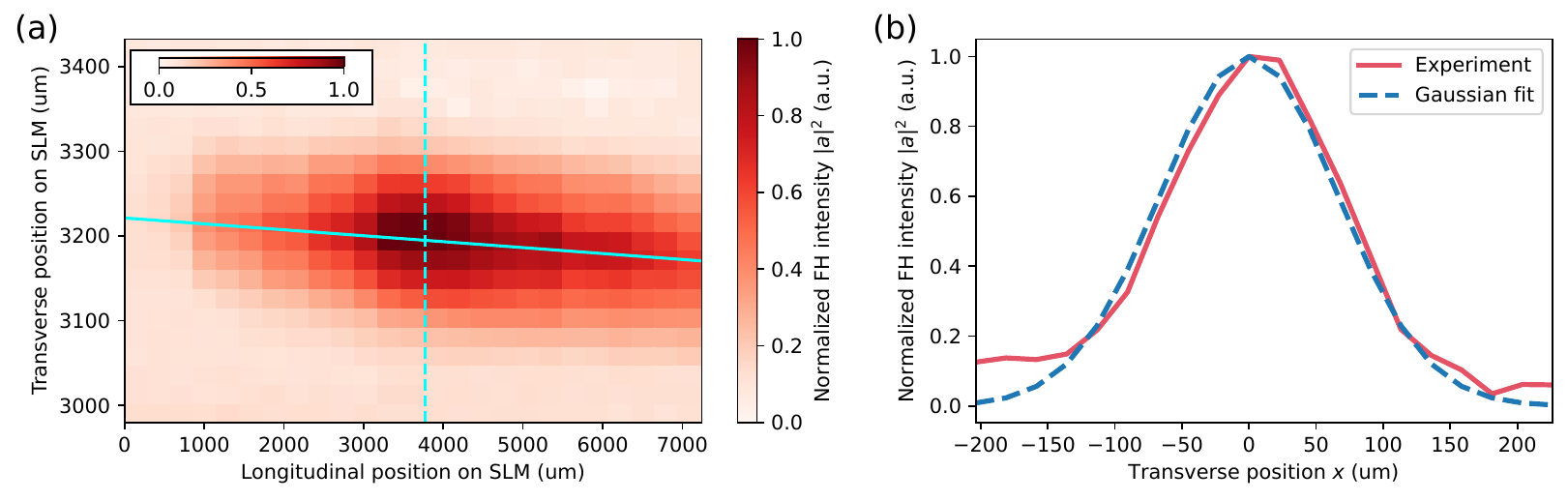}
    \caption{(a) Distribution of the intensity of the pump light mapped using the technique shown in Fig.~\ref{fig:mapping}. The coordinates were defined according to the axis of the SLM. The solid cyan line represents the approximate center of the pump beam, which defines the $z$-axis for the longitudinal coordinate with $x=0$. The dashed cyan line represents the longitudinal position $z=z_\text{0}=\SI{3772}{\micro m}$ around which the SHG was strongest. (b) Cross section of the FH beam profile at $z=z_0$, shown with a Gaussian fit. }
    \label{fig:mapping-SHG-profile}
\end{figure}

In Fig.~\ref{fig:mapping-SHG-profile}(a), we show the results of this measurement, where the experimentally measured pump-intensity distribution $|a|^2$ is presented in SLM coordinates. On the basis of this map, we defined a waveguide coordinate system---a Cartesian coordinate system aligned with the pump beam. When the pump beam was well aligned, the SLM and waveguide coordinates approximately coincided; however, small corrections were usually necessary for accurate system calibration. In Fig.~\ref{fig:mapping-SHG-profile}(a), the solid cyan line represents the centroid of the pump beam, which was tilted by $\SI{0.007}{rad}$ relative to the SLM coordinate. We designated this line as the $z$-axis of the longitudinal waveguide coordinate, which in turn defined the origin for the transverse waveguide coordinate. The origin of the longitudinal coordinate was set at the beginning of the region in which the programming illumination was projected. The dashed cyan line indicates the approximate longitudinal position where the SHG intensity was strongest, $z=z_\text{ref}=\SI{3772}{\micro m}$, which we used as a reference point in the following.

To quantitatively parameterize the beam profiles, we approximated the pump field at the reference point as a Gaussian beam:
\begin{align}
\label{eq:fh-profile-fit}
    a(x,z=z_\text{ref}) \propto e^{iu_\text{ref}x^2/2}\,e^{-x^2/w_\text{FH-ref}^2}.
\end{align}
This was a reasonable assumption since the pump beam was coupled from a single-mode fiber, which cleans up its spatial profile. In Fig.~\ref{fig:mapping-SHG-profile}(b), we show the cross-section of the FH intensity at $z=z_\text{ref}$, along with a Gaussian fit, $|a|^2 \propto \exp(-2x^2/w_\text{FH}^2)$, on the basis of Eq.~\eqref{eq:fh-profile-fit}. The beam waist at the reference position was found to be $w_\text{FH-ref}=\SI{132}{\micro m}$. The SH field generated from a thin slice around $z=z_\text{ref}$ was given by
\begin{align}
\label{eq:sh-profile-fit}
    b(x,z=z_\text{ref}) \propto a^2(x,z=z_\text{ref}) \propto e^{iu_\text{ref}x^2}\,e^{-x^2/w_\text{SH-ref}^2},
\end{align}
with a beam waist of $w_\text{SH-ref}=w_\text{FH-ref}/\sqrt{2}=\SI{93.3}{\micro m}$.

\begin{figure}[h]
    \centering
    \includegraphics[width=0.9\linewidth]{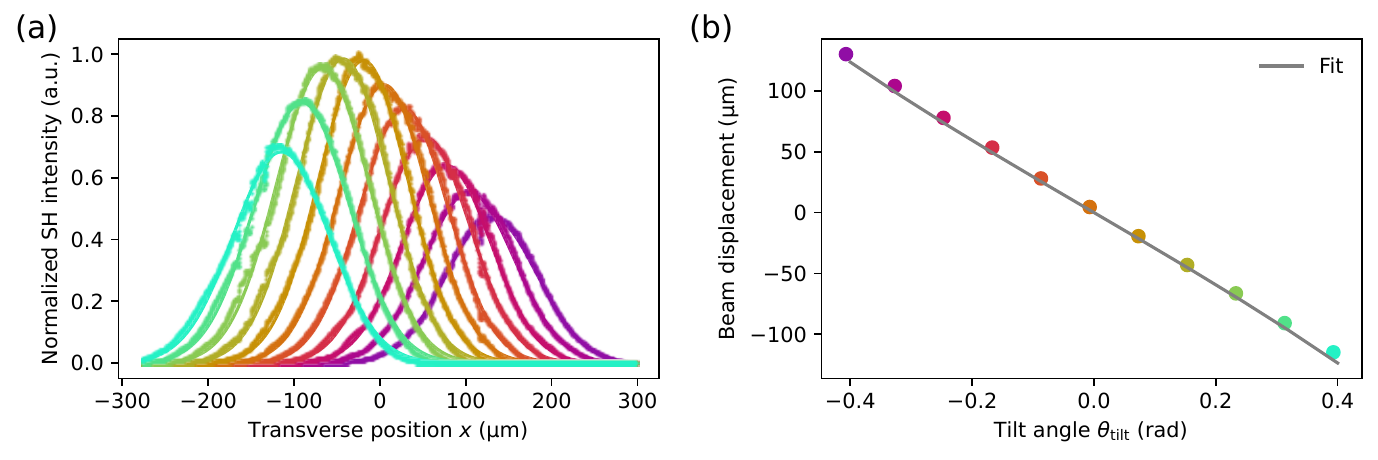}
    \caption{Beam-steering technique used to characterize the beam propagation distance. (a) Circles: Experimentally measured transverse beam profile of the SHG for various tilt angles of the QPM grating with period $\Lambda=\SI{16.75}{\micro m}$. Solid lines: Gaussian fits of the spatial profiles. (b) Circles: The location of the beam centroid based on Gaussian fits for various tilt angles. Solid line: A least-squares fit based on \eqref{eq:disp-fit}, yielding $L_\text{out}=\SI{1.58e4}{\micro m}$. The colors of the markers in (b) provide the legend for the values of $\theta_\text{tilt}$ in (a).}
    \label{fig:beam-steering}
\end{figure}

To design accurate QPM gratings for engineering the spatial profiles of SHGs, it was essential to characterize the overall geometry of the waveguide precisely. In particular, we needed to determine the position of the output facet, $z=L_\text{out}$, relative to the defined coordinates. For this purpose, we employed an approach based on beam steering of the SHG. Specifically, we projected a flat QPM grating with a tilt angle of $\theta_\text{tilt}$ relative to the waveguide coordinate axes and measured the displacement observed in the camera image of the output SH profile. Intuitively, a larger $L_\text{out}$ produced a larger displacement. In Fig.~\ref{fig:beam-steering}(a), we show the output beam profiles for various values of $\theta_\text{tilt}$. The origin $x=0$ in the camera image was set so that the output beam was approximately centered when $\theta_\text{tilt}=0$. The measured displacement, $d_\text{disp}$, was fitted to the following function:
\begin{align}
\label{eq:disp-fit}
    d_\text{disp} = -\frac{2\pi \tan\theta_\text{tilt}}{k_2 \Lambda}(L_\text{out} - z_0),
\end{align}
where $k_2$ is the wavenumber for the SH and the QPM grating period $\Lambda=\SI{16.75}{\micro m}$ was chosen to phase match the SHG at an output wavelength of $\SI{790}{nm}$. The fitting process yielded $L_\text{out}=\SI{1.58e4}{\micro m}$.

The final unknown parameter needed for a full characterization of the experimental setup was the transverse spatial chirp of the field, $u_\text{ref}$. Because this parameter affected only the phase of the electric field, it was difficult to extract it from an intensity distribution (e.g., Fig.~\ref{fig:mapping-SHG-profile}), particularly when the Rayleigh length was large. To determine $u_\text{ref}$, we again exploited the programmability of our nonlinear waveguide. Specifically, we projected a QPM grating with quadratic curvature, parameterized as
\begin{align}
\label{eq:quadratic-QPM}
    I_\text{quad}(x,z)=\frac{1}{2}\Bigl\{\sin\Bigl(2\pi z\Lambda^{-1} -q_\text{quad}\,x^2\Bigr) + 1\Bigr\},
\end{align}
onto a thin slice of the region around the reference point, defined by $z_\text{ref}-\epsilon_z/2\leq z\leq z_\text{ref}+\epsilon_z/2$. The generated SH profile then inherited a spatial chirp and became
\begin{align}
\label{eq:sh-profile-quad}
    b_\text{SH-quad}(x,z=z_\text{ref}) \propto e^{-iq_\text{quad}\,x^2}a^2(x,z=z_\text{ref}) \propto e^{i(-q_\text{quad}+u_\text{ref})x^2}e^{-x^2/w_\text{SH-ref}^2}.
\end{align}
Intuitively, the spatial chirp term acted like a lens applied to a Gaussian beam, where $q_\text{quad}$ controlled the effective curvature of the lens. Thus, tuning $q_\text{quad}$ allowed us to control the focus of the SH beam.

\begin{figure}[h]
    \centering
    \includegraphics[width=0.9\linewidth]{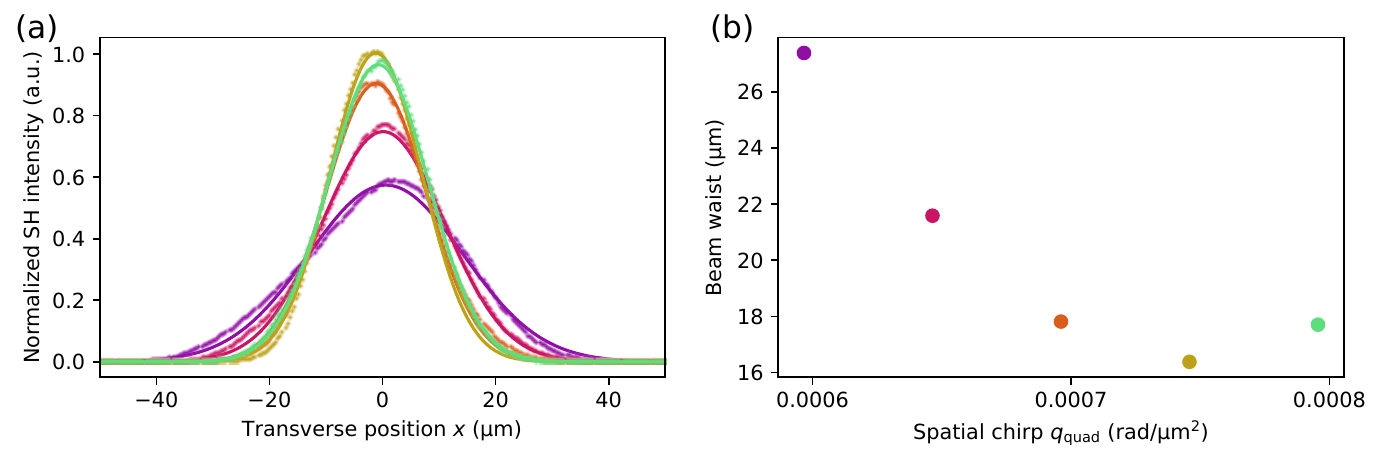}
    \caption{(a) Circles: Experimentally measured spatial profile of the generated SH with a quadratically chirped QPM grating \eqref{eq:quadratic-QPM} with $\epsilon=\SI{377.2}{\micro m}$ and various values of $q_\text{quad}$. Solid lines: Gaussian fits of the spatial beam profile. (b) Beam waists of the SH beams obtained by Gaussian fitting for various chirps $q_\text{quad}$. The colors of the markers provide the legend for $q_\text{quad}$ in (a).}
    \label{fig:beam-focus}
\end{figure}

In Fig.~\ref{fig:beam-focus}, we show experimentally measured output SH beam profiles for various $q_\text{quad}$ values. Specifically, we determined an optimal value $q_\text{quad}=q_\text{opt}$ that minimized the width of the SH beam at the output facet, $z=L_\text{out}$, resulting in a minimum beam waist of $w_\text{SH-opt}=\SI{16.4}{\micro m}$ at the output facet. We assumed that the beam was spatially unchirped at this point, which allowed us to calculate the Rayleigh range of the SH as $z_\text{SH-R}=\SI{2072}{\micro m}$. Next, we backtracked the spatial evolution of the beam profile from the output facet to the reference point, yielding
\begin{align}
\label{eq:sh-profile-opt}
    b_\text{SH-opt}(x,z=z_\text{ref}) \propto e^{i\frac{k_{2}\,x^2}{2\tilde{R}_\text{SH-opt}}} e^{-x^2/\tilde{w}_\text{SH-opt}^2},
\end{align}
with
\begin{align}
    \tilde{R}_\text{SH-opt} &= (z_\text{ref}-L_\text{out})\Biggl[1+\frac{z_\text{SH-R}^2}{(z_\text{ref}-L_\text{out})^2}\Biggr],\\[1ex]
    \tilde{w}_\text{SH-opt}^2 &= w_\text{SH-opt}^2\Biggl(1+\frac{(z_\text{ref}-L_\text{out})^2}{z_\text{R-opt}^2}\Biggr).
\end{align}
Equations \eqref{eq:sh-profile-quad} and \eqref{eq:sh-profile-opt} should be equivalent when $q_\text{quad}=q_\text{opt}$. Solving the equality $\tilde{w}_\text{SH-opt}=w_\text{SH-ref}$ for $L_\text{out}$ yielded $L_\text{out}=\SI{1.54e4}{\micro m}$, which reasonably agreed with the value estimated via the beam-steering method, $L_\text{out}=\SI{1.58e4}{\micro m}$. Owing to the sparsity of the data in Fig.~\ref{fig:beam-focus}, the beam-focusing approach was expected to be less accurate; therefore, we adopted the beam-steering estimate of $L_\text{out}$ as our best value. Finally, by equating the spatial chirp values, we determined $u_\text{ref}=\SI{1.23e-4}{rad/\micro m^2}$.

\subsection{Model for spatially engineered SHG}
\label{appendix:SHG-simulation-model}
In this section, we establish a numerical model for spatially engineered SHG in a programmable nonlinear waveguide using the parameters calibrated in the previous section. To simplify the analysis, we make the following assumptions: (i) the pump field remained undepleted during propagation; (ii) the pump field was monochromatic CW light with a wavelength of $\SI{1580}{nm}$, generating SHG at $\SI{790}{nm}$; (iii) the optical loss in the waveguide was negligible; and (iv) the induced $\chi^{(2)}$ nonlinearity was proportional to the intensity of the programming illumination, $I(x,z)$.

We take the models derived in Appendix~\ref{appendix:SHG-model} as our starting point. The FH field profile at the reference position $z=z_\text{ref}$ is given by
\begin{align}
    a(z_\text{ref},x)=a_0\,e^{iu_\text{ref}x^2/2}\,e^{-x^2/w_\text{FH-ref}^2},
\end{align}
where the parameters $u_\text{ref}$ and $w_\text{FH-ref}=\sqrt{2}\,w_\text{SH-ref}$ are determined in Appendix~\ref{appendix:calibration}. Under assumptions (i), (ii), and (iii), we can solve Eq.~\eqref{eq:FH-eom} to obtain
\begin{align}
\label{eq:approximate-FH-dist}
    a(z,x)=e^{\frac{i}{2k_1}(z-z_\text{ref})\partial_x^2}\,a(z_\text{ref},x).
\end{align}
Assumption (iv) implies that
\begin{align}
    r(x,z)=C\,I(x,z),
\end{align}
where $r(x,z)$ represents the spatial distribution of the $\chi^{(2)}$ nonlinearity and $C$ is a constant. Since only the Fourier component of $I(x,z)$ near the spatial frequency $2\pi/\Delta k$ significantly contributes to the SHG dynamics, we modify Eq.~\eqref{eq:SH-eom} to obtain
\begin{align}
\label{eq:SH-eom-spatial}
    \partial_zb(x,z) \approx \frac{i}{2k_2}\partial_x^2b(x,z) - i\,C\,\kappa\,\mathcal{I}(x,z)\,a^2(x,z),
\end{align}
where the complex illumination pattern $\mathcal{I}$ is defined via
\begin{align}
    I(x,z)=e^{i\Delta k\,z}\,\mathcal{I}(x,z) + e^{-i\Delta k\,z}\,\mathcal{I}^*(x,z).
\end{align}
We numerically obtained the SH field profile, up to an overall scaling factor $C$, by integrating Eq.~\eqref{eq:SH-eom-spatial} with the initial condition $b(x,z)=0$ and using the FH field profile from Eq.~\eqref{eq:approximate-FH-dist}. The output SH beam measured by the camera is proportional to $|b(x,z=L_\text{out})|^2$.

\subsection{Experimental results and simulations}

In this section, we use the calibration results from Appendix~\ref{appendix:calibration} to design QPM grating structures for spatially engineering the SHG and compare them with numerical simulations based on the model developed in Appendix~\ref{appendix:SHG-simulation-model}. In particular, we describe how the results in Fig.~\ref{fig:spatial} in Sec.~\ref{sec:spatial-engineering} were obtained.

For Fig.~\ref{fig:spatial}(b), we projected a monotonic QPM grating with period $\Lambda=\SI{16.75}{\micro m}$, in which $\Delta k=2\pi/\Lambda$ was set. We used this parameter for the remaining experiments as well. The beam propagation simulation was performed with $\mathcal{I}(x,z)=1$. Note that the induced $\chi^{(2)}$ nonlinearity was assumed to be zero for $z>L_\text{img}$ (i.e., outside the region of patterned illumination). The overall scaling and phase of $\mathcal{I}(x,z)$ do not affect the result.

For Fig.~\ref{fig:spatial}(c), we implemented a quadratically chirped grating. The programming illumination pattern was given by
\begin{align}
\label{eq:illumination-pattern-focus}
    I_\text{focus}(x,z)=\frac{1}{2}\Bigl\{\sin\Bigl(2\pi z/\Lambda - q_\text{focus}(z)x^2\Bigr)+1\Bigr\},
\end{align}
where the chirp parameter was defined as
\begin{align}
    q_\text{focus}(z)=q_\text{opt}+q_\text{chirp}(z-z_\text{ref}).
\end{align}
The resulting nonlinearity pattern became
\begin{align}
    \mathcal{I}_\text{focus}(x,z)\propto e^{-iq_\text{focus}(z)x^2}.
\end{align}
The linear chirp of the curvature, $q_\text{chirp}$, compensated for the variation in the optimal transverse chirp at different longitudinal positions. We used $q_\text{chirp}=\SI{5.3e-7}{rad/\micro m^3}$.

\begin{figure}[h]
    \centering
    \includegraphics[width=\linewidth]{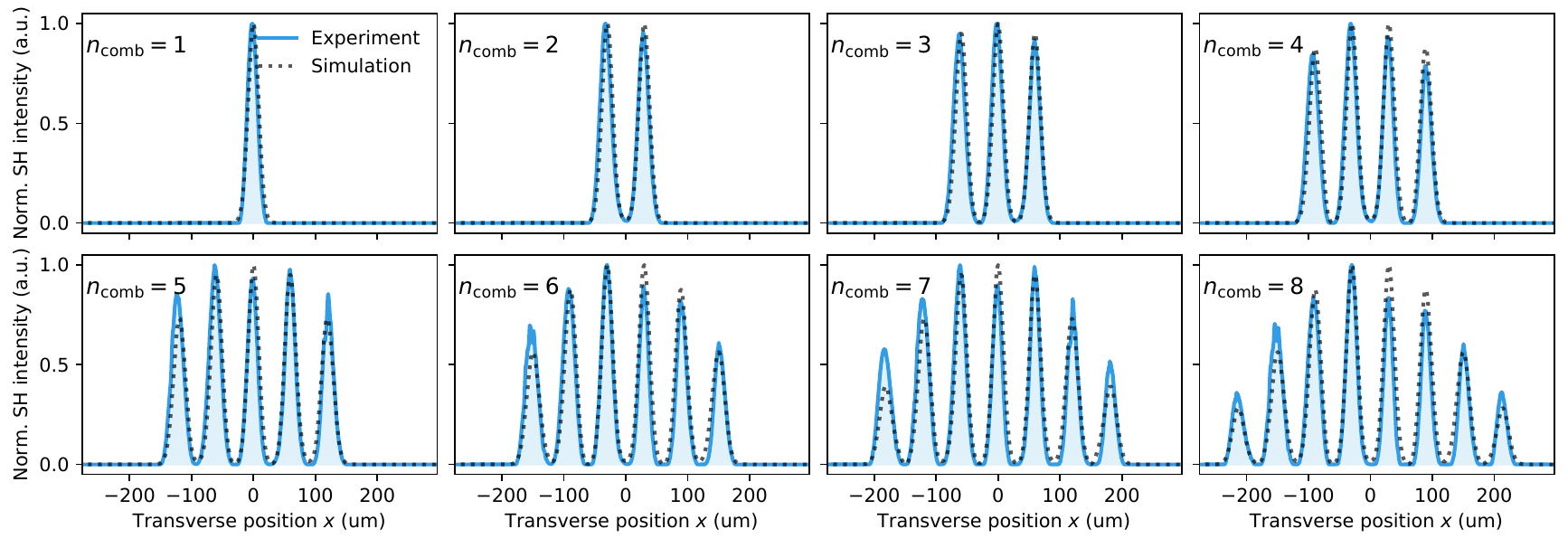}
    \caption{Solid lines: Experimentally measured spatial SH beam profiles for the illumination patterns of \eqref{eq:illumination-pattern-comb} for various numbers of peaks $n_\text{comb}$ with $\Delta x=\SI{52.8}{\micro m}$. Dotted lines: Results of the numerical simulations with no free parameter except for the overall scaling, which was set so that the peak values are unity.}
    \label{fig:comb-simulation}
\end{figure}

In Fig.~\ref{fig:spatial}(d), we superimposed multiple grating patterns, $I_\text{focus}$, with different spatial offsets to produce a comb-like structure with $n_\text{comb}$ peaks. For this purpose, we projected an illumination pattern given by
\begin{align}
\label{eq:illumination-pattern-comb}
    I_\text{comb}(x,z)=\mathcal{N}\sum_{j=1}^{n_\text{comb}}I_\text{focus}\bigl(x-x_j,z\bigr)+\mathcal{C},
\end{align}
where $x_j=\Delta x\left(\frac{n_\text{comb}-1}{2}+j-1\right)$ is the offset for the $j$th pattern. The constants $\mathcal{N}$ and $\mathcal{C}$ were chosen so that the dynamic range of the illumination was unity. The overall distribution of the $\chi^{(2)}$ nonlinearity used in the simulation was
\begin{align}
    \mathcal{I}_\text{comb}(x,z)\propto \sum_{j=1}^{n_\text{comb}}e^{-iq_\text{focus}(z)(x-x_j)^2}.
\end{align}
In addition to the case of $n_\text{comb}=9$ shown in the main text, we present results for different values of $n_\text{comb}$ in Fig.~\ref{fig:comb-simulation}, which uniformly demonstrate good agreement between theory and experiment.

\begin{figure}[h]
    \centering
    \includegraphics[width=\linewidth]{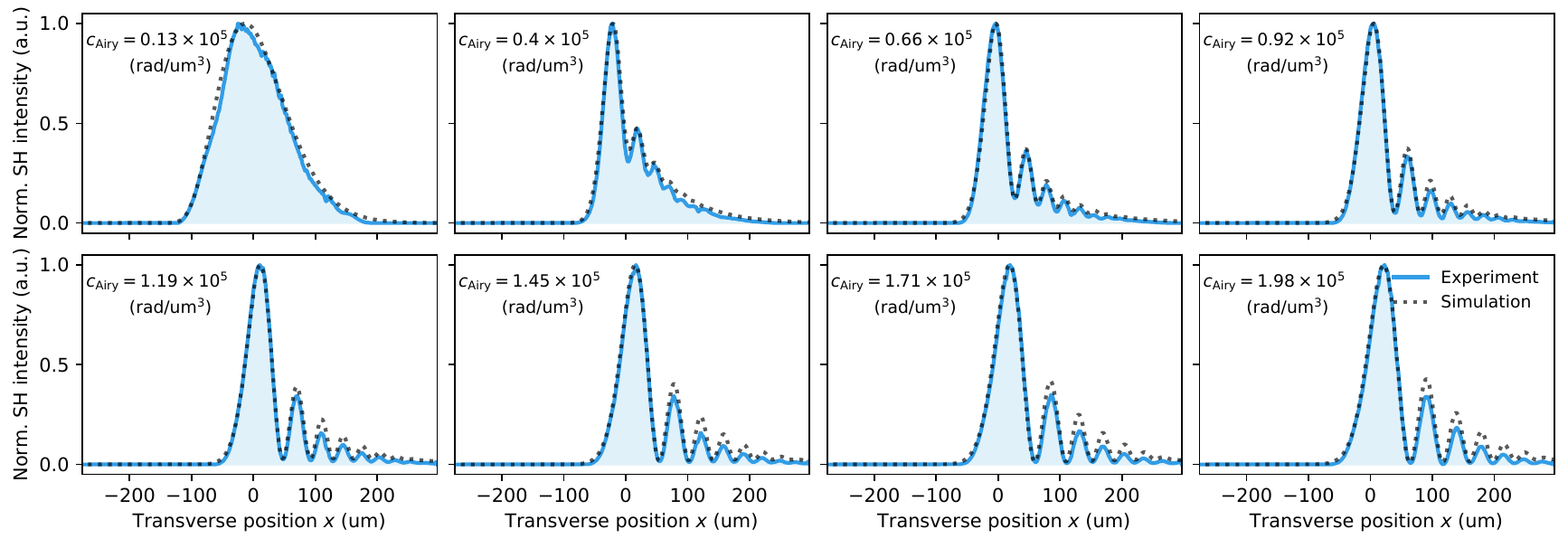}
    \caption{Solid lines: Experimentally measured spatial SH beam profiles for the illumination patterns of \eqref{eq:illumination-pattern-airy} for various values of $c_\text{Airy}$. Dotted lines: Results of the numerical simulations with no free parameter except for the overall scaling, which was set so that the peak values were unity.}
    \label{fig:airy-simulation}
\end{figure}

Finally, in Fig.~\ref{fig:spatial}(d), we demonstrated the generation of an Airy beam via a cubically chirped QPM grating. Specifically, we used
\begin{align}
\label{eq:illumination-pattern-airy}
    I_\text{Airy}(x,z)=\frac{1}{2}\left(\sin\left(2\pi z/\Lambda + c_\text{Airy}x^3\right)+1\right),
\end{align}
which was projected onto a thin slice defined by $z_\text{ref}-\epsilon_z/2\leq z\leq z_\text{ref}+\epsilon_z/2$, where $\epsilon_z=\SI{7242}{\micro m}$. For the simulation, we used
\begin{align}
    \mathcal{I}_\text{Airy}(x,z)\propto e^{ic_\text{Airy}x^3}.
\end{align}
In the main text, we present the experimental results for $c_\text{Airy}=\SI{1.05e-5}{rad/\micro m^3}$, and in Fig.~\ref{fig:airy-simulation}, we present the results for additional values of $c_\text{Airy}$.

\section{Spatio-spectral engineering}
\label{appendix:spatio-spectral}
\begin{figure}[h]
    \centering
    \includegraphics[width=0.9\linewidth]{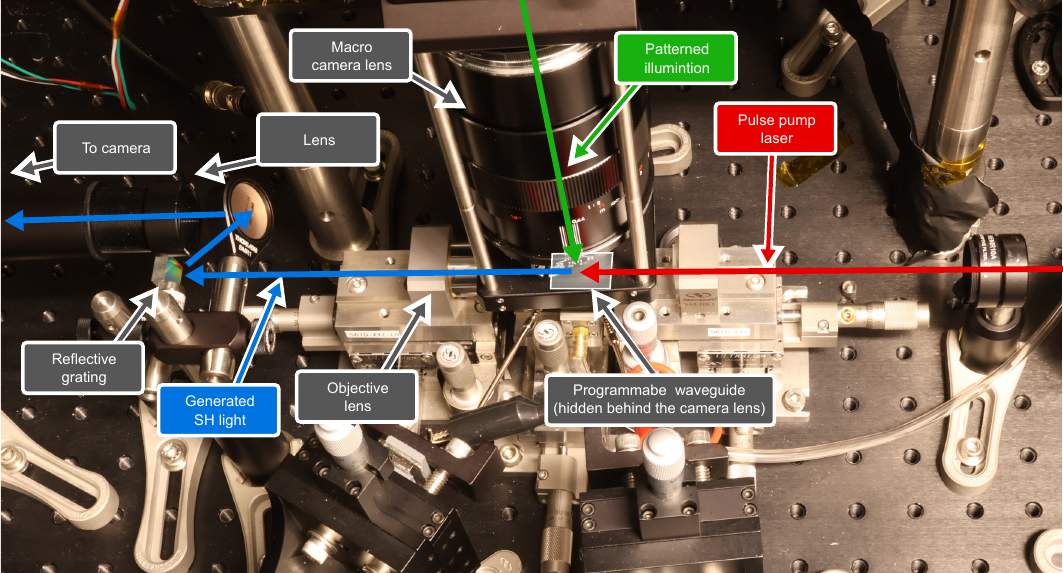}
    \caption{Photograph of the experimental setup used to perform spectrally resolved imaging of the broadband SH output. The setup produced the data for Sec.~\ref{sec:spatio-spectral}.}
    \label{fig:hyperspectral}
\end{figure}

In this section, we describe how we obtained the results presented in Sec.~\ref{sec:spatio-spectral} for the spatio-spectral engineering of the SHG. Figure~\ref{fig:hyperspectral} shows a photograph of the experimental setup used to acquire the data. The optics for the input and programming illumination, as well as the electronics, were the same as those described in Appendices~\ref{appendix:common-part} and \ref{appendix:spatial}; however, we extended the detection system to resolve both spatial and spectral features. We used the bias voltage of $V_\text{tot}=\SI{600}{V}$ for the experiments presented in this section.

In our setup, the objective lens of a microscope was placed one focal distance from the output facet of the programmable waveguide. The beam then impinged on a reflective grating that diffracts the light in the vertical (i.e., $y$) direction, where the diffraction angle depended on the wavelength. A second lens, positioned one focal distance from the grating, focused the output onto a camera for detection. Consequently, the horizontal axis of the camera image resolved the spatial features of the output light, whereas the vertical axis resolved the spectral features.

We calibrated the setup for both the spatial and spectral domains. To characterize the spatial magnification of the imaging system, we followed the procedure described in Appendix~\ref{appendix:spatial}. Calibrating the spectral coordinate---that is, determining which wavelength corresponds to each camera pixel---was more involved. First, we projected monotonic QPM grating patterns with period $\Lambda$ and measured the generated SH wavelength with a spectrometer. This measured relationship between $\Lambda$ and the SH wavelength established the mapping between the vertical coordinate of the camera sensor and the SH wavelength. We also calibrated the pump beam profile on the waveguide via the procedure outlined in Appendix~\ref{appendix:spatial}.

\begin{figure}[h]
    \centering
    \includegraphics[width=\linewidth]{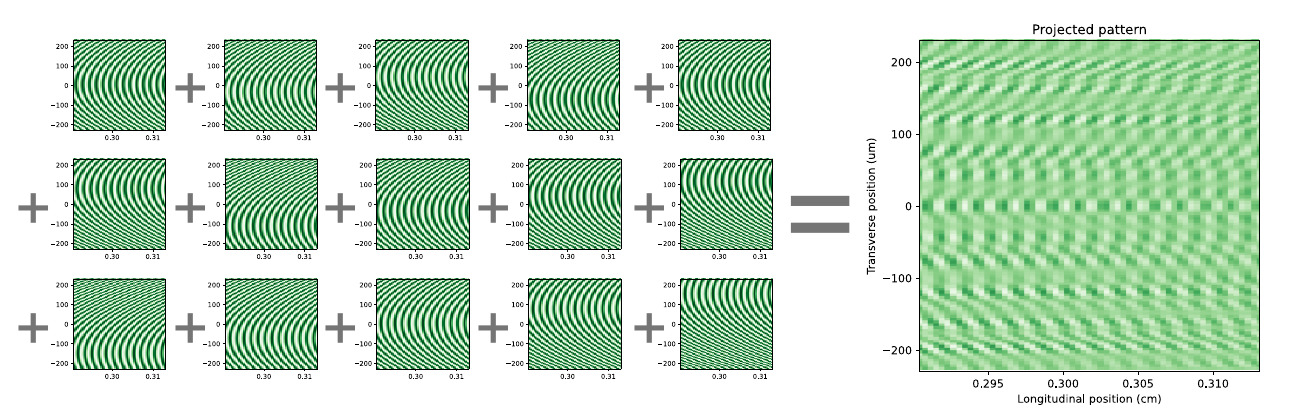}
    \caption{The full illumination pattern \eqref{eq:illumination-pattern-2d-comb}, composed of fifteen base patterns, is shown. Each pattern on the left focused a particular wavelength of the SH onto a particular position. Overall, the total pattern shown on the right focused the SH light into fifteen distinct peaks in the two-dimensional phase space spanned by the transverse position and the SH wavelength. }
    \label{fig:polling-pattern-image}
\end{figure}

\begin{figure}[h]
    \centering
    \includegraphics[width=0.85\linewidth]{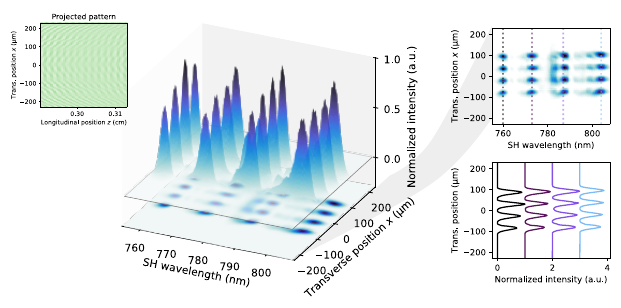}
    \caption{Demonstration of the generation of sixteen-peak spatiotemporal features via SHG on a programmable nonlinear waveguide. The projected programming illumination was characterized by $n_\text{spec}=4$, $\Lambda_1=\SI{16.4}{\micro m}$, $\Lambda_2=\SI{16.583}{\micro m}$, $\Lambda_3=\SI{16.767}{\micro m}$, $\Lambda_4=\SI{16.4}{\micro m}$, and $n_{\text{comb},k}=4$. The weights were $c_1=0.3$, $c_2=0.15$, $c_3=0.12$, and $c_4=0.47$.}
    \label{fig:peaks-16}
\end{figure}

For simultaneous engineering of spatial and spectral features, we exploited the full two-dimensional programmability of the waveguide. In Fig.~\ref{fig:spatio-spectral}(b), we superimposed multiple quadratically chirped grating patterns \eqref{eq:illumination-pattern-focus} with different spatial offsets and base periods. Specifically, we had
\begin{align}
\label{eq:illumination-pattern-2d-comb}
    I_\text{2D-comb}(x,z)=\mathcal{N}\sum_{k=1}^{n_\text{spec}}\sum_{j=1}^{n_{\text{comb},k}}c_k\, I_\text{focus}(x-x_j,z,\Lambda_k)+\mathcal{C},
\end{align}
where $n_\text{spec}$ is the number of spectral bands we addressed independently, $n_{\text{comb},k}$ is the number of comb lines in the $k$th spectral band, and $\Lambda_k$ is the QPM grating period used to phase match the SHG for the $k$th band. In the main text, we used $n_\text{spec}=5$, meaning that we independently controlled 5 different spectral bands. The wavelength of each band was determined by the QPM grating period, which we set as $\Lambda_1=\SI{16.4}{\micro m}$, $\Lambda_2=\SI{16.5375}{\micro m}$, $\Lambda_3=\SI{16.675}{\micro m}$, $\Lambda_4=\SI{16.8125}{\micro m}$, and $\Lambda_5=\SI{16.95}{\micro m}$, with relative weights of $c_1=0.3$, $c_2=0.14$, $c_3=0.13$, $c_4=0.19$, and $c_5=0.45$. The number of peaks we engineer was given by $n_{\text{comb},k}=k$ for $k\in\{1,2,3,4,5\}$. The normalization constant $\mathcal{N}$ was set so that the maximum value of the grayscale illumination was 0.8. In Fig.~\ref{fig:polling-pattern-image}, we display all fifteen base patterns that were summed to produce the full QPM grating pattern $I_\text{2D-comb}$. Finally, in Fig.~\ref{fig:peaks-16}, we present another example of spatio-spectral engineering, which indicated the formation of sixteen distinct peaks.

The spatio-spectral engineering achievable on a programmable nonlinear waveguide is not limited to comb-like structures. As shown in Fig.~\ref{fig:spatio-spectral}(c), we engineered Airy beams with different chirp parameters at different SH wavelengths. This was achieved by projecting the following programming illumination pattern:
\begin{align}
\label{eq:illumination-pattern-2D-airy}
    I_\text{2D-Airy}(x,z)=\mathcal{N}\sum_{k=0}^{n_\text{spec}}c_k(z) I_\text{Airy}(x,z;\Lambda_k, c_{\text{Airy},k})+\mathcal{C}
\end{align}
with $n_\text{spec}=2$, $\Lambda_1=\SI{16.4}{\micro m}$, $\Lambda_2=\SI{16.95}{\micro m}$, $c_\text{Airy,1}=\SI{1.85e-5}{rad/\micro m^3}$, and $c_\text{Airy,2}=\SI{-1.32e-5}{rad/\micro m^3}$. Here, $c_1(z)$ took a value of $0.3$ only when $\SI{1886}{\micro m}\leq z\leq \SI{3019}{\micro m}$. Similarly, $c_2(z)$ took a value of $0.4$ only for $\SI{3019}{\micro m}\leq z\leq \SI{3772}{\micro m}$.

\section{Experimental estimation of electric-field-induced $\chi^{(2)}$ nonlinearity}
\label{appendix:estimated-chi2}

In this section, we estimate the electric-field-induced $\chi^{(2)}$ nonlinearity in a programmable waveguide on the basis of the experimentally measured CW-pumped SHG conversion efficiency. Assuming that the pump field remains undepleted during SHG and that the effects of spatial diffraction are negligible, the evolution of the SH field can be written as
\begin{align}
    \partial_z b = -i\kappa e^{-i\Delta k z}r(x,z)a^2,
\end{align}
where the definitions of $\kappa$ and $r(x,z)$ are provided in Appendix~\ref{appendix:SHG-model}. When a monotonic QPM grating pattern with the correct period is projected to achieve quasi-phase matching, the nonlinearity distribution takes the form of a square wave
\begin{align}
    r_\text{square}(x,z)=\frac{1}{2}\Bigl(\mathrm{sign}(\sin(\Delta k z))+1\Bigr)
    =\frac{1}{2}+\frac{1}{i\pi}\sum_{\ell=1}^\infty \frac{1}{2\ell-1}\Bigl(e^{i(2\ell-1)\Delta k z}-e^{-i(2\ell-1)\Delta k z}\Bigr).
\end{align}
Since only the Fourier components with spatial frequencies close to $\Delta k$ contribute significantly to SHG, if the waveguide is sufficiently long, the equation of motion for the SH field can be approximated as 
\begin{align}
    \partial_z b \approx -\kappa_\text{eff} a^2,
\end{align}
with
\begin{align}
    \kappa_\text{eff}=\frac{1}{\pi}\kappa.
\end{align}
With the undepleted pump approximation, we can analytically integrate this equation to obtain
\[
b(x,z=L_\text{QPM}) = \kappa_\text{eff}\, a^2(x,z=0)\, L_\text{QPM},
\]
where $L_\text{QPM}$ is the quasi-phase-matched distance.

We parameterize the field profile of the input FH field as
\begin{align}
\label{eq:gaussian-fh-profile}
    |a(x,0)|^2 = \frac{\sqrt{2}P_\text{FH}}{\sqrt{\pi}w}\, e^{-2x^2/w^2},
\end{align}
which has a total power flux of $P_\text{FH}$. Under this condition, we obtain
\begin{align}
   |b(x,z=L_\text{QPM})|^2 = \frac{2\kappa^2_\text{eff} L_\text{QPM}^2 P_\text{FH}^2}{\pi w^2}\, e^{-4x^2/w^2}.
\end{align}
Integrating over the transverse coordinate yields
\begin{align}
\label{eq:eta-equation}
    \eta_\text{norm} = \frac{P_\text{SH}}{P_\text{FH}^2} = \frac{\kappa^2 L_\text{QPM}^2}{\pi^{5/2} w},
\end{align}
which establishes the relationship between the experimentally measurable $\eta_\text{norm}$ and the nonlinear coupling $\kappa$.

\begin{figure}[h]
    \centering
    \includegraphics[width=0.7\linewidth]{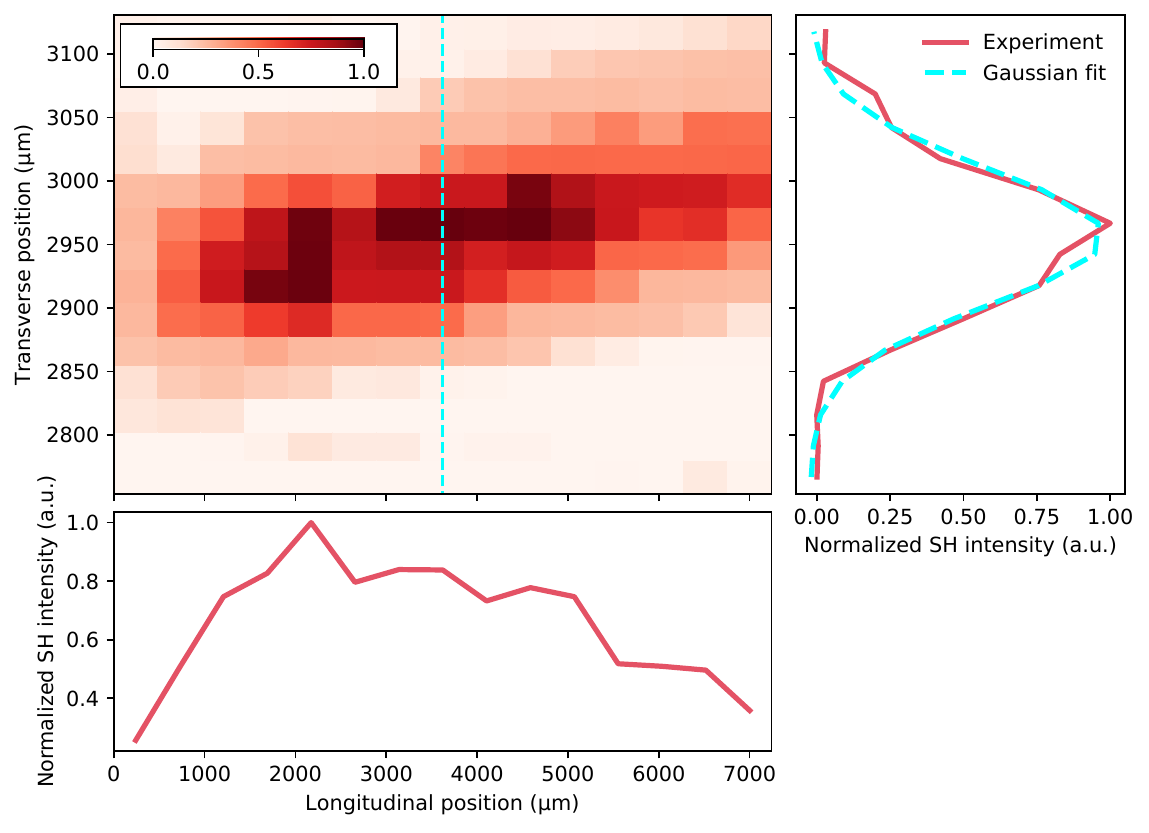}
    \caption{Experimentally measured contributions to SHG from different regions of a programmable nonlinear waveguide. The data were acquired with a pulse pump laser via the measurement technique described in Appendix~\ref{appendix:calibration}. The cyan line on the right top panel is the Gaussian fit (with offset) of the cross section at the vertical cyan line in the left top panel. Similarly, the cyan line on the left bottom panel is the Gaussian fit (with offset) of the cross section at the horizontal cyan line in the left top panel.}
    \label{fig:cw-shg-beam-map}
\end{figure}

In Fig.~\ref{fig:cw-shg-beam-map}, we show the experimentally mapped distributions of the SHG contributions from different regions of the waveguide. From a Gaussian fit to the beam profile, we obtained an SH beam width of $\SI{108}{\micro m}$, which corresponded to $w=\SI{152}{\micro m}$ in Eq.~\eqref{eq:gaussian-fh-profile}.

\begin{figure}
    \centering
    \includegraphics[width=0.45\linewidth]{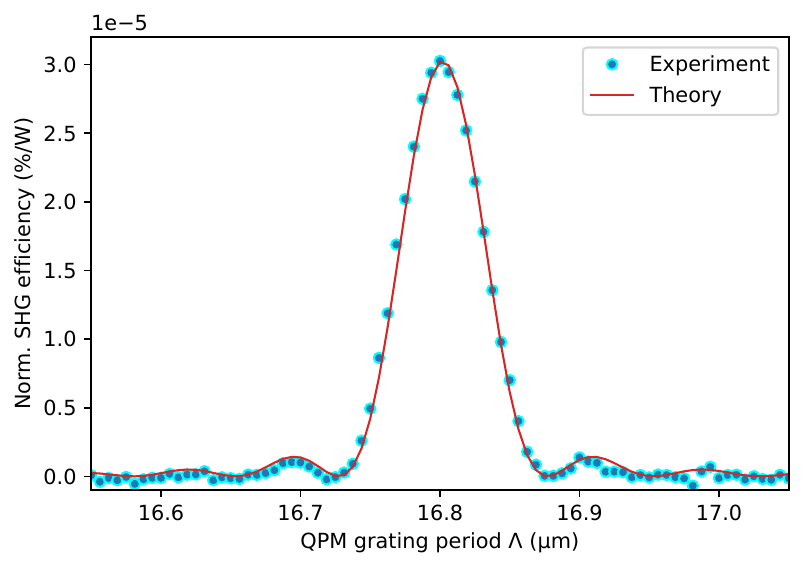}
    \caption{Experimentally measured $\eta_\text{norm}$ with a pump wavelength of $\SI{1580}{\micro m}$. A bias voltage of $V_\text{tot}=\SI{1600}{V}$ was used. The red solid lines represent the theoretical curve $\propto\mathrm{sin}^2(\pi(\Lambda-\Lambda_0)L_\text{QPM})$ with $\Lambda_0=\SI{16.802}{\micro m}$ and $L_\text{QPM}=\SI{3772}{\micro m}$. }
    \label{fig:high-voltage}
\end{figure}

To estimate the maximum $\chi^{(2)}$ nonlinearity inducible on our platform, we projected a monotonic grating pattern with a varying period $\Lambda$ and measured the SHG conversion efficiency when pumping the waveguide with a CW laser at a wavelength of $\lambda=\SI{1580}{nm}$. We set the bias voltage to $V_\text{bias}=\SI{1600}{V}$ at a frequency of $\SI{5}{Hz}$. The measurement result is shown in Fig.~\ref{fig:high-voltage}, where we observed a maximum conversion efficiency of $\eta_\text{norm}=\SI{1.51e-5}{\%W^{-1}}$. Note that in this measurement, we limited the projection of the grating pattern to the region $\SI{1509.6}{\micro m}\leq z\leq \SI{5283.6}{\micro m}$, corresponding to $L_\text{QPM}=\SI{3772}{\micro m}$. This restriction avoided the nonuniform nonlinearity in the longitudinal direction observed in Fig.~\ref{fig:cw-shg-beam-map}. The good agreement between the experimental data and the theoretical sinc curve---assuming $L_\text{QPM}=\SI{3772}{\micro m}$ in Fig.~\ref{fig:high-voltage}---indicates that the $\chi^{(2)}$ nonlinearity was mostly uniform within this window.

Now, we could estimate the value of the nonlinearity by solving \eqref{eq:eta-equation} for $\chi^{(2)}_{yyy}$ with \eqref{eq:kappa-equation}. For this purpose, we used numerically calculated waveguide parameters for the fundamental TM modes: $L_\text{eff}=\SI{1.693}{\micro m}$, $n_\text{eff}^{(\omega)}=1.891$, and $n_\text{eff}^{(2\omega)}=1.938$. Overall, we obtained $\chi^{(2)}_{yyy}=\SI{\chitwo}{pm/V}$ as our best estimate of the measured $\chi^{(2)}$ nonlinearity. As discussed in Appendix~\ref{appendix:electric-properties}, improvements in the photoconductor can increase the induced nonlinearity by a factor of $\mathcal{R}_\text{max}\approx 2.3$, suggesting that $\chi^{(2)}_{yyy}=\SI{\chitwomax}{pm/V}$ could be physically feasible.

Although we present our estimates to two significant figures, it is well known that estimating nonlinear coefficients is notoriously difficult even for bulk materials, so we do not expect our estimates to be accurate to better than a factor of 2. Possible sources of error include uncertainty in the off-chip collection efficiency, contributions from nonlinear tensor elements other than $\chi^{(2)}_{yyy}$, fringing of the bias electric field inside the core (see Ref.~\cite{Onodera2024Linear}), and variations in the spatial beam profiles on the waveguide.

\section{Need for experimental characterization of the electric-field-induced $\chi^{(2)}$ nonlinearity}
\label{appendix:physics-chi2}

In this section, we describe the underlying physics of electric-field-induced $\chi^{(2)}$ nonlinearity, relating its value to a tensor element of the native $\chi^{(3)}$ nonlinearity. Our discussion suggests that identifying an ideal material for a programmable nonlinear waveguide (and for electric-field-induced $\chi^{(2)}$ nonlinearity in general) dedicated experimental efforts are required to directly measure the induced $\chi^{(2)}$ for each material rather than relying solely on tabulated values of the optical $\chi^{(3)}$ nonlinearity.

Below, we adopt the mathematical notation provided in Ref.~\cite{Boyd2008nonlinear}, which differs slightly from that used elsewhere in the manuscript. The electric-field-induced $\chi^{(2)}$ nonlinearity can be understood as follows. The third-order nonlinear polarization of a material is given by
\begin{align}
    P_i^{(3)}(\omega_m+\omega_n+\omega_o)=\epsilon_0\sum_{jkl}\sum_{(mno)}\chi^{(3)}_{ijkl}(\omega_m+\omega_n+\omega_o,\omega_m,\omega_n,\omega_o)E_j(\omega_m)E_k(\omega_n)E_l(\omega_o),
\end{align}
where $(mno)$ denotes all permutations of the indices. Electric-field-induced $\chi^{(2)}$ nonlinearity arises when one of the electric fields is replaced by a bias field $E_\text{bias}$, whose frequency is essentially zero compared with the optical frequency. Focusing on interactions among fields in the $y$ direction, we obtain
\begin{align}
    P_y^{(2)}(\omega_m+\omega_n)&=3\epsilon_0\sum_{(mn)}\chi^{(3)}_{yyyy}(\omega_m+\omega_n,\omega_m,\omega_n,0)E_y(\omega_m)E_y(\omega_n)E_\text{bias}\nonumber\\&=\epsilon_0\sum_{(mn)}\chi^{(2)}_{yyy}(\omega_m+\omega_n,\omega_m,\omega_n)E_y(\omega_m)E_y(\omega_n).
\end{align}
Thus, the effective $\chi^{(2)}$ nonlinear tensor element is given by
\begin{align}
    \chi^{(2)}_{yyy}(\omega_m+\omega_n,\omega_m,\omega_n)=3\chi^{(3)}_{yyyy}(\omega_m+\omega_n,\omega_m,\omega_n,0)E_\text{bias}.
\end{align}
To achieve better performance for a programmable nonlinear waveguide, it is therefore desirable to find a material with a large value of $\chi^{(3)}_{yyyy}(\omega_m+\omega_n,\omega_m,\omega_n,0)$.

One might assume that optimal materials can be identified via tabulated values of the $\chi^{(3)}$ optical nonlinearity, such as the nonlinear index $n_2$. However, this approach is inadequate because the tabulated values refer to $\chi^{(3)}(\omega_m+\omega_n+\omega_o,\omega_m,\omega_n,\omega_o)$, where all the frequencies are optical. In contrast, we require $\chi^{(3)}(\omega_m+\omega_n,\omega_m,\omega_n,0)$, which involves one DC field. Since the optical frequency $\omega_o$ is vastly different from DC, inferring the value of $\chi^{(3)}(\omega_m+\omega_n,\omega_m,\omega_n,0)$ from the values of $\chi^{(3)}(\omega_m+\omega_n+\omega_o,\omega_m,\omega_n,\omega_o)$ in the literature is difficult.

This difficulty can be intuitively understood by considering the example of $\chi^{(2)}$ nonlinearity. For a material with a native $\chi^{(2)}$ nonlinearity, the strengths of the optical nonlinearity and the electro-optic effect are given by $\chi^{(2)}(\omega_m+\omega_n,\omega_m,\omega_n)$ and $\chi^{(2)}(\omega_m,\omega_m,0)$, respectively. The former is not generally a good indicator of the latter, or vice versa. For example, III-V semiconductor materials exhibit very large optical nonlinearities but only moderate electro-optic coefficients. On the other hand, for example, lithium niobate has a smaller nonlinear-optical coefficient and larger electro-optic coefficient than typical III-V materials.

\section{Potential future applications of programmable poling}
\label{appendix:applications}

In this section, we provide a speculative list of potential uses of programmable on-chip nonlinear photonics. Here, we focus on the topics that warrant further details beyond the discussions in the main text.

\subsection{Widely tunable on-chip light sources with unlimited phase-matching bandwidths}

Integrated photonics leverages the tight confinement of optical fields to achieve efficient nonlinear-optical processes, enabling highly compact and functional light sources. However, compared with their free-space counterparts, integrated photonic devices typically offer much less flexibility in modifying phase-matching conditions after fabrication, which imposes severe limitations on the bandwidth of light that can be phase matched.

For example, consider an SHG process pumped by a tunable laser, where the goal is to create an SH light source with a tunable wavelength. In free-space optics, there are numerous techniques for continuously tuning the phase-matching condition to achieve maximum SHG efficiency at any pump wavelength. For example, one can simply rotate the angle of a nonlinear optical crystal to vary its phase-matching conditions via the material's birefringence~\cite{Hobden1967Birefringence, Giordmaine1962Birefringence} or use a ``fanned-out'' periodically poled structure that allows continuous tuning of the QPM grating period by translating the nonlinear crystal~\cite{Powers1998FanOut}. Additionally, temperature serves as an independent fine-tuning knob.

In contrast, in integrated photonics, the optical modes are fixed by the waveguides, and there are no mechanically movable components, in contrast to the angle and position adjustments available in free-space optics. These constraints remove many powerful tuning degrees of freedom. Although temperature can still be used as a fine-tuning parameter for SHG, the achievable tuning bandwidth is quite limited. A popular approach for addressing this limitation is adiabatic SHG~\cite{Margules2021-Adiabatic-SHG}, in which the period of the QPM grating is continuously chirped so that the SHG for a given pump wavelength is phase matched for at least one point along the waveguide. However, this approach inherently trades off the phase-matching bandwidth against the nonlinear conversion efficiency because a larger bandwidth requires a faster chirp in the QPM grating period, which in turn reduces the effective length over which SHG is phase matched. This efficiency--bandwidth tradeoff generally applies to other nonlinear optical processes as well, including sum-frequency, difference-frequency, and optical parametric generation.

The ability to program the QPM grating in a nonlinear waveguide offers a fundamental solution to this problem. As with the free-space fanned-out QPM grating, continuously tuning the QPM grating period on-chip avoids the efficiency--bandwidth tradeoff, allowing maximum conversion efficiency at any pump wavelength while preserving the unique benefits of integrated photonics, such as tight transverse field confinement and the absence of spatial diffraction.

As another concrete example, consider a tunable optical parametric oscillator (OPO). On-chip realization of such a tunable light source is nontrivial owing to the difficulty of tuning the phase-matching conditions over a wide range. 
Several innovative approaches have been explored to address this challenge. For example, Refs.~\cite{Lin2008TunableOPO, Sayson2017TunableOPO, Sayson2019TunableOPO} used sophisticated dispersion engineering to make the phase-matching condition for optical parametric oscillation highly sensitive to the pump wavelength; thus, even a small change in tuning the pump wavelength results in a large change in the signal wavelength, yielding a highly tunable OPO. However, this approach still requires a tunable pump laser. In Ref.~\cite{Pidgayko2023HeatTunableOPO}, a coupled resonator system was designed to achieve wide tunability of the phase-matching condition via temperature control, thereby obtaining a tunable OPO without the need for a tunable pump light source.

\begin{figure}[h]
    \centering
    \includegraphics[width=0.65\linewidth]{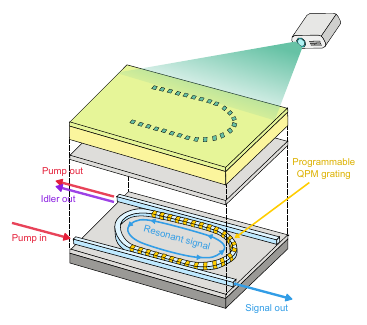}
    \caption{Illustration of a possible realization of a programmable OPO. Instead of using the slab waveguide geometry employed in the main text, we consider etching waveguides and resonator structures on the core layer. The resonator and the bus waveguides are designed to resonate with the signal light, whereas the idler and the pump lights couple out before a full round trip. On top of the core layer, we have a top cladding layer, a photoconductor layer, and a transparent electrode. The entire stack enables $\chi^{(2)}$ nonlinearity to be programmed on the resonator by programming illumination, yielding a programmable QPM grating. While it is difficult to perform an efficient QPM on curved waveguide structures with a conventional approach, the programmability of the device could enable in situ optimization to circumvent this problem. The color scheme in the figure follows that in Fig.~\ref{fig:figure1}.
    }
    \label{fig:opo}
\end{figure}

The ability to program the QPM grating on a nonlinear waveguide therefore provides a unique solution to these challenges. In Fig.~\ref{fig:opo}, we illustrate how a tunable OPO could be obtained with programmable periodic poling. In this OPO, the pump light with frequency $\omega_\text{p}$ generates light at the signal frequency $\omega_\text{s}$ and idler frequency $\omega_\text{i}$. We assume that only the signal light resonates with the resonator. Note that energy conservation, $\omega_\text{p}=\omega_\text{s}+\omega_\text{i}$, still leaves one degree of freedom for the signal wavelength, $0\leq\omega_\text{s}\leq\omega_\text{p}$, which is determined by the phase matching and resonant conditions.

For a fixed pump wavelength, we can vary the QPM grating period to adjust the phase-matching condition and maximize the gain for a desired signal wavelength $\omega_\text{s}$. The temperature remains an independent parameter, which we can use to ensure that the signal frequency $\omega_\text{s}$ is resonant with the cavity. This programmable OPO is expected to offer several unique advantages over existing approaches. First, it eliminates the need for sophisticated dispersion engineering to tune the phase-matching condition, thereby broadening the range of suitable materials and device geometries. Second, our approach enables more continuous tuning of the signal wavelength because it does not rely on coupled resonators, which can lead to forbidden wavelength windows due to mode crossings~\cite{Pidgayko2023HeatTunableOPO}. Finally, by engineering advanced QPM grating structures---as demonstrated for SHG in Sec.~\ref{sec:spectral}---we can simultaneously phase match various nonlinear-optical processes. This ability could provide a unique OPO that outputs multiple colors of light with tunable power ratios and relative phases.

\subsection{$100\,\%$-yield periodic poling}

Fabrication errors in QPM gratings are often the dominant limiting factor in the yield of nonlinear photonic devices. In practice, it is frequently necessary to fabricate tens of waveguides with slightly different QPM grating periods to achieve the correct phase-matching condition. Moreover, for devices with longer propagation distances, a one-dimensional sweep of the QPM grating period is insufficient to achieve maximum efficiency. Even small local variations in the film thickness can affect the phase-matching condition, necessitating careful local modifications of the QPM grating periods~\cite{Chen2023AdaptedPoling, Xin2024Polling}. This challenge is further compounded in more sophisticated QPM grating structures designed to engineer complex broadband nonlinear-optical functions, such as quantum pulse gates~\cite{Laura2023QPM}, pulse compression~\cite{Arbore1997PulseCompression,Arbore1997PulseCompressionExperiment}, Gouy phase compensation~\cite{Major2008Gouy}, and pulse shaping~\cite{Imeshev1998PulseShaping}.

The low yield not only restricts the range of possible applications but also makes it nearly impossible to build large-scale nonlinear photonic circuits, as the overall system yield decreases exponentially with the number of components. This severe exponential scaling underscores the need for a fundamentally different approach. Notably, a programmable nonlinear waveguide allows us to optimize the QPM grating period in situ on the basis of real-time experimental feedback, effectively achieving near-unit efficiency in periodic poling. This capability opens new pathways for obtaining large-scale nonlinear photonic systems that would be impractical under the conventional paradigm.

\section{Methods for inverse design}
\label{appendix:inverse-design}

In this section, we discuss methods for training a programmable nonlinear waveguide to achieve the desired function. Formally, the behavior of a photonic device, including a programmable nonlinear waveguide, can be described by an input--output mapping
\begin{align}
\mathbf{y}=f_\mathbf{w}(\mathbf{x}),
\end{align}
where the vectors $\mathbf{x}$ and $\mathbf{y}$ denote the optical fields of the input and output, respectively, and $f_\mathbf{w}$ is a generally nonlinear function parameterized by the device parameters $\mathbf{w}$. The performance of the device is characterized by a loss function
\begin{align}
    \mathcal{L}(\bar{\mathbf{y}},\mathbf{y}),
\end{align}
which quantifies the deviation of the output $\mathbf{y}$ from the desired output $\bar{\mathbf{y}}$ for a given input $\mathbf{x}$. When the desired input--output mapping is perfectly implemented, $\mathcal{L}=0$.

\emph{Inverse design} is a powerful paradigm in which one first specifies the desired device function by choosing an appropriate loss function $\mathcal{L}$ and then searches for the optimal device parameters $\mathbf{w}$ that minimize $\mathcal{L}$, thereby realizing the desired function. As demonstrated in various works in photonics~\cite{Lee2024InverseSCG, Molesky2018InverseDesign}, highly nontrivial functions can be obtained via inverse design, especially for devices with many degrees of freedom that are challenging to optimize manually.

In the spectral engineering experiment presented in Fig.~\ref{fig:spectral}(b) in Sec.~\ref{sec:spectral}, we implemented an inverse design by defining $\mathcal{L}$ as the distance between the desired and experimentally measured SH spectra. In this case, the parameters $\mathbf{w}$ were represented by 20 scalar values that parameterized the pattern of the programming illumination. As shown in Sec.~\ref{sec:spectral} and Appendix~\ref{appendix:spectral}, this approach enabled us to identify complex QPM grating structures that yielded highly nontrivial SH spectra.

A key distinction between our approach and conventional inverse design is that our approach is performed \emph{in situ} and in real time on the basis of experimental feedback---an advantage that is uniquely enabled by the programmability of the device. In contrast, conventional inverse design is performed \emph{in silico} (via digital simulations), and the resulting design is then implemented during fabrication. Below, we review these two paradigms of
inverse design, discussing their advantages and limitations.

\subsection{In silico inverse design}
If we have a digital model $\tilde{f}_\mathbf{w}$ that accurately approximates the behavior of a device $f_\mathbf{w}$, we can optimize $\mathbf{w}$ much more efficiently via gradient-based methods. Specifically, we apply the chain rule to calculate the gradient of the loss function with respect to the parameters, i.e.,
\begin{align}
    \frac{\partial\mathcal{L}(\bar{\mathbf{y}},\tilde{f}_\mathbf{w}(\mathbf{x}))}{\partial \mathbf{w}},
\end{align}
which indicates the direction in which the loss decreases. For example, a simple gradient descent algorithm uses the update rule
\begin{align}
    \mathbf{w} \mapsto \mathbf{w} - \alpha\,\frac{\partial\mathcal{L}(\bar{\mathbf{y}},\tilde{f}_\mathbf{w}(\mathbf{x}))}{\partial \mathbf{w}},
\end{align}
where $\alpha$ is the learning rate. More sophisticated algorithms may further improve the optimization performance~\cite{Kingma2015ADAM}.

Note that these optimizations are performed \textit{in silico}, i.e., entirely digitally. An optimal $\mathbf{w}$ found through this approach is expected to perform well when deployed on a real physical device if our digital model $\tilde{f}_\mathbf{w}$ accurately reflects reality. Conversely, any discrepancy between the digital model and reality results in suboptimal device performance. In this sense, the paradigm of in silico inverse design faces inherent challenges in handling fabrication imperfections, miscalibration of the experimental setup, and environmental drift.

\subsection{Hybrid in situ--in silico inverse design} 
In the paradigm of in situ inverse design, we optimize the device parameters on the basis of real-time feedback from the actual device. The simplest in situ inverse design method employed in this work is random optimization~\cite{Matyas1965random}. In each iteration, we update the parameters as follows:
\begin{align}
    \mathbf{w} \mapsto \mathbf{w} + \delta\mathbf{w},
\end{align}
where $\delta\mathbf{w}$ is randomly generated; we then experimentally measure the loss function $\mathcal{L}$. If the update reduces the loss relative to the previous value, we accept the new parameters; otherwise, we retain the previous parameters and proceed to the next iteration. A key advantage of this approach is that it does not require any prior knowledge of $f_\mathbf{w}$, which increases the robustness of the optimization. However, the optimization process tends to slow as the dimensionality of $\mathbf{w}$ increases.

Physics-aware training (PAT) is a hybrid in situ--in silico training method that enables efficient gradient-based optimization and leverages the robustness provided by experimental feedback~\cite{Wright2022PNN}. In PAT, similar to in silico inverse design, we construct a digital differential numerical model $\tilde{f}_\mathbf{w}$ that mimics the input--output map of the system. However, instead of using the digital model for the forward path, we use the actual physical system to obtain an experimental result and calculate the error vector $\frac{\partial \mathcal{L}}{\partial\mathbf{y}}$. The digital model $\tilde{f}_\mathbf{w}$ is then used to backpropagate the error and compute the gradient $\frac{\partial \mathcal{L}}{\partial\mathbf{w}}$.

\end{document}